\documentstyle[11pt,epsf]{article}
\input psfig.sty

\newcommand{\be}{\begin{equation}}
\newcommand{\ee}{\end{equation}}
\newcommand{\bea}{\begin{eqnarray}}
\newcommand{\eea}{\end{eqnarray}}
\newcommand{\ba}{\begin{eqnarray}}
\newcommand{\ea}{\end{eqnarray}}
\def\cb{{\sc cobe--dmr\,}}
\def\ah{{\bf H}^3}
\def\es{{\bf S}^3}
\def\eu{{\bf E}^3}
\def\cq{C(\hat n, \hat n^\prime)}
\def\spc{{\xi_{\Phi}}}
\def\vx{{\bf \vec x}}
\def\vxp{{\bf {\vec x}^\prime}}
\def\pow{{{\cal P}_\Phi}}
\def\hn{\hat n}
\def\hnp{\hat n^\prime}
\def\gta{\mathrel{
\hbox to 0pt{\lower
3pt\hbox{$\mathchar"218$}\hss}\raise 2.0pt\hbox{$\mathchar"13E$}}}

\def\lta{
\mathrel{\hbox to
0pt{\lower 3pt\hbox{$\mathchar"218$}\hss}\raise 2.0pt\hbox{$\mathchar"13C$}}}

\renewcommand{\theequation}{\arabic{section}.\arabic{equation}}

\begin{document}

\begin{titlepage}

\begin{flushright}
\end{flushright}

\vspace{.5in}

\begin{center}
\huge
{\bf Topology and the Cosmic Microwave Background}

\vspace{.5in}

\large{
Janna Levin}

\normalsize
\vspace{.2in}

{\em DAMTP, CMS, Cambridge University,  \\ 
Wilberforce Rd, Cambridge CB3 0WA, U.K.\\
j.levin@damtp.cam.ac.uk \\
and\\
Theoretical Physics Group, Imperial College, London}

\vspace{.5in}

\begin{abstract}

Nature abhors an infinity.  The limits of general 
relativity are often signaled by infinities:
infinite curvature as in the center of a black hole, the infinite energy
of the singular big bang.  We might be inclined to add an infinite universe 
to the list of intolerable infinities.  Theories that move beyond general
relativity naturally treat space as finite.  In this review we discuss
the mathematics of finite spaces and our aspirations to observe the finite
extent of the universe in the cosmic background radiation.

\end{abstract}

\end{center}

\end{titlepage}

\tableofcontents

\newpage

\section{Introduction}
\label{introduction}
\setcounter{equation}{0}

\def\theequation{\thesection.\arabic{equation}}

Is the universe infinite?  The universe had a birth and will
have a death, whether from old age
or a more catastrophic demise through
a big crunch.  With the advent 
of general relativity,  cosmology has accepted space as an
evolving geometry.  Although the universe is believed to
begin and end a dynamical life, we continue to ascribe 
a property to the universe that we would never assign to any other
natural object, and that is the property of being spatially infinite.

The universe is home to a plethora of structures 
all of which must be finite.  It would be
untenable to suggest that any physical structure be infinite, yet 
cosmologists often cavalierly assume space itself is infinite.
Yet an infinite universe means an infinite number of stars and 
galaxies and people.  By the 
law of probability an infinite universe
accommodates an infinite number of events each happening
an infinite number of times \cite{{lum},{css_cqg}}.  Somewhere else
in the cosmos, you are there.  In fact there are an infinite number
of  you littering space.
An infinite number of us living the same lives and
an infinite number living slightly different lives and all range of
variants.
While this concept of infinity is manageable to some, it
reeks of something pathological to others.
Is the universe a surreal object, unlike any of its inhabitants, or is it 
real and physical and limited and finite?

If the universe is finite, then light will wrap around the finite 
space decorating the sky with ghost images, sometimes called clone images.
In principle we should be able to see copies of ourselves at different
ages in different directions.  A kaleidescopic version of `This is your
life'.  In practice, the universe does not appear to be
so small.  The universe,
if finite, must at least be big.  Big enough to house clusters of galaxies
and perhaps even bigger than the entire light travel time since the big
bang.  Therein lies our question.  If the universe is finite, how small
is it, and how do we measure its shape?

The question of the infinite extent of space is often answered
dynamically by Einstein's theory. 
If the 
amount of matter and energy 
is underdense then the universe will expand forever, 
if overdense then the universe will ultimately
recollapse to a big crunch, and if critical then
the universe will expand forever.
To each of these dynamical possibilities corresponds a geometry 
of spacetime:
the underdense cosmos is negatively curved and infinite,
the overdense cosmos is positively curved and finite 
and the critical cosmos is flat and infinite.

Each of these three geometries can support topologies with finite volumes
without altering the dynamics or the curvature.  Multiconnected
topologies are often overlooked in favor of the simply connected
possibilities.  This oversight is encouraged
by a limitation of Einstein's theory which elevates gravity to 
a theory of geometry but does not provide a theory of topology.
If unification of gravity with the other forces ever succeeds,
topology will inevitably be integrated as a predictive feature of 
any cosmology.  The topology of additional compact dimensions is
already understood to be a crucial feature in superstring theories
and, in general, 
discretization of space into topologically finite bundles is a natural
consequence of quantization \cite{lum}.  No one is good enough at 
quantum gravity or string theory 
to make a prediction for the geometry and shape of 
the whole universe.  Still the requirement of extra topologically 
compact dimensions has become a commonplace notion in fundamental physics
even if it seems esoteric in astronomy.
The first suggestion of compact extra dimensions began with Kaluza back
in Einstein's day and have been dressed up in the modern guise of 
superstring theories since.  Even if we can never see the extent of the
large three dimensions, we may be able to look for small extra dimensions
to find clues about the geometry of the cosmos.

In the absence of a predictive theory we can 
still explore the mathematical possibilities for a finite universe.
This review covers
topologies consistent with
the standard cosmological 
geometries 
and  
our aspirations to observe
the global extent of space in patterns in the oldest historical 
relic accessible, the cosmic microwave background (CMB).
Methods to determine topology from the CMB range from the geometric such
as the search for circles
in the sky \cite{{css1},{css2},{css_cqg}}and topological pattern formation 
\cite{{conf},{pat},{imogen}}
to the brute force statistical methods
such as the method of images \cite{{bpsI},{bpsII}}. These methods will
be explained in turn.
This review is intended both for mathematicians interested in cosmology
and cosmologists interested in the mathematics of topology.
Not much will be taken for granted and we will take the time
to explain even the standard cosmology.

Topology
can be observed with the CMB or with the distribution of collapsed objects
such as quasars or clusters of galaxies.
There are reviews on galactic methods
\cite{{galreviews1},{galreviews2}}.  
To keep the scope of this article manageable we
will limit our discussion to searches for artifacts of topology 
in the CMB and
defer a discussion of galactic methods to the reviews 
in Ref.\ \cite{{galreviews1},{galreviews2},{roukreview}}.
(see also Refs. \cite{{ghosts1},{ghosts2}}.)
Also, much of the
history on topology 
can be found in an earlier review \cite{lum}.
Other accessible reviews can be found in Refs.
\cite{{thursweeks},{glennpop}}.
Collections of papers for two international workshops can be found in
\cite{cqg} and \cite{ctp98}.
The observational section in this review will focus on recent
methods only, starting where Ref.\ \cite{lum} left off.

\newpage

\section{Topology and Geometry}
\label{mathsection}

\def\theequation{\thesection.\arabic{equation}}
\label{topintro}

Cosmic topology aims 
to deduce the global shape of the universe
by experimentally observing
a pattern in the distribution of astronomical objects.
To do so, it is helpful to first understand the possible topologies from
a purely mathematical perspective.  The isotropy of the cosmic
microwave background radiation implies that the curvature of the observable
universe is very nearly constant, so in the present article we consider
only manifolds of constant curvature.  The homogeneity of the observable
universe does not, of course, exclude the possibility that the curvature
of space varies enormously on a global scale beyond our view, but because
that hypothesis is presently untestable we do not pursue it here.
By the same reasoning, we are most interested in multiconnected spaces that are
small enough to witness, although 
theories beyond general relativity, such as string
theory, may help us to push beyond those limitations 
(see \S \ref{extradsect}).
There is a previous Physics Reports on cosmic topology which provides
an excellent overview of the mathematical principles
\cite{lum}.  The reader is encouraged to consult that article
as well as Refs.\ \cite{{wolf},{ellis}}.  We will not repeat these
detailed reviews but do survey the same topological methods for 
completeness and extend the discussion to include some additional
modern methods.

The global shape of any space, including the ultimate outerspace,
is characterized by a {\it geometry} and a {\it topology}.
The term {\it geometry}, as used by mathematicians,
describes the local curves while {\it topology} describes global features which
are unaltered by smooth deformations.  
A continuous transformation is known as a {\it homeomorphism}, a smooth
continuous and invertible map which deforms one manifold into another
without cutting or tearing.  General relativity is not invariant
under homeomorphisms but is invariant under 
{\it diffeomorphisms}, which amount to a change of coordinates.
It is this covariance of classical gravity which underlies
the principles of general relativity.  Covariance ensures
that all observers experience the same laws of physics regardless of 
the worldline along which they travel and therefore regardless of the
coordinate system within which they interpret the world.  
Relativity does not invoke such a principle with respect to topology.
For this reason we do not have a fundamental principle to guide us when 
contemplating the topology of the universe.  We might hope that
such a principle or at least a prediction might precipitate
from a complete theory of gravity
beyond classical.  In the meantime, we consider any mathematically
allowed topology, restricting ourselves to the constant curvature
spaces as indicated by observational cosmology.

\subsection{Overview of Geometry}
\label{ovgeom}

\begin{figure}
\centerline{\psfig{file=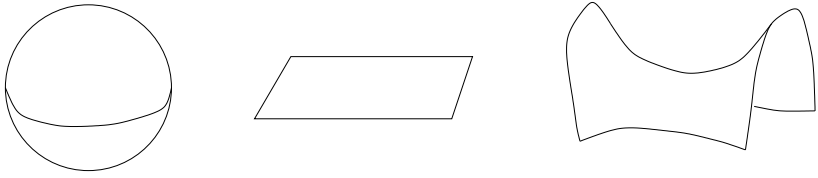,angle=-90,width=4.25in}}
\vskip 5truept
\caption{A homogeneous isotropic space may be spherical (positive curvature),
flat (zero curvature), or hyperbolic (negative curvature).}
\label{N.2.1.1}
\end{figure}

A cosmological principle asserts that the Earth should not be
special in location or perspective.  Cosmological 
observations on the largest scales indicate we live in a 
space with the symmetries of {\it homogeneity}, space is the same in 
all directions, and {\it isotropy}, space looks the same in all directions.
Homogeneous and isotropic manifolds have constant curvature.  The
$n$-dimensional geometries of constant curvature are the everywhere
spherical 
${\bf S}^n$,
flat ${\bf E}^n$, and hyperbolic 
${\bf H}^n$ geometries (fig.\ \ref{N.2.1.1}).  
One natural consequence of curved geometries worth remembering is that
the sum of the interior angles of a triangle are greater than 
$\pi$ on ${\bf S}^n$, exactly $\pi$ on ${\bf E}^n$, and less than $\pi$
on ${\bf H}^n$.  

In $2D$, all Riemannian surfaces are homeomorphic to the
three constant curvature geometries ${\bf S}^2, {\bf E}^2$,
and ${\bf H}^2$.
The infinite hyperbolic plane ${\bf H}^2$ cannot be drawn properly
in $3D$ and so is difficult to visualize.  It can be drawn schematically
as a surface which everywhere has the curvature of a saddle.  A nice
way to treat ${\bf H}^2$ is as a pseudosphere embedded in $(2+1)$-Minkowski
space.  The pseudosphere has the equation 
$-z^2+x^2+y^2=-1$.  The full isometry group of the hyperboloid is
$PSL(2,{\bf R})\equiv SL(2,{\bf R})/Z_2$ with 
$SL(2,{\bf R})$ the special Lorentz group of real 
$2\times 2$ matrices with unit determinant.  The embedding in Minkowski
space makes the symmetry group of 
${\bf H}^2$ more obvious.

The sphere ${\bf S}^2$ can be embedded in 
Euclidean 3-space
as a surface with radius $x^2+y^2+z^2=1$.  The isometry group for the sphere
is $O(3)$, all orthogonal $3\times 3$ matrices with the absolute value
of the determinant equal to one.

The plane ${\bf E}^2$ is the infinite surface $(x,y)$ with
the Minkowski time coordinate $z$ fixed at unity
(see fig.\ \ref{embed}).  The Euclidean plane has the
full Galilean group of isometries:  translations, reflections, rotations
and glide reflections.
A glide reflection involves a translation with 
a reflection along the line parallel to the translation.

In $3D$, all manifolds are not homeomorphic to the constant
curvature manifolds.  Instead there are $8$ homogeneous geometries some
of which are anisotropic.  These $8$ geometries were classified by Thurston
in the mathematical literature \cite{{thurclass},{Thurston}}.  Cosmologists
are more familiar with the equivalent classification of 
Bianchi into $8$  homogeneous manifolds \cite{Bianchi}.
Out of respect for the cosmological principle we will consider the
fully homogeneous and isotropic spaces of constant curvature 
${\es},{\eu}$, and ${\ah}$.  Similar to $2D$, ${\ah}$ can be embedded 
as a pseudosphere in $(3+1)$-Minkowski space, $\es$ as the sphere and
$\eu $ as the plane at fixed Minkowski time.
This isometry group 
of $\es$ is $SO(4)$.
The isometry group of
$\eu $ 
is the product of translations and
the special orthogonal $3\times 3$ matrices.  
The isometry group of $\ah$ is
$PSL(2,{\bf C})\equiv SL(2,{\bf C})/Z_2$.

\subsection{Overview of Topology}
\label{ovtop}

Topology is the pursuit
of equivalence classes of spaces; that is, the classification of 
homeomorphic manifolds.  To a topologist, all surfaces with one
handle for instance are equivalent.  Thus the doughnut and the coffee cup,
in addition to providing a dubious breakfast, provide the quintessential
example of equivalent topologies, despite their obviously different
geometries (i.e. curves).

\begin{figure}
\centerline{\psfig{file=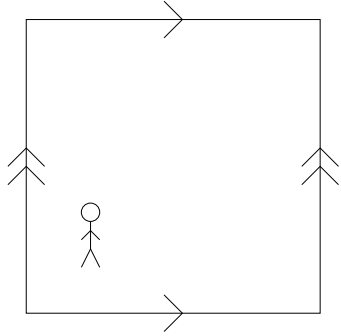,angle=-90,width=1.75in}}
\vskip 5truept
\caption{The flat 2-dimensional torus is defined by abstractly
gluing opposite edges of a square.}
\label{N.1.1.1}
\end{figure}

\begin{figure}
\centerline{\psfig{file=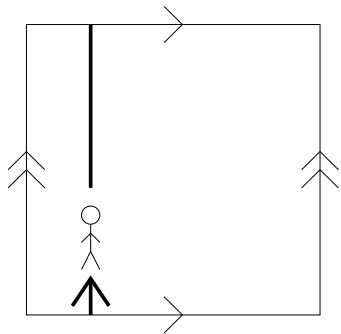,angle=-90,width=1.75in}}
\vskip 5truept
\caption{An inhabitant of the 2-torus looks forward and sees
herself from behind.}
\label{N.1.1.2}
\end{figure}

\begin{figure}
\centerline{\psfig{file=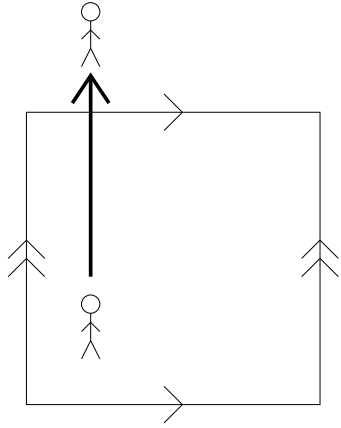,angle=-90,width=1.75in}}
\vskip 5truept
\caption{The inhabitant of the 2-torus has the illusion of seeing
a copy of herself.}
\label{N.1.1.3}
\end{figure}

\begin{figure}
\centerline{\psfig{file=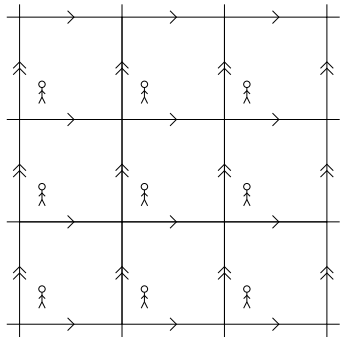,angle=-90,width=1.75in}}
\vskip 5truept
\caption{Indeed, the inhabitant of the 2-torus sees a lattice
of images of herself.}
\label{N.1.1.4}
\end{figure}

The doughnut and coffee cup are both manifestations of
the 2-dimensional torus (fig.\ \ref{N.1.1.1}), 
a prototypical multiply connected
space.  The torus has a finite area, yet has no boundary.  When an inhabitant
of the torus looks forward (fig.\ \ref{N.1.1.2}), her line of sight wraps around
the space and she sees herself from behind.  She has the illusion of seeing
another copy of herself directly in front of her (fig.\ \ref{N.1.1.3}).  
Indeed she
also sees herself when she looks to the side, or along a 45 degree line,
or along any line of rational slope.  She thus has the illusion of seeing
infinitely many copies of herself (fig.\ \ref{N.1.1.4}).  In other words, even
though
her universe is finite, she has the illusion of living in an infinite
universe containing an infinite lattice of repeating objects.

\begin{figure}
\centerline{\psfig{file=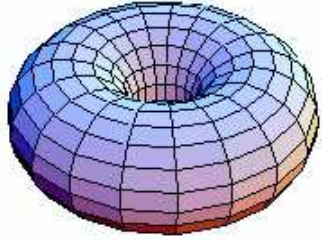,width=1.75in}}
\vskip 5truept
\caption{The torus of revolution has the same topology as the flat torus
but different curvature.}
\label{torus}
\end{figure}

As described below, all multiconnected surfaces can be built by gluing
the edges of a fundamental polygon.  The torus is built by identifying
opposite edges of a parallelogram.
If we begin
with a flat rectangle, we can bend this flat 
sheet into three-dimensions to glue the top to the bottom and the left
edge to the right.  In doing so we have made a torus of revolution
as in fig.\ \ref{torus}
which is neither flat nor constant curvature.  The curvature
of the torus clearly varies over the surface.  This bending of the
torus is an artifact of embedding the surface in $3$-dimensions.  A truly
flat torus, better known as $T^2=S^1\times S^1$, is content to live in 
$2D$ with no such bending and projecting into $3D$.  The flat torus
is akin to the video game with periodic boundary conditions where
the left edge is identified with the right edge so that a flat explorer
could stick their hand out the right edge only to have it appear poking
out of the left edge.  The edge in fact is as meaningless to the inhabitant
of a flat torus as it is to an inhabitant of the torus of revolution.
$T^2$ is truly flat and so has a different geometry from the torus of 
revolution although they are topologically equivalent, being characterized
by the one handle.
The genus $g$ of the manifold is a topological invariant which counts 
the number of handles and holes.  The torus has genus $g=1$.

\begin{figure}
\centerline{\psfig{file=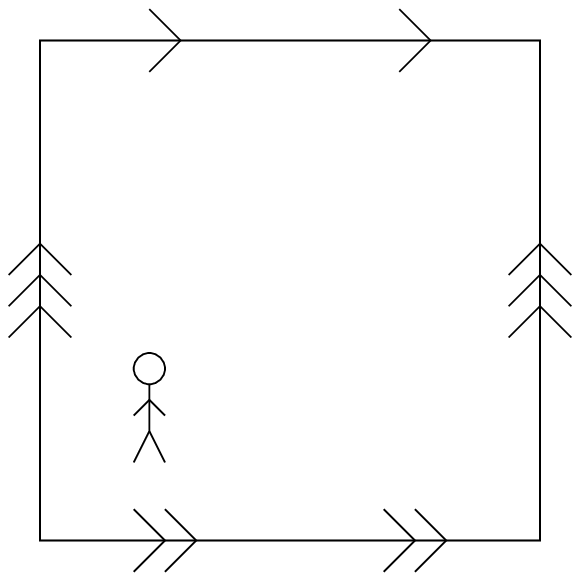,angle=-90,width=1.75in}\quad\quad
\psfig{file=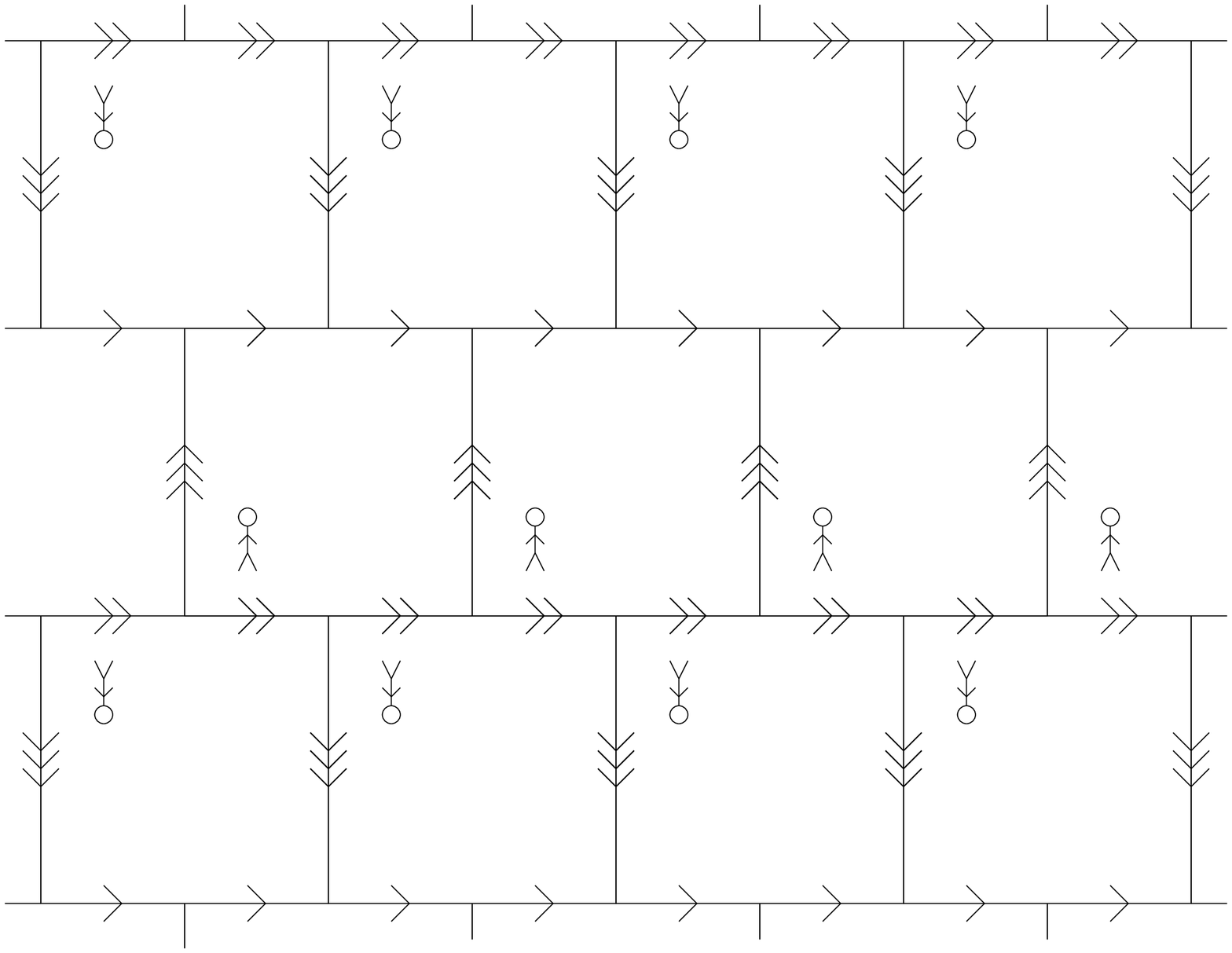,angle=-90,width=2.5in}}
\vskip 5truept
\caption{Left: The Klein bottle is similar to the torus, only the square's
top and bottom edges are each glued to themselves, not to each other.
The lattice of images in the Klein bottle is different from
that in the torus because it contains glide reflections as well as
translations.}
\label{N.1.1.5}
\end{figure}

\begin{figure}
\centerline{\psfig{file=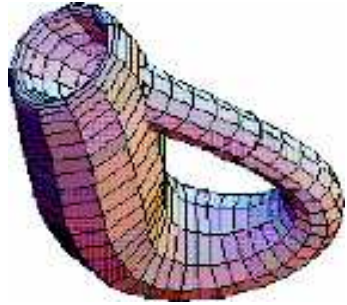,width=1.75in}}
\vskip 5truept
\caption{A surface with the topology of the Klein bottle.}
\label{klein}
\end{figure}

The 2-dimensional Klein bottle (fig.\ \ref{N.1.1.5}) 
is similar to the torus, only
now the top and bottom of the 
rectangle are glued not to each other, but each to itself with
a shift of half a unit.  
The traditional way to make a Klein bottle is
to start with the
rectangle, glue top to bottom but glue the left to the right after a 
rotation through $\pi$.
In the present article we use a different (but equivalent) 
construction (namely gluing the top and bottom each to itself with
a shift of half a unit)
because it will simplify the construction of flat 3-manifolds
in section \ref{threed}.
As in the torus an inhabitant has the illusion
of seeing an infinite lattice of objects, only now the structure of the
lattice is different:  it contains glide reflections as well as
translations.  
The truly flat Klein bottle is topologically equivalent to the
bottle immersed in $3D$ and pictured in fig.\ \ref{klein}.  The embedding again
leads to a curved surface and in this case to a self-intersection of the
surface.  The flat Klein bottle, content to live in $2D$, has no such curvature
or self-intersection.

\begin{figure}
\centerline{\psfig{file=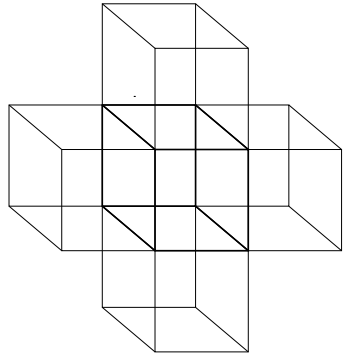,width=1.75in}}
\vskip 5truept
\caption{3-dimensional lattice of cubes.}
\label{cube}
\end{figure}

These examples generalize readily to three dimensions.  A cube (fig.\
\ref{cube}) with opposite
faces glued is a 3-dimensional torus, or 3-torus.
Its inhabitant has the illusion of seeing an infinite 3-dimensional
lattice of images of himself and of every other other object in the space.
Varying the gluings of the faces 
varies the structure of the lattice of images.
Cosmic topology aims to observe these distinctive lattice-like patterns.

\subsection{Tools of Topology}
\label{tools}

\begin{figure}
\centerline{\psfig{file=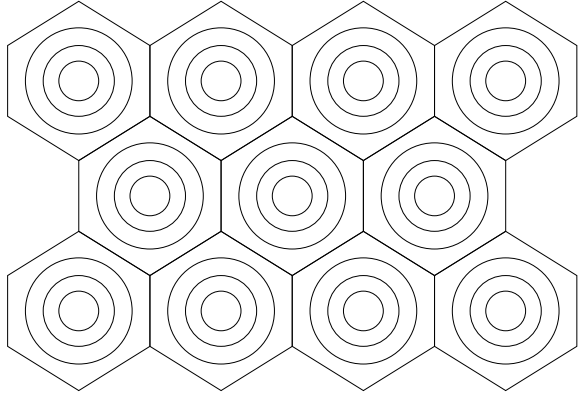,angle=-90,width=3.in}}
\vskip 5truept
\caption{Start inflating a balloon, with its center point fixed, and let
it expand until it fills the entire space.  The resulting polygon
(in two dimensions) or polyhedron (in three dimensions) is called
a Dirichlet domain.  In some manifolds the shape of the Dirichlet domain
depends on the choice of the center point;  in other manifolds it does not.
}
\label{N.1.2.1}
\end{figure}

In this preliminary discussion several important concepts have already
begun to emerge.  The first is the fundamental domain.
A {\it fundamental domain} is a polygon or polyhedron from which a manifold
may be constructed.  For example, fig.\ \ref{N.1.1.1} shows a square fundamental
domain for a 2-torus, and fig.\ \ref{cube} 
shows a cube fundamental domain
for a 3-torus.
A {\it Dirichlet domain} is a special type of fundamental domain.
To construct a Dirichlet domain, pick an arbitrary point in the manifold
of interest to serve as a {\it basepoint}.  Start inflating a balloon
with center at the basepoint (fig.\ \ref{N.1.2.1}) and let the balloon expand
uniformly.  Eventually different parts of the balloon will bump into
each other.  When this happens, let them press flat against each other,
forming a flat (totally geodesic) boundary wall.  Eventually the balloon
will fill the whole manifold, at which point it will have the form
of a polygon in two dimensions or polyhedron in three dimensions
whose faces are the aforementioned boundary walls.  This polyhedron is
the Dirichlet domain.  Gluing corresponding faces recovers the original
manifold.  
Notice that the fundamental rectangular domain for the Klein bottle
when glued after a shift of half a unit is not Dirichlet, but when glued
after a rotation through $\pi$ is Dirichlet.
In some manifolds the Dirichlet domain depends on the choice
of basepoint while in others it does not -- more on this later.

By identifying the edges of the fundamental domain, a multiconnected space
is constructed.
Figures \ref{N.1.1.4} and \ref{N.1.1.5}
show how an inhabitant of a finite multiply
connected space may have the illusion of living in an infinite simply
connected space.  This apparent space is called the {\it universal
cover} and can be thought of as the simply connected manifold with
the constant curvature geometry.  The fundamental domain is cut from this cloth
and glued together while preserving the geometry of the universal cover.
So the flat torus preserves the geometry of the universal cover while the
torus of revolution does not.

The group of {\it covering transformations} is the group of symmetries
of the universal cover.  That is, it is the group of motions of the universal
cover that take images of a given object to other images of the same object.
For example, in the case of a torus the group of covering transformations is
the obvious lattice of translations.  In the case of the Klein bottle the
group
of covering transformations contains glide reflections as well as
translations.

\begin{figure}
\centerline{\psfig{file=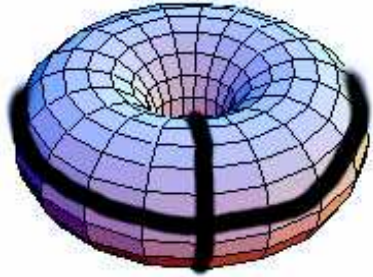,width=2.in}}
\vskip 5truept
\caption{Two homotopically distinct loops around the torus of revolution.}
\label{torusloops}
\end{figure}

In two-dimensions all topologies have been completely 
classified.  We give that classification in \S \ref{twod}.
However, in
three-dimensions the classification of hyperbolic manifolds remains 
incomplete although huge advances were made last century.
There is a collection of topological invariants which can be used
to recognize the equivalence classes of spaces.  
We have already encountered the {\it genus} of the manifold which
counts handles.
A group structure can 
be used to capture handles and thereby detect
the connectedness of the space.  The most important
of these is the loop group also known as the
{\it fundamental group} or the first {\it homotopy group}.  
All loops which can be smoothly deformed into each other
are called {\it homotopic}.
A loop 
drawn on the surface of a sphere can be contracted to a point.  In 
fact all loops drawn on the surface of a sphere can be contracted a 
point.  Since all loops are homotopic, the sphere is 
{\it simply connected}
and in fact does not have any handles.  If however we draw a loop
around the torus of revolution as in fig.\ \ref{torusloops}, 
there is no way to smoothly
contract that loop to a point, similarly for a loop drawn around the
circumference of the toroid
also shown in fig.\ \ref{torusloops} .  
The torus is therefore {\it multiply connected}.
The group of loops
is the fundamental group $\pi_1({\cal M})$ and is a topological invariant.
In more than $2D$, the fundamental group is not sufficient to determine
the connectedness of the manifold.  Nonetheless, the fundamental group
is a powerful tool.
P\" oincar\'e has conjectured that
if the fundamental group is trivial, then
the $n$-manifold is topologically equivalent to $S^n$.

A closely related structure is the
{\it holonomy group}. 
The holonomy group can be thought of as the set of instructions
for identifying the faces of the fundamental domain.  
The holonomies act discretely and without fixed point.
The holonomy group
is 
a subset of the full
isometry group of the covering space.  
The full symmetries of the universal cover are broken by
the identifications and it is customary to represent
the compact manifold
as a quotient space $G/\Gamma $ where $G$ is the
isometry group of the universal cover and $\Gamma $ is the holonomy 
group.  
The holonomy group is of extreme utility in the game of cosmic
topology.  The group provides the set of rules for distributing
images throughout the universe and thereby provides
the mathematical means by which to determine the geometric distribution
of astronomical images in the cosmos.  The group is synonymous with
the boundary conditions for the manifold and as we will see this is
critical for building a prediction for
the spectrum of the cosmic microwave background.

There can be many different representations for the holonomy group.
There is often a simplest presentation $\{\gamma_1...\gamma_{n};r\}$ 
where $\gamma_i \in \Gamma$
are elements of the group.  Implicitly 
$r=1$ following the semicolon and 
is a list of all of the relations among the group elements.
The most natural presentation for cosmology 
has a group element, and so a letter, 
for every identification rule.
Such a presentation
has  many
letters and many relations.

The holonomies 
can map the end-point of an orbit to the start of that orbit so
that it becomes periodic in the compact space.
In other words, a periodic orbit has
$x_{\rm end}=\gamma x_{\rm start}$ where $\gamma $ can be a composite
word
$\gamma=\prod^n_i\gamma_{k_i}$.  
Each word has a corresponding periodic orbit.  

\begin{figure}
\centerline{\psfig{file=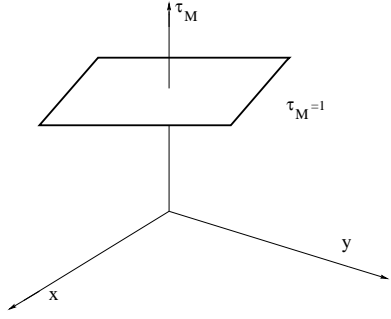,width=2.in}}
\vskip 5truept
\caption{The embedding of ${\bf E}^2$ into $(2+1)$-Minkowski
space.  The manifold appears
as an infinite sheet fixed at $\tau_M=1$.}
\label{embed}
\end{figure}

\begin{figure}
\centerline{\psfig{file=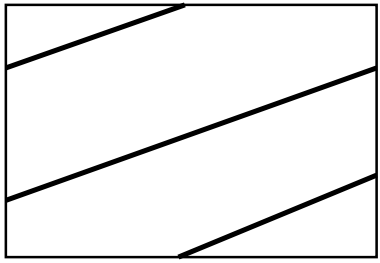,width=2.in}}
\vskip 5truept
\caption{A 
periodic orbit on the torus
which corresponds to 
$x_{\rm end}=T_yT_x^2x_{\rm start}$.
}
\label{fund}
\end{figure}

As an explicit example we return to the canonical torus.
We can make use of the embedding of ${\bf E}^2$ in a $(2+1)$-dimensional 
space.
Specifically, the $2$-dimensional coordinate
is replaced with the $(2+1)$-dimensional coordinate
	\be
	x^a=\pmatrix{\tau_M \cr x \cr y}
	\ee
where $\tau_M$ is fixed at unity as in fig. \ref{embed}.  
In this coordinate system
the holonomy group for the torus is the set of 
generators, 
	\be
	T_x=
	\pmatrix{1 & 0 & 0  \cr
	  L_x & 1 & 0\cr
	0 & 0 & 1 },
\quad \quad
	T_y=
	\pmatrix{1 & 0 & 0  \cr
	 0&1 & 0 \cr
	L_y &  0  & 1}.
	\ee
The boundary condition is then $x^a\rightarrow T^a_b x^b$ 
so that identification
of the left edge to the right edge results in
	\be
	\pmatrix{\tau_M \cr x\cr y\cr}
	\rightarrow 	\pmatrix{1 & 0 & 0  \cr
	  L_x & 1 & 0\cr
	0 & 0 & 1 }	\pmatrix{\tau_M \cr x\cr y\cr}=
	\pmatrix{\tau_M \cr x+L_x\cr y\cr}
	\label{leftright}
	\ee
The holonomy group for the torus in this presentation is
$\{T_x,T_y;T_xT_yT_x^{-1}T_y^{-1}\}$.  The relation
$T_xT_yT_x^{-1}T_y^{-1}=1$ is simply the commutivity of the
generators and prunes the number of unique periodic orbits
and therefore the number of clone or ghost images.  Because
the generators commute, the periodic orbits (homotopic loops)
can be counted symbolically
in terms of the number of windings taken around $x$ and around $y$,
$(m_x,m_y)$.  
The number of loops and therefore the number of
clone images grows as a polynomial with length.
For example, consider the periodic orbit of fig.\ \ref{fund} 
which has $x_{\rm end}=T_yT_x^2x_{\rm start}$
and so winding number $(m_x=2,m_y=1)$.

The non-orientable Klein bottle
first twists the $z$-faces through $\pi$ before 
identification.  The generators of $\Gamma$ for the twisted space
are $T_x,R(\pi)T_y$ with $R(\theta)$ the rotation matrix:
	\be
	R(\theta)=\pmatrix{1 & 0 & 0 \cr
	0 & \cos \theta & \sin \theta  \cr
	0 & -\sin \theta & \cos \theta}
	\  \ .
	\ee
All of the multiconnected, flat topologies can be built out of a
combination of rotations and these translations.

In hyperbolic space, the generators cannot easily be written in functional
form but can nonetheless be found with numerical entries in the matrix.
The program
SnapPea provides the generators explicitly in this embedded coordinate
system 
for a census of known hyperbolic manifolds and orbifolds
\cite{snappea}.
(Orbifolds can have singular points.)
The number of periodic orbits and therefore of ghost images grows 
exponentially in a compact hyperbolic manifold.  The exponential proliferation
of images is a manifestation of the chaotic flows supported on compact 
hyperbolic spaces and discussed in \S \ref{chaossect}.

\subsection{Tessellations}
\label{tessellations}

The modest tools we have introduced go a long way in constructing
cosmological models.  To summarize we will characterize a manifold by:\par
	$\bullet $  The {\bf fundamental domain}
-- the shape of space around a given basepoint.\par
	$\bullet $  The {\bf universal cover} -- the simply connected
manifold with the same constant curvature geometry.\par
	$\bullet $  The {\bf holonomy group} -- 
face pairing isometries, i.e.,
the rules for identifying the faces
of the fundamental domain.\par

Armed with these tools, we can begin to visualize life in a finite universe.
An important visualization technique comes from a tessellation of the
universal cover.  Consider the torus again.
The flow of an observer or of light can be followed through the compact
flat space by conscientiously respecting the boundary conditions and drawing
the motion within the fundamental polygon as in fig.\ \ref{fund}.  
The Dirichlet domain is the set of points
	\be \{y\in {\cal M}\}\nonumber
	\ee
such that 
	\be d(y,x)\le d(y,\gamma x), \forall
	\gamma \in \Gamma \ \  .\label{fd}\ee
In $T^2$, light
travels along straight lines as do all inertial observers.
Another faithful representation of $T^2$ which does not distort the everywhere
flat nature of the geometry is provided by a tiling picture.  
Beginning with the fundamental polygon, the periodic boundary condition
which identifies the left edge of the rectangle to the right edge for
instance can be represented by gluing an identical copy of the fundamental
polygon left edge to right edge.  
This amounts to moving an entire copy of the fundamental domain with
$T_x(FD)$.
Similarly identical copies can be
glued top to bottom 
$T_y(FD)$
ad infinitum until the entire infinite flat plane
is tiled with identical copies of the rectangle.  
In general, the generators of the holonomy group can be thought of as defining
an alphabet of $2n$ letters, for $n$ generators and their $n$
inverses.
The alphabet will vary with different representations, as will the grammar.
Usually, though not always, the longer the word
built out of the letters $\gamma_i\in \Gamma$,
the more distant the corresponding image or tiling
in the tessellation picture.  
Figures \ref{N.1.1.4}, \ref{N.1.1.5}, and \ref{N.1.2.1} 
demonstrate tessellations.

These copies are truly
identical.  If someone stands in the center of one rectangle
he will see images of himself standing at the center of every other
rectangle.  If he moves to the left he will see every copy of himself
move left.  Because of the finite light travel time, it will take longer
to see the more distant images movings.  The light from these more distant
images takes a longer path around the compact space before
reaching the observer, the source of the very image. 
This creates the pattern of images already described so that the universe
is an intricate hall of mirrors.

Notice that the tiling completely fills the plane without overlaps or gaps.
We can also imagine tiling the plane with a perfect square or even a 
parallelogram without overlaps or gaps.  A flat torus can therefore be 
made by identifying the edges of these other polygons.  All 
have the same topology and the same curvature but some have different
metrical features in the sense that the rectangle is truly longer than it is
wide.  An observer could line rulers up in space and unambiguously
measure that their space was longer than it was wide while an observer
in the perfect square would know that their space was equally long and wide.
The torus is thus globally anisotropic
even though the geometry
of the metric is still locally isotropic.  
The torus is still globally homogeneous.  
The center could be moved arbitrarily and the
fundamental polyhedron redrawn around that center to give an equally valid
tiling. Most topologies globally break both homogeneity and isotropy.

The holonomies provide the list of rules for identifying the edges of the 
polygon.  So the Klein bottle, while it can begin with the same fundamental
polygon as the torus and has the same universal cover, it has a different set 
of rules for identifying the edges and therefore a different holonomy group.
The tiling will be correspondingly different with identical copies 
glued top edge to bottom edge but left edge to right only after
rotating the tile by $\pi$.  So the tessellation creates the reflected 
and translated images specific to the Klein bottle (fig.\ \ref{N.1.1.5}).

\begin{figure}
\centerline{\psfig{file=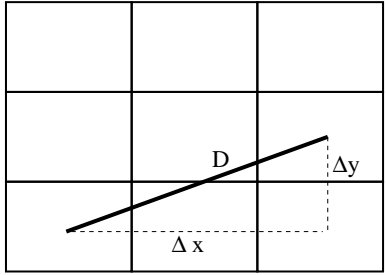,angle=-90,width=2.in}}
\vskip 5truept
\caption{
The compact hypertorus can be represented as an identified
parallelepiped.  Alternatively, the compact topology can be 
represented
by tiling flat space 
with identical copies of the fundamental 
parallelepiped.  In the tiling picture above
only the $(x,y)$ directions are shown.
A particular
periodic orbit is drawn which corresponds
to 
$x_{\rm end}=T_yT_x^2x_{\rm start}$.
}
\label{tile}
\end{figure}

\begin{figure}
\centerline{\psfig{file=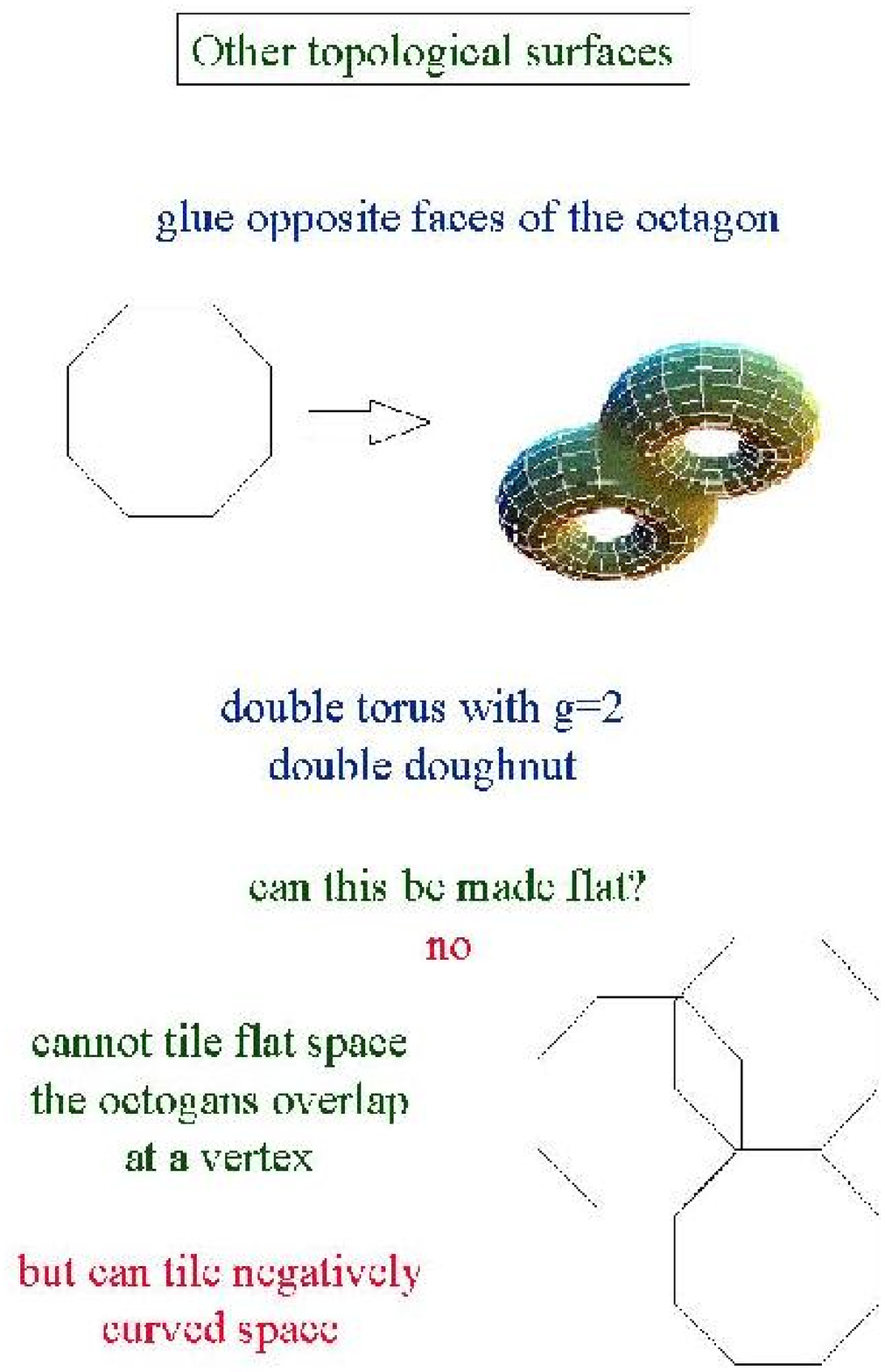,width=2.5in}}
\vskip 5truept
\caption{}
\label{oct}
\end{figure}

The flat plane can also be tiled by hexagons 
(fig.\ \ref{N.1.2.1}) as many a bathroom floor
demonstrates.  The hexagon therefore also reproduces a flat torus.
However the
general parallelogram and hexagon
exhaust the possible convex polygons which can
tessellate flat space.  If we tried to fill the plane with octagons,
we would find that they overlapped at a vertex and could not smoothly
fill the plane, as in fig.\ \ref{oct}.  If we glue the octagons together without
confining them to the plane, they will create a curling surface 
that tries to have a hyperbolic structure.  In fact, the
$2D$ hyperbolic plane ${\bf H}^2$ can be tiled by regular octagons. 
Recall that the angles of a triangle sum
to less than $\pi$ in hyperbolic geometry. 
Since the 
interior angles of the polygon narrow on the hyperbolic plane, they
can be drawn just the right size relative to the curvature scale 
so that precisely the right number fit around a vertex to fill the 
negatively curved plane without gaps or overlaps.
Notice also, that there is no scale in flat space and there is no 
intrinsic meaning to the area of the compact torus.  It can be made arbitrarily
large or small relative to any unit of measure.  This is not the case when
the surface is endowed with a curved metric.  The area is then
precisely determined by the topology.  The area of the octagon has to 
be precise relative to the curvature scale to properly tessellate the universal
cover.  

\begin{figure}
\centerline{\psfig{file=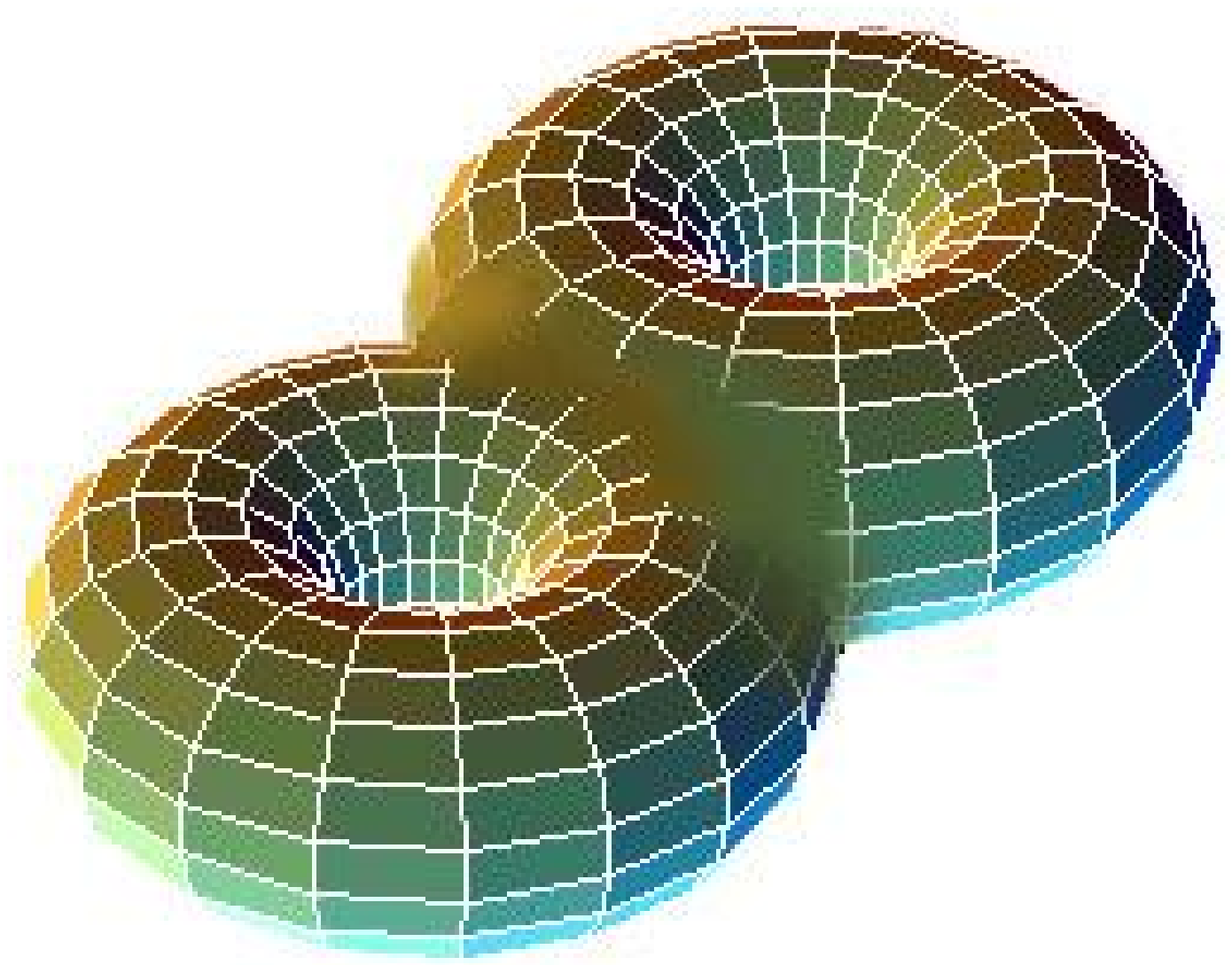,width=2.5in}}
\vskip 5truept
\caption{}
\label{double}
\end{figure}

The topology created by the compact octagon is a double torus
$T_2$ where the $2$ refers to the number of handles
or genus, g, of the manifold.  Hyperbolic $T_2$ is topologically equivalent
to the double doughnut of fig.\ \ref{double} which lives
in $R^3$ although again $T_2$ has a different
geometry, being flat, while the double doughnut is clearly curved.

A connected sum of tori can be constructed to make $T_g$,
a flat $2D$ compact surface with $g$ handles.  In order to make 
a $g$ handled object, a polygon with $4g$ edges is needed.  
The polygons
have to be drawn larger relative to the curvature scale the more
edges they have in order
that the interior angles be thin enough 
to fit together around a vertex and tile $H^2$.  Consequently
a relationship emerges between the area of the surface and the topology.
The area of the hyperbolic surface can be related to the 
Poincar\'e-Euler characteristic, $\chi=2(1-g)$ through the 
Gauss-Bonnet theorem
	\be 
	A=- 2\pi\chi=4\pi(g-1).
	\ee
Although the area is a topological invariant in $2D$, the lengths need
not be fixed.  For instance, the asymmetric octagon and the
regular octagon have the same topology, namely that of the double 
doughnut, and the same negative curvature,
but they have a different spectrum of periodic orbits.

The tessellation of the universal cover is of particular importance when
we consider three-dimensional spaces.  We simply do not have four dimensions
available to us in which to bend the three-dimensional volume and that
visualization technique fails us.  However, we can still visualize
filling the three-dimensional volume with $3D$ tiles which preserve
the local geometry of the manifold.  Therefore we rely heavily on the 
universal cover, the fundamental polyhedron and the holonomy group
when investigating the topologies of these compact spaces.

\subsection{Hyperbolic topologies and Volumes}

All 2-dimensional manifolds have been classified, as have
flat and spherical 3-manifolds, but hyperbolic 3-manifolds
are not yet fully understood.
These continue to resist classification although enormous advances were
made last century.
An important realization was the rigid connection between metrical quantities
and topological features in hyperbolic three-manifolds.
Metrical quantities describe lengths and would ordinarily fall under the 
domain of geometry.  Yet in three-dimensions the Mostow
\cite{mp} rigidity
theorem ensures that once the topology is specified, then all metrical 
quantities
are fixed for $3$-D hyperbolic manifolds.  This means that not only the volume
but also the lengths of the shortest geodesics are immutable for a given
topology.
This is to be compared with $2D$ where a given topology can support an
infinite number of metrics as already described.  

Although the topology fixes the compact hyperbolic
volume, the volume does not uniquely
specify the topology.  In other words there can be different topologies
with the same volume.
Because of this powerful link between geometry and topology, compact
hyperbolic manifolds can be ordered according to their volume.
A remarkable result is that compact hyperbolic manifolds form 
a countably infinite number of countably infinite sequences ordered
according to volume.  A given sequence shows an accumulation of
manifolds near a limiting volume set by a cusped manifold.  A cusped 
manifold has finite volume but is not compact, having 
infinitely long cusped corners which taper infinitely thin. 
The cusps are topologically a $2$-torus which is conformally shrunk to 
zero down the narrowing throat.  The sequence of compact manifolds
are built systematically through the process of Dehn surgery which is used
in the next section to organize the spectrum of topologies.  Dehn 
surgery is a formal process whereby a torus is drilled out of the cusped 
manifold and replaced with a solid torus.
The surgery is taken along a periodic orbit of $T^2$ identified
by the number of windings $(m_x,m_y)$ the orbit takes around 
the torus before closure.
The program SnapPea names manifolds according to the number of 
windings so that the manifold m003(3,-1) is a compact space constructed by 
Dehn filling the orbifold m003 along the periodic orbit with 
winding numbers (3,-1).

The space m003(3,-1) is also known as the Weeks space after the mathematician 
who discovered it and coincidentally has contributed to this article
\cite{weekspace}.
The Weeks space is of particular relevance to cosmology since it is 
the smallest known manifold with a volume 
${\cal V}\sim 0.94$ in curvature units.  The Weeks space displaced 
the Thurston space m003(-2,3) which has volume ${\cal V}\sim 
0.98$.
Although the Weeks space is the smallest known manifold it may or may not be
the smallest possible.  The current bound on the minimum volume
for any compact hyperbolic manifold is about 0.3 in units of the
curvature radius.
Contrast this with the
flat hypertorus which can be made infinitesimally small.
Again, the rigidity of compact hyperbolic spaces is at work.

Because of the rigidity of $3D$ hyperbolic manifolds, we can also 
characterize a topology by certain volume measures.  Very useful quantities
in cosmological investigations are the {\it in-radius} $r_-$ and the
{\it out-radius} $r_+$.  The in-radius is the radius of the largest geodesic
ball centered at the origin 
which can be inscribed within the Dirichlet domain. 
The out-radius is the radius of the smallest geodesic ball
centered at the origin which can
encompass the Dirichlet domain.  
The disparity between $r_{-}$ and $_+$ can be taken as a rough indication
of the global anisotropy introduced by the topological identification.
The {\it injectivity radius} is the radius of the largest ball,
centered
at the given point, whose interior embeds the manifold.
The Dirichlet domain given by default in SnapPea is centered at a local 
maximum of the injectivity radius.  In this way it often provides the
most symmetric form for the Dirichlet domain.
Moving the observer away from this basepoint will change the appearance of the
global shape of space observationally.
A somewhat ambiguous but sometimes useful quantity is an estimate 
of the 
diameter of the manifold.  Since the manifold is not isotropic
this is not precisely defined.  Nonetheless, one estimation is given by 
	\be
	d_{\cal M}=sup_{x,y\in {\cal M}}d(x,y);
	\ee
that is, it is the largest distance between any two points in the Dirichlet
domain.  
Notice that it is not a geodesic distance so no periodic 
orbit lies along the diameter.  
A more well defined diameter is the maximum
distance between two points in the manifold, where the distance is
measured
in the manifold itself, not in the Dirichlet domain:
	\be
	diam=sup_{x,y} (\inf_{{\rm paths \ connecting}\ x\ {\rm to}\ y}
({\rm length\ of\ path})).
	\ee

The periodic geodesics are quite important and are closely related
both to the eigenmodes in the space and the distribution of ghost
or clone images.  The periodic geodesics can be found with the
holonomy group.  The eigenvectors of the holonomy group span a plane
in the $4D$ embedding which intersects the pseudosphere $\ah$.
The line of intersection occurs along the periodic orbit corresponding
to that group element.  
A composite element can be thought of as a word formed from 
the fundamental alphabet of the presentation.
The fact that the number of words grows exponentially
with length means an exponential growth of long periodic orbits.
The proliferation of periodic orbits, dense in phase space, are the
canonical mark of chaotic dynamics.
This leads us to remark that compact hyperbolic spaces are not just
complicated mathematically but they are truly complex, supporting
chaotic flows of geodesics.  Since most cosmological investigations
try to circumvent the chaotic nature of the motions, we mention
chaos only occasionally as we go along.  We reserve a final section for a 
brief discussion  of this fascinating feature.  A review of 
chaos on the pseudosphere can be found in Ref.\ \cite{bv}.

\newpage
\section{Survey of Compact Manifolds}
\label{overview}

Having described the topological principles at work and some
examples, we can compile a list of known manifolds, their geometries
and topologies.  The classification is clearly presented in Refs.\
\cite{{lum},{wolf},{ellis}}.  We present a different mathematical
approach for variety.  The classification is organized by 
J. Weeks \cite{jwp} in relation
to four
questions:

$1$. What constant curvature manifolds exist?\par

$2$. How flexible is each manifold?  That is, to what extend can we change
a manifold's shape without changing its topology or violating
the constant curvature requirement?\par

$3$. What symmetries does each manifold have?\par

$4$. Does a Dirichlet domain for the manifold 
depend on the 
choice of basepoint?\par

It turns out that the answers to these questions depend strongly on both
the curvature and the dimension.  The remainder of \S \ref{overview}
answers these
questions in each of the six possible cases:  that is, in spherical, flat,
or hyperbolic geometry in two dimensions (\S \ref{twod}) or three dimensions
(\S \ref{threed}).


\subsection{Two-dimensional manifolds}
\label{twod}

\subsubsection{Overview of two-dimensional manifolds}
\label{twodover}

\begin{figure}
\centerline{\psfig{file=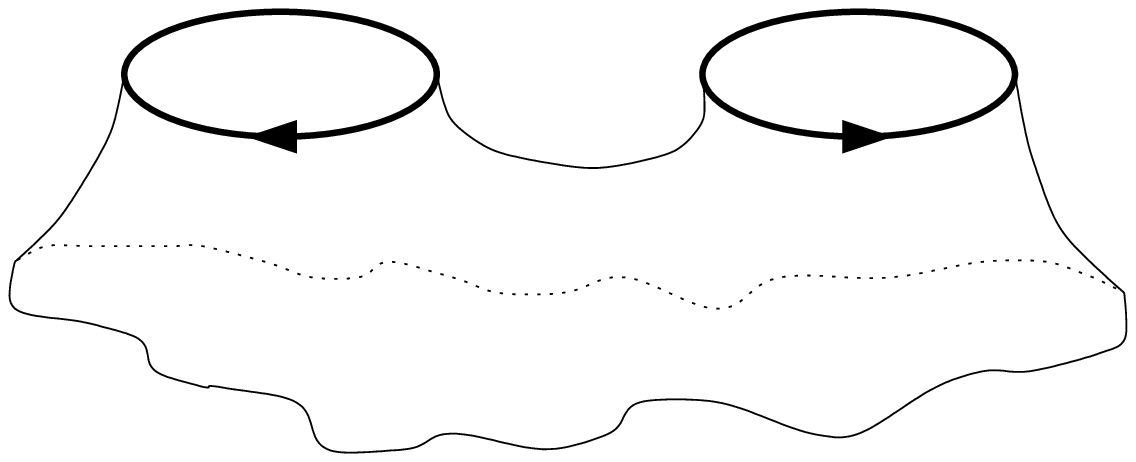,angle=-90,width=1.75in}
\psfig{file=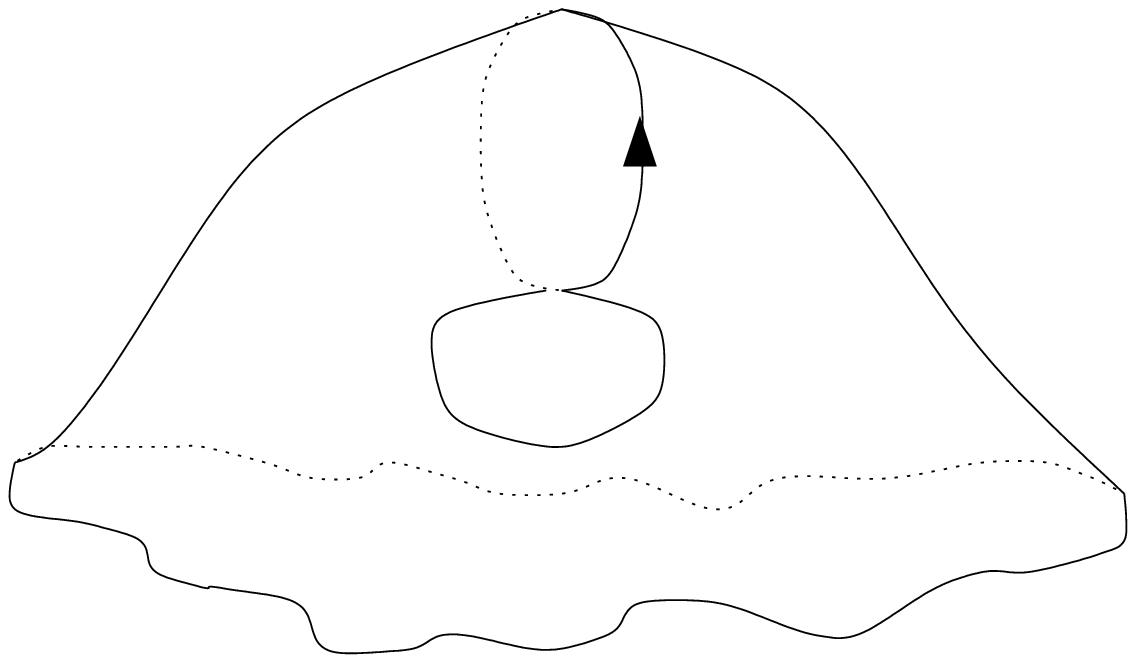,angle=-90,width=1.75in}}
\vskip 5truept
\caption{To add a handle to a surface, remove two disks and glue the
remaining boundary circles to each other.}
\label{N.2.1.2}
\end{figure}
\begin{figure}
\centerline{\psfig{file=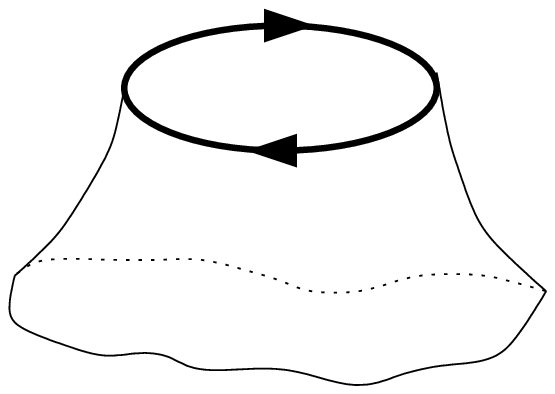,angle=-90,width=1.75in}
\psfig{file=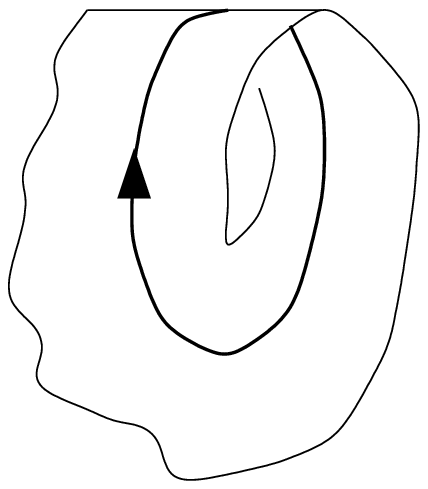,angle=-90,width=1.75in}}
\vskip 5truept
\caption{To add a cross cap to a surface, remove one disk and glue
the remaining boundary circle to itself by identifying antipodal points.
The construction cannot be physically carried out in $R^3$ without
self-intersections, but abstractly the resulting surface has a seam
that looks locally like the centerline of a M\"obius strip.
}
\label{N.2.1.3}
\end{figure}

Every closed surface
may be obtained by starting with a sphere and adding handles 
(fig.\ \ref{N.2.1.2})
or crosscaps (fig.\ \ref{N.2.1.3}).  
Each handle is constructed by removing two disks 
from the sphere and gluing
the resulting boundary circles to each other (fig.\ \ref{N.2.1.2}), and each
crosscap
is constructed by removing one disk and gluing the resulting boundary circle
to itself (fig.\ \ref{N.2.1.3}).
Conway found a particularly simple and elegant proof
of this classification;  for an illustrated exposition see 
Ref.\ \cite{Francis-Weeks}.

\begin{figure}
\centerline{\psfig{file=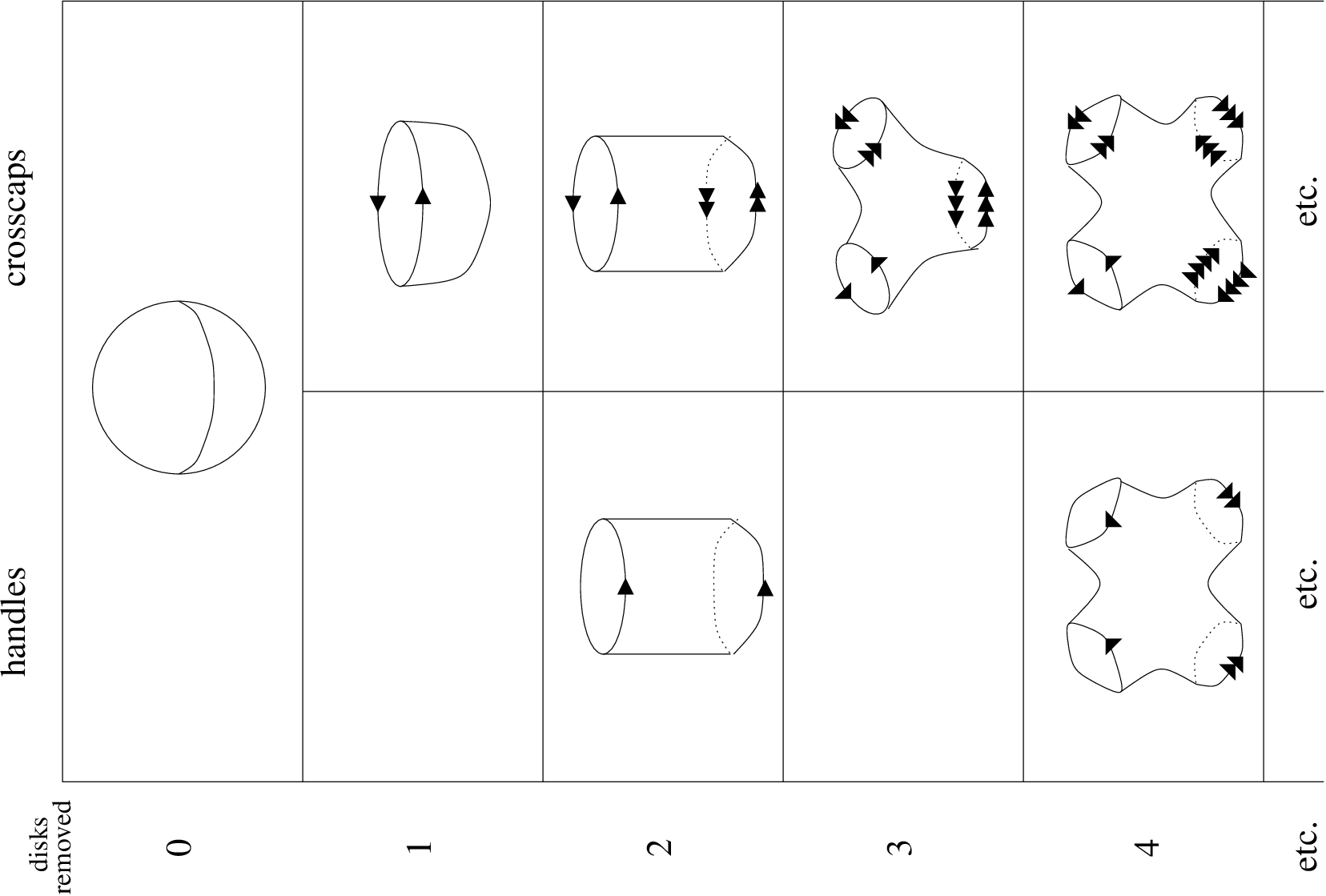,angle=-90,width=2.75in}}
\vskip 5truept
\caption{Topological classification of closed 2-manifolds.
Every closed 2-manifold may obtained from a sphere by
adding either handles (left column) or cross caps (right column).
}
\label{N.2.1.4}
\end{figure}

The interaction between geometry and topology is a primary theme in our
understanding of 2-manifolds.
The promised interaction between geometry and topology is that each closed
topological surface may be given a constant curvature geometry.
Each closed 2-manifold admits a unique geometry.  That is, a surface
which admits a spherical geometry cannot also admit a flat geometry, and so
on.
For an elementary proof see Ref.\ \cite{Shape}.  Infinite 2-manifolds,
by contrast, may admit more than one geometry.  For example, the Euclidean
plane and the hyperbolic plane are topologically the same 2-manifold.
We organize the surfaces according to the total
number of disks that get removed during the construction 
(fig.\ \ref{N.2.1.4}).

In the case that no disks are removed, we have the sphere.

In the case that one disk is removed, we draw the sphere-minus-a-disk
as a hemisphere, so that when the boundary circle gets glued to itself
the spherical geometry continues smoothly across the seam from one side
to the other, giving the resulting sphere-with-a-crosscap a uniform
spherical geometry.

In the case that two disks are removed we may have either
a sphere-with-a-handle, better known as a torus, or
a sphere-with-two-crosscaps, better known as a Klein bottle.
We draw the sphere-minus-two-disks as a cylinder.  The cylinder has
an intrinsically flat geometry. 
You can construct one from a sheet of paper
without stretching, and when you glue its top and bottom of the
rectangle to each other
(for the torus) or each to itself (for the Klein bottle) the flat geometry
continues uniformly across the seams.  Thus the torus and the Klein bottle
each has a flat geometry.

\begin{figure}
\centerline{\psfig{file=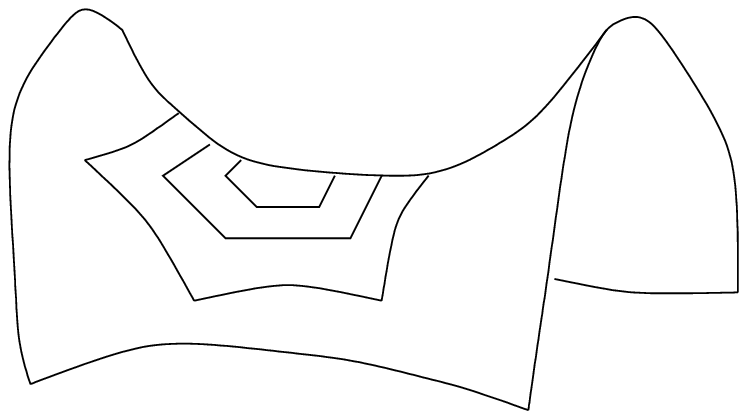,angle=-90,width=1.75in}
\psfig{file=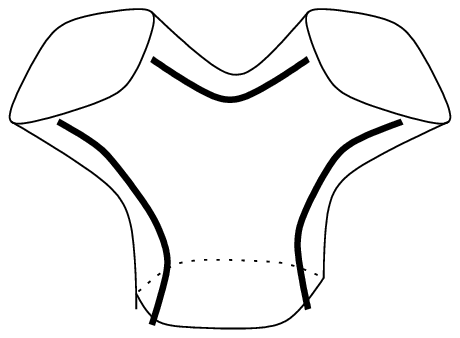,angle=-90,width=1.75in}}
\vskip 5truept
\caption{A pair of pants can be made from two hexagons.  For best results
use hexagons cut from the hyperbolic plane, with all interior angles
being right angles.}
\label{N.2.1.5}
\end{figure}

In the case that three disks are removed we have a
sphere-with-three-crosscaps,
constructed from a three-way cylinder better known as a {\it pair of
pants}.
The pair of pants may in turn be constructed from two hyperbolic hexagons
(fig.\ \ref{N.2.1.5}), giving it an intrinsically hyperbolic geometry.
When the pants' boundary circles (the cuffs) are glued to themselves
to make the sphere-with-three-crosscaps, the hyperbolic geometry continues
uniformly across the seams.  Thus the sphere-with-three-crosscaps also gets
a hyperbolic geometry.

All remaining surfaces -- those for which four or more disks are removed in
the construction -- may be built from pairs of pants, so all get a
hyperbolic
geometry.  In summary, we see that of the infinitely many closed
2-manifolds,
two admit spherical geometry (the sphere and the sphere-with-crosscap),
two admit flat geometry (the torus and the Klein bottle), and all the rest
admit hyperbolic geometry.

The next three sections will examine the flat, spherical, and hyperbolic
closed 2-manifolds in greater detail, in each case answering
the four questions raised in \S \ref{ovtop}.

\subsubsection{Flat two-dimensional manifolds}
\label{flattwo}

Manifolds:  torus and Klein bottle.

Flexibility:
   The torus has three degrees of freedom in its construction:  the height
   of the cylinder, the circumference of the cylinder, and the twist
   with which the top gets glued to the bottom.
   The Klein bottle has only two degrees of freedom in its construction:
   the height and circumference of the cylinder.  Each end of the cylinder
   gets glued to itself, identifying diametrically opposite points,
   so there is no twist parameter.

\begin{figure}
\centerline{\psfig{file=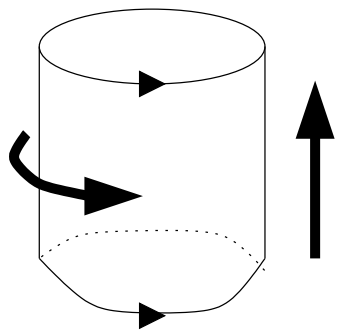,angle=-90,width=1.75in}}
\vskip 5truept
\caption{The torus has a two-parameter continuous family of symmetries,
consisting of rotations and vertical translations.
}
\label{N.2.2.1}
\end{figure}

\begin{figure}
\centerline{\psfig{file=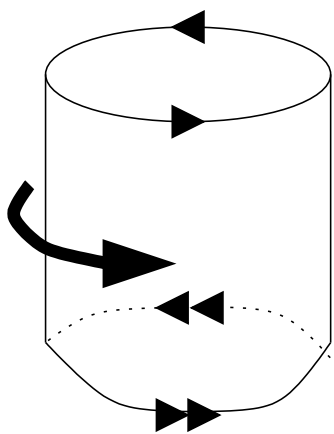,angle=-90,width=1.75in}}
\vskip 5truept
\caption{The Klein bottle has continuous rotational symmetry, but
any attempt at vertical translation is blocked by the cross caps.}
\label{N.2.2.2}
\end{figure}

Symmetry:
   Continuous symmetries.  The torus has a two-parameter continuous family
   of symmetries, consisting of rotations and vertical translations
   (fig.\ \ref{N.2.2.1}).  
   This shows that the torus is {\it globally homogeneous}
   in the sense that any point may be taken to any other point
   by a global symmetry of the manifold.
   The Klein bottle has only a one-parameter continuous family of
symmetries,
   consisting of rotations (fig.\ \ref{N.2.2.2}).  It is globally inhomogeneous.
   When diametrically opposite points on a boundary circle are glued to make
   a crosscap, the seam is a closed geodesic that looks locally like
   the centerline of a M\"obius strip.  Points lying on this centerline are
   fundamentally different from other points in the manifold.

\begin{figure}
\centerline{\psfig{file=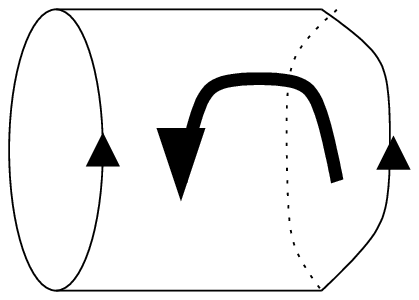,angle=-90,width=1.in}
\psfig{file=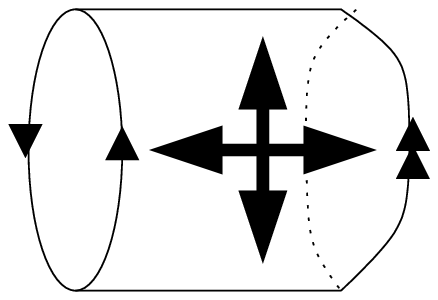,angle=-90,width=1.in}}
\vskip 5truept
\caption{The torus always has a discrete Z/2 symmetry given by a half turn,
even if the cylinder's top and bottom circles are glued to each other
with a twist.  The Klein bottle has a discrete Z/2 + Z/2 symmetry,
which interchanges and/or inverts the cross caps.}
\label{N.2.2.3}
\end{figure}

\begin{figure}
\centerline{\psfig{file=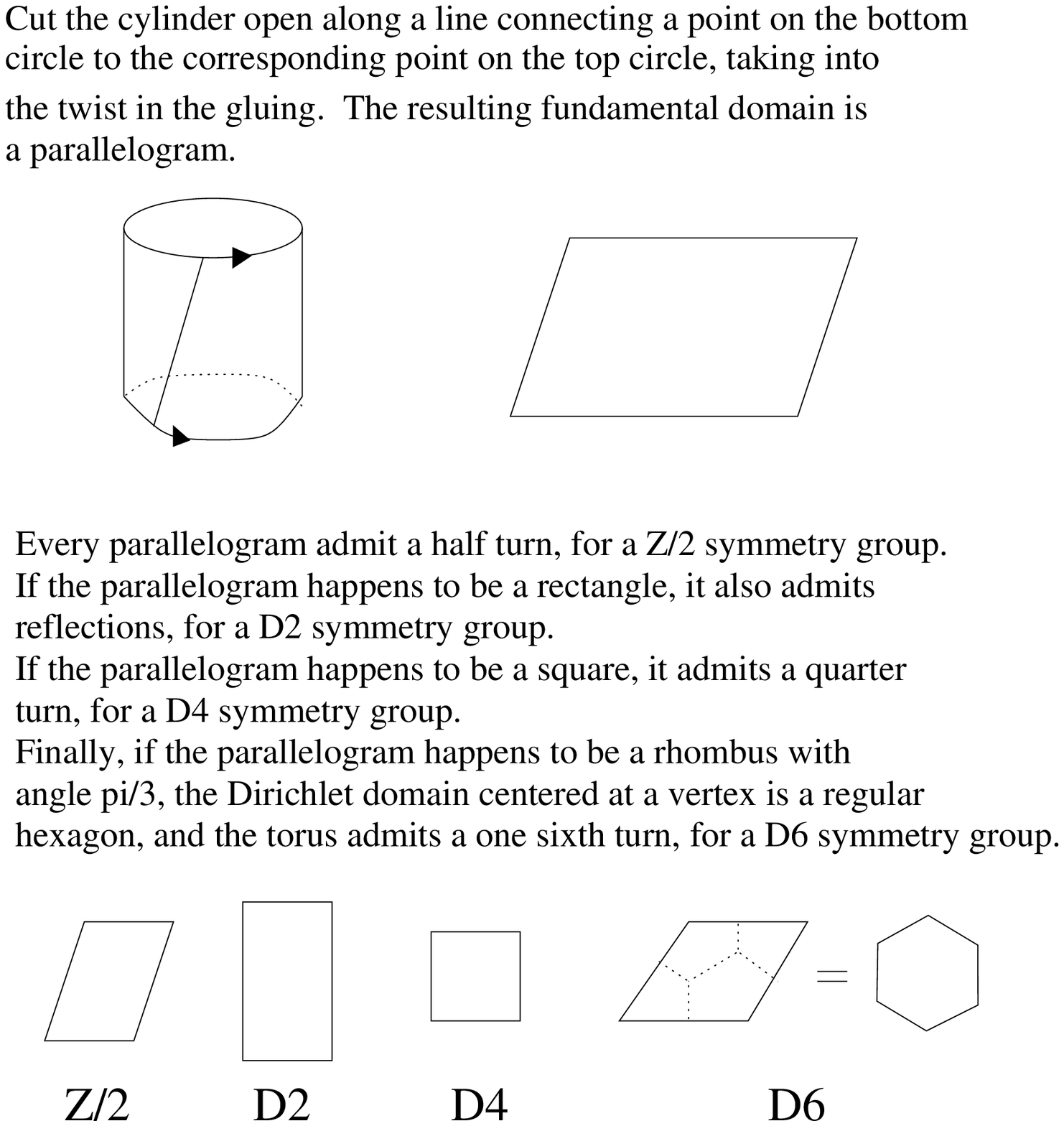,angle=-90,width=4.in}}
\vskip 5truept
\caption{}
\label{N.2.2.4}
\end{figure}

   Discrete symmetries.  As well as its continuous family symmetries, the
torus
   has a discrete Z/2 symmetry given by a 180 degree rotation interchanging
   the cylinder's top and bottom (fig.\ \ref{N.2.2.3}).  The discrete and
continuous
   symmetries may of course be composed.  The best way to think of this is
that
   the complete symmetry group of the torus consists of two disconnected
   components:  symmetries that interchange the boundary circles and
   symmetries that do not.  In special cases the torus may have additional
   discrete symmetries (fig.\ \ref{N.2.2.4}).
   The Klein bottle's discrete symmetry group is Z/2 + Z/2, generated
   by vertical and horizontal reflections of the cylinder.  Vertical
reflection
   interchanges the two crosscaps.  Horizontal reflection takes each
crosscap
   to itself, but reverses the direction of its centerline.

\begin{figure}
\centerline{\psfig{file=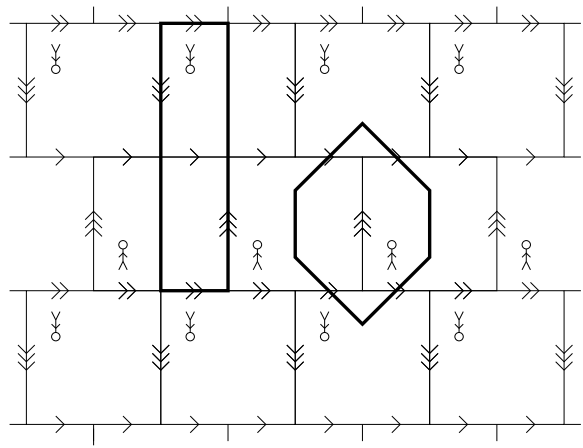,angle=-90,width=3.in}}
\vskip 5truept
\caption{In a Klein bottle, the shape of the Dirichlet domain depends
on the choice of basepoint.  For example, a Dirichlet domain centered
at the triple arrowhead is a hexagon, while a Dirichlet domain
centered at a single arrowhead is a rectangle.}
\label{N.2.2.5}
\end{figure}

Dirichlet domain:
   We have seen that the torus is globally homogeneous, so the shape of the
   Dirichlet domain does not depend on the choice of basepoint.
   On the other hand, the torus has three degrees of flexibility, and
   the Dirichlet domain's shape does depend on the shape of the torus.
   Typically the Dirichlet domain is an irregular hexagon, but in
   special cases it may be a square or a regular hexagon.
   The Klein bottle is globally inhomogeneous, so even for a fixed geometry
   of the Klein bottle, the Dirichlet domain's shape depends on the choice
   of basepoint (fig.\ \ref{N.2.2.5}).

\subsubsection{Spherical two-dimensional manifolds}
\label{sphertwo}

Manifolds:  sphere and sphere-with-crosscap.

Flexibility:
   The sphere and the sphere-with-crosscap have no flexibility whatsoever.
   Their geometry is completely determined by their topology.
   Here and throughout we make the assumption that spherical geometry
   refers to a sphere of radius one.  Otherwise these two manifolds would
   have the flexibility of rescaling -- they could be made larger or
smaller.

Symmetry:
   The sphere has a symmetry taking any point to any other point, with any
   desired rotation.  These symmetries form a 3-parameter continuous family.
   It also has a single discrete symmetry, given by reflection through
   the origin.
   The sphere-with-crosscap has a 3-parameter continuous family of
symmetries
   taking any point to any other point with any desired rotation and either
   reflected or not reflected.  It has no discrete symmetries.
   Indeed the continuous family already takes a neighborhood of any point
   to a neighborhood of any other point in all possible ways, leaving no
   possibilities for discrete symmetries.

Dirichlet domain:
   The sphere and the sphere-with-crosscap both have somewhat degenerate
   Dirichlet domains (try the inflating-the-balloon experiment in each),
   but these manifolds are globally homogeneous so the degenerate Dirichlet
   domains do not depend on the choice of basepoint.

\subsubsection{Hyperbolic two-dimensional manifolds}
\label{hyptwo}

Manifolds:
   There are infinitely many closed hyperbolic 2-manifolds,
   illustrated in fig.\ \ref{N.2.1.4}.

Flexibility:
   All closed hyperbolic 2-manifolds are flexible.
   The shape of each manifold is parameterized by the circumferences
   of the cuffs of the pairs of pants used to construct it, and
   by the twists with which distinct cuffs are glued to each other.
   A cuff that is glued to itself to make a crosscap has no twist
parameter.

Symmetry:
   Closed hyperbolic 2-manifolds never have continuous families of
symmetries.
  The reason is that each seam (where cuffs are joined together)
   is the shortest loop in its neighborhood,
and therefore cannot be slid
   away from itself.  A seam cannot slide along itself either, because
   a pair of pants admits no continuous rotations.
   A generic closed hyperbolic 2-manifold also has no discrete symmetries,
   but nongeneric ones may have discrete symmetries if the cuff lengths
   and twists are chosen carefully.

Dirichlet domain:
   The Dirichlet domain of a closed hyperbolic 2-manifold always depends
   on the choice of basepoint.  For carefully chosen basepoints it may
   exhibit some of the manifold's symmetries.

\subsection{Three-dimensional manifolds}
\label{threed}

\subsubsection{Overview of three-dimensional manifolds}
\label{overthree}

Unlike two-manifolds, not all three-manifolds admit a constant curvature
geometry.  For a survey of the geometry of 3-manifolds, see 
Ref.\ \cite{Tbull}.
Roughly speaking,  any closed 3-manifold may be cut
into pieces in a canonical way, and each piece admits a geometry that
is homogeneous but not necessarily isotropic.  For a very elementary
exposition of homogeneous anisotropic geometries, see 
Ref.\ \cite{Shape}
For a more detailed but still readable exposition see
\cite{pscott}.  The observed isotropy of the microwave background
radiation implies that at least the observable portion of the universe
is approximately homogeneous and isotropic, so cosmologists restrict
their attention to the three homogeneous isotropic geometries:  spherical,
flat and hyperbolic.  However, as noted earlier, an approximately flat
observable universe is also consistent with a very large universe of
varying curvature.

\subsubsection{Flat three-dimensional manifolds}
\label{flatthree}

\begin{figure}
\centerline{\psfig{file=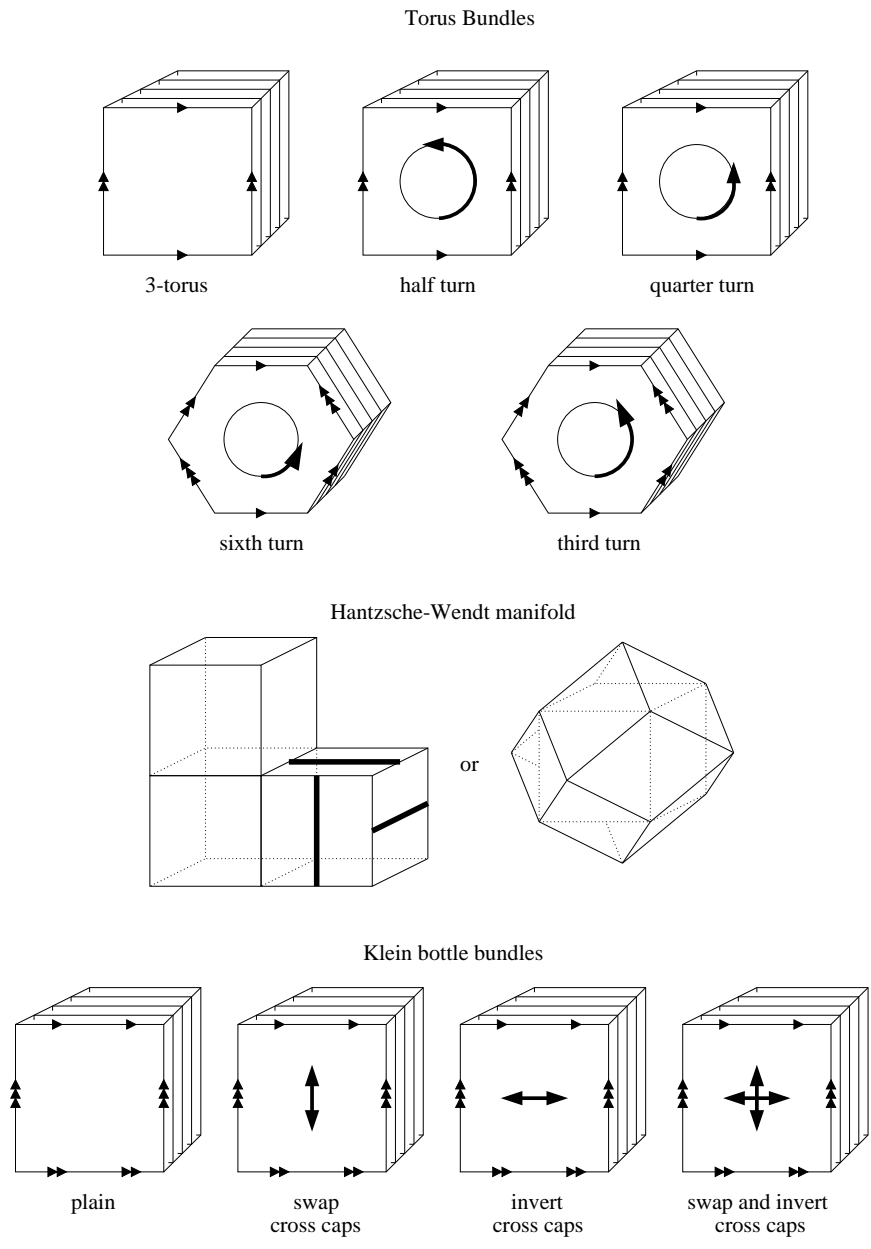,angle=-90,width=4.5in}}
\vskip 5truept
\caption{The ten closed flat 3-manifolds.}
\label{N.3.2.1}
\end{figure}

Manifolds:
   Figure \ref{N.3.2.1} illustrates the ten closed flat 3-manifolds.

   The first is the 3-torus introduced earlier -- a cube with
   pairs of opposite faces glued.

   The second and third manifolds are variations on the 3-torus in which
   the front face is glued to the back face with a half (respectively quarter)
turn,
   while the other two pairs of faces are glued as in the 3-torus.
Other possibilities for gluing the faces of a cube -- for example gluing
   all three pairs of opposite faces with half turns -- fail  to yield flat
   3-manifolds because the cube's corners do not fit together properly
   in the resulting space.  See Ref.\
\cite{Shape} for a more complete
   explanation.

   The fourth and fifth manifolds are similar in spirit to the first three,
   only now the cross section is a hexagonal torus instead of a square
torus.
   That is, the cross section is still topologically a torus, but has
   a different shape (recall fig.\ \ref{N.2.2.4}).  The hexagonal cross section
allows
   the front of the prism to be glued to the back with a one-sixth or
   one-third turn, producing two new flat 3-manifolds.  The side faces are
all
   glued to each other in pairs, straight across.

   The sixth manifold, the famous Hantzsche-Wendt manifold, is the most
   interesting of all.  Unlike the preceding five manifolds, which we
defined
   by constructing fundamental domains, the most natural way to define the
   Hantzsche-Wendt manifold is to start in the universal cover and define
its
   group of covering transformations.  Specifically, its group of covering
   transformations is the group generated by screw motions about a set
   of orthogonal but nonintersecting axes (indicated by heavy lines
   in fig.\ \ref{N.2.2.4}).  Each screw motion consists of a half turn about an
axis
   composed with a unit translation along that axis.  Note that this group
   of covering transformations does {\it not} take a basic cube to all
other
   cubes in the cubical tiling of 3-space.  Rather, the images of a basic
cube
   fill only half the cubes in the tiling, checkerboard style.  Thus a
complete
   fundamental domain would consist of two cubes, the basic cube and any one
   of its immediate neighbors;  images of the basic cube would fill the
black
   cubes in the 3-dimensional checkerboard, while images of its neighbor
would
   fill the remaining white cubes.  But we would really prefer a fundamental
   domain consisting of a single polyhedron.  To get one we employ the
   balloon construction for a Dirichlet domain introduced in Section N.1.2.
   Let the Dirichlet domain's basepoint be the center of a basic cube.
   As the balloon expands its fills that basic cell entirely, and also fills
   one sixth of each of its six immediate neighbors.  The resulting
Dirichlet
   domain is a rhombic dodecahedron (fig.\ \ref{N.3.2.1}).  The face gluings are
given
   by the original screw motions along the axes.
Note that this construction of the Hantzsche-Wendt manifold
          corrects an error, appearing elsewhere in the cosmological
          literature, which takes the three screw axes to be coincident.
          This erroneous construction leads to a cube with each pair
          of opposite faces glued with a half turn.  The cube's corners
          are therefore identified in four groups of two corners each
          instead of a single group of eight corners.  The resulting
          space has four singular points and is thus an orbifold instead
          of a manifold.

   The seventh through tenth manifolds are again defined by fundamental
domains.
   They are similar to the first five manifolds, except now the cross
section
   is a Klein bottle instead of a torus.  That is, the fundamental domain's
   left and right faces are glued to each other as in a torus, but the top
face
   (resp. bottom face) is glued to itself with a horizontal shift of half
the
   width of the fundamental domain, taken modulo the full width, of course.
   This ensures that every cross section is a Klein bottle, which may be
   understood as a cylinder with one crosscap at the top and another at the
   bottom.  The fundamental domain's front face, which is now a Klein
bottle,
   may be glued to its back face in one of four ways:  plain (seventh
manifold),
   interchanging the crosscaps (eighth manifold), inverting the crosscaps
   (ninth manifold), or interchanging and inverting the crosscaps
   (tenth manifold).  Unlike the first six manifolds, these four are all
   nonorientable.

Flexibility:
   All ten closed flat 3-manifolds have at least two degrees of flexibility:
   the depth of each fundamental domain (fig.\ \ref{N.3.2.1}) may be chosen
   independently of its width and height.  For some of the manifolds,
   for example the one-sixth turn manifold, this is the only flexibility.
   Others have greater flexibility:  for example the cross section of the
   half turn manifold may be any parallelogram, and the fundamental domain
   for the 3-torus may be any parallelepiped.

Symmetry:
   All closed flat 3-manifolds except the Hantzsche-Wendt manifold have
   a continuous one-parameter family of symmetries that pushes into the
page
   in the fundamental domains in fig.\ \ref{N.3.2.1}.  These symmetries take each
   cross section to a cross section further along, with the cross sections
   that fall off the back of the stack reappearing cyclically at the front.
   Some of the manifolds have additional continuous families of symmetries
   as well.  The 3-torus is the most symmetrical, with a 3-parameter family
   of symmetries taking any point to any other point.
   All ten manifolds have discrete symmetries as well, which
   vary from case to case and may depend on the manifold's exact shape
   as well as its topology.  For example, a cubical 3-torus has a 3-fold
   symmetry defined by a one-third turn about one of the cube's long
diagonals,
   but a 3-torus made from an arbitrary parallelepiped lacks this symmetry.

Dirichlet domain:
   The 3-torus is exceptional in that its Dirichlet domain does not depend
   on the choice of basepoint.  It is always the same, no
matter
   what basepoint you choose, because there is a symmetry taking any point
   to any other point.  For the nine remaining closed flat 3-manifolds,
   the Dirichlet domain always depends on the choice of basepoint.

Observational status:
   All ten closed flat 3-manifolds are of course completely consistent
   with recent evidence of a flat observable universe.
   However, if the fundamental domain is larger than our horizon radius,
   the topology may, in practice, be unobservable.

\subsubsection{Spherical three-dimensional manifolds}
\label{spherthree}

Manifolds:
   This section requires only a minimal understanding of the 3-sphere,
   or hypersphere, defined as the set of points one unit from the origin
   in Euclidean 4-space.  Readers wanting to learn more about the 3-sphere
may
   consult 
Ref.\ \cite{Shape}
for an elementary exposition or
Ref.\ \cite{Thurston}
for a deeper understanding.

   Spherical manifolds fall into two broad categories.  The manifolds
   in the first category have very different properties from those
   in the second.

   First category:

   The first category manifolds are most easily defined in terms of their
   groups of covering transformations, although it is not hard to work out
   their Dirichlet domains afterwards.  Amazingly enough, their groups
   of covering transformations -- which are symmetries of the 3-sphere --
   come directly from finite groups of symmetries of the ordinary 2-sphere!
   In other words, each finite group of symmetries of the 2-sphere
   (that is, the cyclic, dihedral, tetrahedra, octahedral, and icosahedral
   groups) gives rise to the group of covering transformations of
   a spherical 3-manifold!

   The bridge from the 2-sphere to the 3-sphere is provided by the
quaternions.
   The quaternions are a 4-dimensional generalization of the
   familiar complex numbers.  But while the complex numbers have a single
   imaginary quantity $i$ satisfying $i^2 = -1$, the quaternions have three
   imaginary quantities $i, j,$ and $k$ satisfying
	\ba
       i^2 &=& j^2 = k^2 = -1  \nonumber \\
       ij &=& k = -ji \nonumber \\
       jk &=& i = -kj \nonumber \\
       ki &=& j = -ik
	\ea
   The set of all quaternions $a + bi + cj + dk$ (for $a,b,c,d$ real)
   defines 4-dimensional Euclidean space, and the set of all unit length
   quaternions, that is, all quaternions $a + bi + cj + dk$ satisfying
 $  a^2 + b^2 + c^2 + d^2 = 1$, defines the 3-sphere.

   Before proceeding, it is convenient to note that there is nothing
   special about the three quaternions $i, j,$ and $k$.

     Lemma N.3.3.1 (quaternion change of basis).  Let $c$ be a 3x3
     orthogonal matrix, and define
	\ba
        i^{\prime} &=& c_{ii} i  +  c_{ij} j  +  c_{ik} k \nonumber \\
        j^{\prime} &=& c_{ji} i  +  c_{jj }j  +  c_{jk} k \nonumber \\
        k^{\prime} &=& c_{ki} i  +  c_{kj}j  +  c_{kk} k 
	\ea
     then $i^{\prime}$, $j^{\prime}$, and $k^{\prime}$ 
satisfy the usual quaternion relations
	\ba
       i^{\prime 2} &=& j^{\prime 2} = k^{\prime 2} = -1 \nonumber \\
       i^{\prime}j^{\prime} &=& k^{\prime} = -j^{\prime}i^{\prime}\nonumber \\
       j^{\prime}k^{\prime} &=& i^{\prime} = -k^{\prime}j^{\prime}\nonumber \\
       k^{\prime}i^{\prime} &=& j^{\prime} = -i^{\prime}k^{\prime}
	\ea
   Lemma N.3.3.1 says that an arbitrary purely imaginary quaternion
   bi + cj + dk may, by change of basis, be written as $b^{\prime}i^{\prime}$.
   If the purely imaginary quaternion bi + cj + dk has unit length,
   it may be written even more simply as $i^{\prime}$.  
A not-necessarily-imaginary
   quaternion $a + bi + cj + dk$ may be transformed to 
$a^{\prime} + b^{\prime}i$.
   If it has unit length it may be written as 
$\cos(\theta) + i^{\prime} \sin(\theta)$ for
   some $\theta $. 

   The unit length quaternions, which we continue to visualize as the
3-sphere,
   may act on themselves by conjugation or by left-multiplication.

     Lemma N.3.3.2 (conjugation by quaternions).  Let $q$ be an arbitrary
     unit length quaternion.  According to the preceding discussion, we may
     choose a basis $\{1, i^{\prime},j^{\prime},k^{\prime}\}$ such that 
$q = \cos(\theta) + i^{\prime} \sin(\theta)$ for some
$\theta$.
     It is trivial to compute how q acts by conjugation on the basis
$ \{1, i^{\prime},j^{\prime},k^{\prime}\}$.
	\ba
        (\cos(\theta) + i^{\prime} \sin(\theta)) 1  (\cos(\theta) - i^{\prime} \sin(\theta)) &=& 1\nonumber \\
        (\cos(\theta) + i^{\prime} \sin(\theta)) i^{\prime} (\cos(\theta) - i^{\prime} \sin(\theta)) &=& i^{\prime}\nonumber \\
        (\cos(\theta) + i^{\prime} \sin(\theta)) j^{\prime} (\cos(\theta) - i^{\prime} \sin(\theta)) &=&  j^{\prime} \cos(2\theta) + k^{\prime}
\sin(2\theta)\nonumber \\
        (\cos(\theta) + i^{\prime} \sin(\theta)) k^{\prime} (\cos(\theta) - i^{\prime} \sin(\theta)) &=& -j^{\prime} \sin(2\theta) + k^{\prime}
\cos(2\theta)
\ea
     We see that conjugation always fixes 1 (the north pole), so all
     the action is in the equatorial 2-sphere spanned by 
$\{i^{\prime},j^{\prime},k^{\prime}\}$.
     The equatorial 2-sphere itself rotates about the $i^{\prime}$ axis through
     an angle of $2\theta $.

   Compare the preceding action by conjugation to the following action
   by left multiplication.

     Lemma N.3.3.3 (left multiplication by quaternions).  Let $q$ be an
arbitrary
     unit length quaternion.  According to the preceding discussion, we may
     choose a basis $\{1, i^{\prime},j^{\prime},k^{\prime}\}$ such that 
$q = \cos(\theta) + i^{\prime} \sin(\theta)$ for some
t.
     It is trivial to compute how q acts by left multiplication on the basis
$   \{1, i^{\prime},j^{\prime},k^{\prime}\}$.
	\ba
        (\cos(\theta) + i^{\prime} \sin(\theta)) 1  &=&
  1  \cos(\theta) + i^{\prime} \sin(\theta)\nonumber \\
        (\cos(\theta) + i^{\prime} \sin(\theta)) i^{\prime} &=& -1  \sin(\theta) + i^{\prime} \cos(\theta)\nonumber \\
        (\cos(\theta) + i^{\prime} \sin(\theta)) j^{\prime} &=&  j^{\prime} \cos(\theta) + k^{\prime} \sin(\theta)\nonumber \\
        (\cos(\theta) + i^{\prime} \sin(\theta)) k^{\prime} &=& -j^{\prime} \sin(\theta) + k^{\prime} \cos(\theta)
	\ea
     We see that left multiplication rotates 1 towards $i^{\prime}$ while
     simultaneously rotating $j^{\prime}$ towards $k^{\prime}$.  
The result is a screw motion
     known as a Hopf flow.  What is not obvious from the above computation
     is that the Hopf flow is completely homogeneous -- it looks the same
     at all points of the 3-sphere.

   All the tools are now in place to convert finite groups of symmetries
   of the 2-sphere to finite groups of symmetries of the 3-sphere.
   The algorithm is as follows:

     Step \#1.  Choose a finite group of symmetries of the 2-sphere.
               (The only such groups are the cyclic groups, the dihedral
               groups, the tetrahedra group, the octahedral group,
               and the icosahedral group.  You might also expect a cubical
               and a dodecahedral group, but they coincide with the
               octahedral and icosahedral groups because the octahedron
               is dual to the cube and the dodecahedron is dual to
               the icosahedron.)

     Step \#2.  Write down the quaternions whose action by conjugation
               gives the chosen symmetries.  Note that each symmetry
               corresponds to two quaternions $q$ and $-q$, because
               $q$ and $-q$ act identically when conjugating,
               i.e. $q r q^{-1} = (-q) r (-q)^{-1}$.

     Step \#3.  Let the quaternions found in Step \#3 act by multiplication.
               This action defines the group of covering transformations.
               Note that although a quaternion's action by conjugation
               always has a pair of fixed points on the 2-sphere,
               its action by left multiplication on the 3-sphere is
               always fixed point free.

   The preceding algorithm is most easily understood in a concrete example.

     Example N.3.3.4.  Consider the so-called {\it Klein four group},
     the group of symmetries of the 2-sphere consisting of half turns
     about the $i$-, $j$-, and $k$-axes.  (Remember that we are in the 3-space
     spanned by $\{i,j,k\}$, so the $i$-, $j$-, and $k$-axes play the role of
     the traditional $x$-, $y$-, and $z$-axes.)  By Lemma N.3.3.2 a half turn
     about the $i$-axis corresponds to conjugation by the quaternions 
$\pm i$.  That is, when $\theta = \pi/2$, the quaternion $i = \cos(\pi/2) 
+ i \sin(\pi/2)$
     acts by conjugation as a rotation about the $i$-axis through an angle
     $2\theta = \pi$.  
Similarly, the half turn about the $j$-axis corresponds to
     the quaternions $\pm j$, and the half turn about the $k$-axis corresponds
     to $\pm k$.  The trivial symmetry corresponds to conjugation by 
$\pm 1$.
     Thus the complete set of quaternions is $\{\pm 1, \pm i, \pm j, \pm k\}$.
  If we now
     let these quaternions act by left multiplication on the 3-sphere,
     they give a group of covering transformations of order 8.
     (Exercise for the reader:  compute the action of each of those
     eight quaternions on the basis $\{1,i,j,k\}$.)
     It is not hard to see that the fundamental domain is a cube.
     The action of the covering transformations shows you that each face
     of the cube is glued to the opposite face with a one-quarter turn.
     Eight copies of the cube tile the 3-sphere like the eight faces
     of a hypercube.


The following table shows the results of applying the algorithm to
each 
finite group of symmetries of the 2-sphere:

\vskip 15truept

{\offinterlineskip
\halign{ \vrule # & \strut\ # & \vrule # &
	\ # \ & \vrule # \cr
\multispan{5}\hrulefill\cr
height2pt & \omit & & & \cr
& symmetry group of 2-sphere & & spherical 3-manifold &\cr
height2pt & \omit & & &  \cr
\multispan{5}\hrulefill\cr
height2pt & \omit & & & \cr
& cyclic $Z/n$              & & lens space $L(2n,1)$          & \cr
&    & & fundamental domain is lens-shaped solid & \cr
&   & & &\cr
& dihedral $D_n$ & & prism manifold & \cr
&  & & fundamental domain is prism with $2n$-gon base & \cr
&   & & &\cr
& tetrahedral (order 12)  & & octahedral space & \cr
&  & & fundamental domain is octahedron  & \cr 
&  & & 24 of which tile the 3-sphere     &\cr
&  & &	in the pattern of a regular 24-cell & \cr
&   & & &\cr
& octahedral (order 24)  & & snub cube space & \cr
&   & & fundamental domain is snub cube & \cr 
&   & & 48 of which tile the 3-sphere& \cr
&   & & &\cr
& dodecahedral (order 60)  & & P\" oincar\' e dodecahedral space & \cr
&   & & fundamental domain is dodecahedron &\cr
&   & &  120 of which tile the 3-sphere & \cr
&   & & in the pattern of a regular 120-cell & \cr
height2pt & \omit & & &  \cr
\multispan{5}\hrulefill\cr }}

\vskip 10truept

   Second category:

   Second category manifolds do not arise from simple left
   multiplication by groups of quaternions.
   The simplest second category manifolds are generic
{\it lens spaces}.  The group of covering
   transformations of a lens space is cyclic, generated by a single
   screw motion.  For the lens space $L(p,q)$, with $p$ and $q$ 
relatively prime
   integers satisfying $p > q > 0$, the screw motion translates a distance
$2\pi/p$
   along a given geodesic, while rotating through an angle $q(2\pi/p)$ about
that
   same geodesic.  The fundamental domain is a lens-shaped solid, $p$ copies
   of which tile the 3-sphere in much the same way that $p$ sectors tile
   the surface of an orange.  In the special case that $q = 1$,
   the screw motion corresponds to left multiplication by the quaternion
   $\cos(2\pi/p) + i \sin(2\pi/p)$, so $L(p,1)$ is a first category manifold.
   Otherwise $L(p,q)$ is a second category manifold.  (Warning:  $L(p, p-1)$
   is the mirror image of $L(p,1)$ and therefore has the properties of a first
   category manifold.  It corresponds to right multiplication
   by $\cos(2\pi/p) + i \sin(2\pi/p)$ instead of left multiplication.)

The remaining second category manifolds are defined by simultaneous
      left and right multiplication by quaternions.  That is, the holonomy
      group of each such manifold is a subgroup of a product 
	$H_1 \times H_2  \times {\pm 1}$,
      where $H_1$ and $H_2$ are finite groups of unit length quaternions.
      An element $(h_1, h_2, \pm1)$ acts according to the rule 
	$q \rightarrow \pm h_1 q h_2^{-1}$.
      For details, Section 4.4 of \cite{Thurston} and Sections 3 and 4
of cite{newsphere}.

Flexibility:
   Spherical 3-manifolds are rigid;  they have no flexibility.
   (As in 2-dimensions we make the assumption that spherical geometry
   refers to a sphere of radius one.  Otherwise all spherical manifolds
would
   have the flexibility of rescaling -- they could be made larger or
smaller.)

Symmetry:
   First category manifolds are always globally homogeneous, that is,
   each has a continuous 3-parameter family of symmetries taken any point
   to any other point.

      Proposition N.3.3.4 (global homogeneity of first category manifolds).
      For any two points P and Q in a first category manifold M,
      there is a symmetry of M taking P to Q.

      Proof.  Think of M as the quotient of the 3-sphere under the action
      of a finite group of unit length quaternions $\{g_1 = 1, g_2, ...,
g_{n}\}$.
      The point P has $n$ images represented by the quaternions
      $\{g_0 x, ..., g_n x\}$, and the point Q has n images represented
      by the quaternions $\{g_0 y, ..., g_n y\}$, for some x and y.
      Right multiplication by $(x^{-1} y)$ is a symmetry of the 3-sphere taking
      the images of P to the images of Q.  More generally, right
multiplication
      takes every set of equivalent quaternions to some other set of
equivalent
      quaternions, so it defines not only a symmetry of the 3-sphere but
also
      a symmetry of the original manifold M.  Q.E.D.

   Second category manifolds have smaller continuous families of symmetries.
   For example, the lens space $L(p,q)$ has a 2-parameter family of continuous
   symmetries defined by translations along, and rotations about, its
central
   geodesic.  Both first and second category manifolds may have discrete
   symmetries, which depend on the manifold.

Dirichlet domain:
   The Dirichlet domain of a first category manifold never depends
   on the choice of basepoint.  Proof:  By Proposition N.3.3.4 the manifold
   has a symmetry taking any point to any other point, so the manifold
   looks the same at all basepoints, and the Dirichlet domain construction
   must yield the same result.
   By contrast, the Dirichlet domain for a second category manifold
   does depend on the choice of basepoint.

Observational status:
   Recent evidence of an approximately flat observable universe implies that
   if the universe is spherical its curvature radius may be 
   comparable to or larger than
   our horizon radius.  So even though all spherical manifolds are viable
   candidates for the topology of the universe, the simplest ones, which
have
   large fundamental domains, would be much larger than the horizon radius
   and couldn't be observed directly.  A more complicated topology, with
   a larger group of covering transformations and a smaller fundamental
domain,
   would be more amenable to direct observation.

\subsubsection{Hyperbolic three-dimensional manifolds}

Manifolds:
   In 1924 Alexander Friedmann published a paper 
\cite{Friedmann1924}
complementing
   his earlier model of an expanding spherical universe with a model
   of an expanding hyperbolic universe.  It is a tribute Friedmann's
foresight
   that he explicitly allowed for the possibility of a closed hyperbolic
   universe at a time when no closed hyperbolic 3-manifolds were known.
   Fortunately examples were not long in coming.  L\"obell published the first
   in 1929 \cite{Löbell1929} and Seifert and Weber published a much simpler
   example soon thereafter \cite{SW1932?}.  Nevertheless, during the first half
   of the twentieth century few closed hyperbolic 3-manifolds were know.
   The situation changed dramatically in the mid 1970's with the work
   of Thurston, who showed that most closed 3-manifolds admit a hyperbolic
   geometry.  For an overview, see 
Ref.\ \cite{Tbull}.  To study
   known low-volume closed hyperbolic 3-manifolds, see Ref.\ 
\cite{snappea}.

Flexibility:
   Unlike closed hyperbolic 2-manifolds, closed hyperbolic 3-manifolds
   are rigid;  they have no flexibility.  As usual we assume
   curvature radius one, to suppress the trivial flexibility of rescaling.

Symmetry:
   Closed hyperbolic 3-manifolds never have continuous families of
symmetries.
   Truly generic closed hyperbolic 3-manifolds appear to have no discrete
   symmetries either, although the symmetry groups of low-volume manifolds
   are typically small cyclic or dihedral groups.  Nevertheless, Kojima
   has shown that every finite group occurs as the symmetry group of a
closed
   hyperbolic 3-manifold \cite{Kojima??}.

Dirichlet domain:
   The Dirichlet domain of a closed hyperbolic 3-manifold always depends
   on the choice of basepoint.  For carefully chosen basepoints, the
   Dirichlet domain may display some of the manifold's symmetries.
   The Dirichlet domain may also display symmetries beyond those
   of the manifold itself;  these so-called hidden symmetries are actually
   symmetries of finite-sheeted covering spaces.

\newpage
\section{Standard Cosmology and The Cosmic Microwave Background}
\label{cmbsection}

\setcounter{equation}{0}

\def\theequation{\thesection.\arabic{equation}}

In standard big bang theory, the universe is created in a hot energetic state,
a giant primordial soup.  The plasma is opaque to light as photons scatter
off hot charged particles.
As the
universe expands, the soup cools and some 300,000 years into our history, the
cosmos cools enough so that light no longer scatters
efficiently and the universe becomes transparent to radiation.
This transition, known as decoupling, marks the time when the
primordial radiation can free stream throughout the universe unimpeded. 
Decoupling happens quickly but not instantaneously and the small spread in
time will introduce some complication in our attempts to measure topology as
discussed in later sections.  This ancient radiation filling 
all of space has cooled today to a mere $2.73{}^\circ$ K, in the 
microwave range from which it has acquired the name
the cosmic microwave background (CMB).
As the CMB carries information almost from
the beginning of time, it is a
relic of our deep past.

\begin{figure}
\centerline{\psfig{file=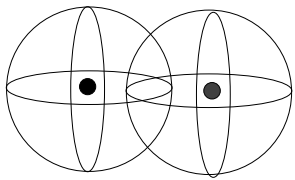,angle=-90,width=1.5in}}
\vskip 5truept
\caption{Two different surfaces of last scattering, each centered on 
a different galaxy.}
\label{alien}
\end{figure}

The light received at the Earth has traveled 
the same distance in
all directions since last scattering
and therefore defines a sphere.
This sphere is known as the surface of last scatter (SLS)
and is only defined relative to a given observer.  An alien civilization
in a neighboring cluster of galaxies, if coincidentally also
conducting observations of the CMB today, would receive photons from
its own SLS as shown in fig. \ref{alien}.
This will be particularly relevant to
the circles method of detecting topology described in section
\S \ref{obscirc}.

The famed CMB
carries much coveted information about the
universe.  It has been hoped that all local geometric quantities such as
the local curvature, the expansion rate, the nature of the
dark matter, and of the luminous matter, can be extracted from this one source. 
Clearly though the data will leave the full set of parameters underdetermined
and a great deal of model dependence is unavoidable.  What exactly the CMB can
tell us is being put to the test with the recent high resolution experiments
such as TOCO \cite{toco}, Boomerang and MAXIMA \cite{exper}.
These experiments are consistent with an approximately flat observable
universe.  If the universe is hyperbolic, its curvature radius may be comparable
to or larger than the horizon radius.  So even though all hyperbolic 
manifolds are viable candidates for the topology of the universe, they
may be too large to see, therein is the challenge.
The CMB can also be used to
decipher the global topology as well as fitting these many local parameters.
Though statistical measures of topology are inherently model dependent,
geometric methods are model independent.
The pattern based searches of circles in the sky and topological pattern
formation require no assumptions about the primordial spectrum nor the nature
of the dark matter.  However, these searches will likely be plagued by
difficulties of their own and it is unclear if these methods will succeed
without a great deal of refitting.

The standard big bang theory also has a powerful successor,
the standard inflationary theory.  The standard inflationary theory suggests
there was a period of accelerated expansion very early in the universe's 
history
which dilutes all fluctuations, all matter, and all energy so that the cosmos
suddenly goes from a tumultuous beginning to a nearly flat, empty, and biggish
universe.  Inflation is exited when it becomes energetically favorable for the
energy driving the expansion to be released into a renewed soup of high energy
particles and light.  This heating of the universe mimics the original 
primordial soup
and the standard history of the universe resumes. 
There are proponents of inflation who 
would revise the big bang still further,
essentially removing the moment of genesis altogether so that inflation is
eternal \cite{{vilenkin},{linde}}.  
The universe can be viewed as a ginger root of inflationary patches
each with its own details.  
As inflation ends in any given patch, the energy is released to heat that
region.
As far as any given observer is 
concerned, the heating epoch on exit from inflation looks like a consequence
of a big bang event.
Since this is all we can ever observe
with the CMB our discussion will apply to any of these early universe 
scripts.  

While
theorists still find it tricky to identify a
specific working model of inflation
and to devise a graceful exit to inflation, 
it is the favored story to tell and some of the
reasons for this are good.
Inflation elegantly resolves the mysterious homogeneity and isotropy of our
observable universe.  Without incredibly precise tuning of the initial
conditions, the universe would tend to be extremely lumpy and would
quickly cool 
if underdense or would quickly collapse if overdense.  The observational
fact that the universe is billions of years old and still just marginally
curved, if not outright flat, seems special and so unsettling.  A more rigorous
and compelling argument is made by Guth in his original paper
\cite{guth0} and many subsequent articles \cite{guth1}.
Our homogeneous and isotropic space becomes remarkably unlikely,
and yet here we
are.

On the other hand lumpiness seems natural 
since the many regions of the universe would not be
in causal contact and therefore should have very different local properties
such as 
temperature, density etc..  Yet when we look at the CMB we see that in fact the
temperature appears identical to better than 1 part in $10^5$ even for regions
which would be separated far beyond causally connected 
distances.  This would
be like finding two ancient civilizations on opposite sides of the Earth with
nearly identical languages.
The civilizations must have been in causal contact.
Likewise two regions of the universe which seem to have equilibrated to 
precisely the same temperature must have been in causal contact.
Inflation resolves this situation by
taking a small causally connected region and stretching it so large that it
exceeds the extent of the observable universe today.  
The CMB is essentially the same
temperature throughout our patch of the cosmos precisely because it was in
causal contact.  The stretching also naturally renders the universe flat.

Another motivation for inflation is that, while it makes the universe on
average homogeneous and isotropic, it also naturally generates 
fluctuations about this average. 
These minute perturbations are critical for initiating the collapse
of matter into galaxies and clusters.
These seemingly insignificant fluctuations become the catalysts
for all the order and structure we observe in the universe.
Because of the enormous stretching of scales involved, it manages to place
fluctuations even on the largest possible scales and therefore gives a causal
explanation for the seeds of structure formation.
(Another mechanism for generating the all important initial density 
fluctuations
is via topological defect models although these have lately fallen into
disrepute \cite{topstrings}.)

To emphasize, the beauty is that inflation both explains the average
homogeneity and isotropy as well as giving a prediction for the 
deviations from homogeneity, observed both in the CMB
and in the form of galaxies.
It seems a good question to ask , Is
topology consistent with inflation?  Precisely as inflation 
drives the curvature
scale beyond observational reach, so too will it drive any topological scale
beyond observational reach.  The universe may still be finite, only
we'll never know it.  Certainly a model of inflation could be concocted which
exited just at the critical moment so that the topology of the universe, and
presumably the curvature too, is just within observation now,
when we also happen to be here to look.  However, such a
contraption would be a perversion against the spirit of inflation as a restorer
of naturalness.  If an observable topology scale and inflation are to be
conjoined, the reasons must be profound and topology must be an integral part
of the inflationary mechanism.  This is conceivable.  For instance the cosmic
Casimir effect is the contribution to the vacuum energy due to any topological
boundary.  If this effect were to dilute as the topology scale expanded then
one can envision inflation driven by this contribution to the vacuum and ending
precisely as the topology scale became some natural big size and the vacuum
energy became some correspondingly natural small size.  To really wrap it all
into one clean package one could even try to explain the seemingly unnatural 
$\Omega_\Lambda=0.7$ this way \cite{imogen}
(see also Refs. \cite{{sokolov},{fagundes1}}.  Attempts to use compact
topology to aid $\Omega<1$ inflation have also been suggested \cite{css3}.
A related mechanism has been put to use in a model with topologically
compact extra dimensions \cite{sst}.
Another recent argument to justify flatness and local isotropy with
topology and without inflation can be found in Ref.\ \cite{bark}.
Barrow and Kodama found that compact topology can severely restrict 
the anisotropies which an infinite universe will allow.  Roughly speaking,
the anisotropic modes will not fit in the compact space.

In any case, the inflationary predictions for the statistical distribution
of fluctuations are often used in fixing the initial fluctuations and so will
be relevant here.
This assumption is a weakness of all statistical
constraints on topology and we will discuss these issues in due course.  The
topic for this section is really, regardless of their origin, how do
fluctuations in spacetime translate into temperature fluctuations in the CMB. 
And once this has been established, how does topology alter the standard
predictions leaving 
an archeological imprint for us to uncover billions of years
later.

\subsection{Standard Cosmological Equations}
\label{standardeq}

This section provides a
quick pedestrian review of the CMB and theories for the generation of its
perturbations.  There is nothing specific to topology in this section
unless explicitly stated.
Many detailed reviews on the CMB have been written and the reader is
referred to these \cite{{mkb},{hu},{ll},{whu}}.  Observing topology in the CMB
is discussed in sections \ref{obstop}, \ref{obsflat}, \ref{obshyp}.

The gravitational
field in general relativity is determined by the local Einstein
equations
	\be
	G_{\mu \nu}=8\pi G T_{\mu \nu}
	.
	\ee
The tensor $G_{\mu \nu}$ describes the curvature and evolution of 
the metric $g_{\mu \nu}$ while $T_{\mu \nu}$ is the energy-momentum
tensor and accounts for the matter fields.
(For thorough discussions on general relativity see \cite{{waldbk},
{mtw},{swein}}.)
It is important to note here
that the Einstein equations are local and therefore only
determine the curvature of spacetime and do not fix the topology.
Furthermore the symmetries of $g_{\mu \nu}$ are also local and are 
nearly always broken in the manifold by topological identifications.
Today the universe appears to be homogeneous and isotropic on the
largest-scales. The CMB gives tremendous confirmation of these
symmetries since the temperature appears identical in every
direction to better than 1 part in $10^5$.  
Homogeneity and isotropy imply that the Earth is not in a privileged position
in the cosmos.
The symmetries of homogeneity and isotropy severely restricts
the class of possible solutions to the Friedman-Robertson-Walker (FRW)
models defined by the gravitational metric
	\ba
	ds^2 &=&g_{\mu \nu}dx^\mu dx^\nu \nonumber \\
	&=&-dt^2+a^2(\eta)
	\left [{dr^2\over 1-\kappa r^2}+r^2(d\theta^2+\sin^2\theta d\phi^2)
	\right  ]
	\label{stanmet}
	\ea
where $\kappa=0,-1,1$ corresponds to a flat, negatively curved (hyperbolic)
and positively
curved (elliptical) space respectively (see fig.\ \ref{N.2.1.1}).
Negatively curved space is often termed
``open'' and positively curved space ``closed'' in reference to their
simply connected geometries.  Since we are interested in constructing
compact, multiconnected 
spaces, we will avoid the ``open'', ``closed'' terminology.

The overall scale factor, $a(\eta)$, describing the 
expansion of the universe is determined by the energy of matter
through the remaining Einstein equations.  
We can always operate in comoving units where the manifold is treated as a 
static constant curvature manifold and all of the dynamics is hidden in the 
conformal scale factor $a(\eta)$. 
An alternative expression for the metric is 
	\be
	ds^2=a^2(\eta)\left [-d\eta^2 +d\sigma^2 \right ]
	\ee
with the spatial part of the metric
	\be
	d\sigma^2=d\chi^2+f(\chi)(d\theta^2+\sin^2\theta d\phi^2)
	\label{cosmet}
	\ee
where we have used conformal time $d\eta=dt/a(t)$ and
	\be
	f(\chi) = \left\{ \begin{array}{lll}
	{\chi}^2 & r=\chi & {\rm flat }  \\
	 \sinh^2\chi & r=\sinh\chi &
	{\rm hyperbolic}\\
	 \sin^2 \chi & r=\sin\chi &
	{\rm spherical}  \end{array}\right.
	\ee
See also Appendix 
\ref{AppendixA}. 

The dynamical evolution of the space is given by
the Einstein equation determining the scale factor,
	\be
	H^2+\kappa/a^2={8\pi G\over 3}\rho.
	\label{fried}
	\ee
The different curvatures correspond to different values of the
global energy density.  Traditionally, $\Omega=\rho/\rho_c$ is defined 
with $\rho_c$ the critical density required to render the universe flat,
so that
$\Omega>1$ corresponds to $\kappa=1$, $\Omega=1$ corresponds to $\kappa=0$,
and $\Omega<1$ corresponds to $\kappa=-1$.
The curvature radius is $R_{\rm curv}=a/|\kappa |^{1/2}$.  
From eqn (\ref{fried}), $8\pi G\rho_c/3=H^2$ so that 
$H^2a^2(\Omega-1)=\kappa$ and the curvature radius can be be expressed in 
terms of $\Omega$ as
	\be
	R_{\rm curv}={1\over H\left |\Omega-1\right |^{1/2}} .
	\ee
When working in comoving
units we take the comoving curvature radius to be unity for curved
space or $\infty$ for 
flat space.  

Conservation of energy requires 
	\be
	\dot \rho+3H(\rho +p)=0 \label{consv}
	\ee
for the perfect fluid energy-momentum tensor
$T_\mu^\nu=(\rho,p,p,p )$. 	
Decoupling occurs during the matter dominated error for which 
$p=0$ and so 
$\rho\propto 1/a^3$ by eqn (\ref{consv}).  
The solution for $a(\eta)$ during matter domination is
	\be
	a(\eta)\propto \left \{\begin{array}{ll}
	 \cosh \eta -1 & \kappa =-1 \\
	 \eta^2/2 & \kappa = 0  \\
	 1-\cos\eta & \kappa =1\end{array}\right.
	\label{aevolv}
	\ee
with the conformal Hubble expansion ${\cal H}=a^{\prime}/a=aH$,
	\be
	{\cal H}\propto  \left \{ \begin{array}	{ll}
	\sinh \eta/(\cosh \eta -1)
	& \kappa=-1 \\
	 2/\eta & \kappa=0\\
	 \sin\eta/(1-\cos\eta) & \kappa=1 \end{array}\right. .
	\ee

In comoving units, the universe is static.  In physical units, the universe 
expands and the cosmic background radiation redshifts.  The physical
wavelength is shifted according to 
	\be
	{\lambda(t)\over \lambda_{\rm initial}}={a(t)\over a(t_{\rm initial})}
	\ee
from which the cosmological redshift is defined:
	\be
	{1+z}={a_0\over a(t)}.
	\ee
A subscript
$0$ will always be used to denote values today.  
The value of $\eta $ during matter domination
can be written in terms of $\Omega_0$ and the redshift 
using $\eta=\int dt/a=\int da/(\dot a a)$ and eqn (\ref{fried}):
	\be
	\eta =\left\{\begin{array}{ll}
	{\rm arccosh}\left (1+{2(1-\Omega_0)\over \Omega_0(1+z)}\right )
		& \Omega_0<1\\
	{3\over (1+z)^{1/2}} &
	\Omega_0=1\\
	\cos^{-1}\left (1-{2(\Omega_0-1)\over \Omega_0(1+z)}\right )
		& \Omega_0>1\end{array}\right.
	\ee
\cite{kolbturner}.

If curvature or topology are to be observable, the geometric scales must 
be smaller than the diameter of the SLS.
In flat space, there is no natural scale since $R_{\rm curv}=\infty$ and 
it would seem a random coincidence if the topology scale just happened 
to be observable.  In curved space, one might expect the curvature and 
topology scales to be comparable.  This may not make it any more natural to 
observe geometric effects, but it would mean that if curvature is 
observable, then topology may be also.

For the purpose of observational topology it is useful to list
some relevant scales.
The radius of the surface of last scatter is defined as the
the distance light travels between the time of decoupling and today,
	\be
	D_\gamma=a(t)\int_{t_{d}}^{t_0} {dt\over a(t)}.
	\ee
The distance depends on the curvature and evolution of the universe
through eqn (\ref{aevolv}).  
In comoving units the
photon travels a distance
	\be
	d_\gamma={D_\gamma\over a(t)}=\int_{t_{d}}^{t_0} {dt\over a(t)}
	=\int_{\eta_{d}}^{\eta_0}d\eta=\Delta \eta
	\ee
with $\Delta \eta\equiv \eta_0-\eta_{d}$.
The diameter of the SLS, $2\Delta \eta$, essentially defines the 
extent of the observable universe.  A loose criterion that can be used
to gauge if
topology will influence the CMB is that 
the in-radius be less than $\Delta \eta$.

The volume enclosed by the SLS is the integral over
$V_{SLS} =\int \sqrt{-g}drd\theta d\phi $.  For a radius of $\Delta \eta$,
	\ba
	V_{SLS} &=& \pi\left (\sinh(2\Delta \eta-2\Delta \eta)\right )
	 \Omega_0<1 \nonumber \\
	&= &{4\pi \over 3}\Delta \eta^3 \quad\quad\quad \Omega_0=1\nonumber \\
	&=&\pi\left (2\Delta \eta-\sin(2\Delta \eta)\right ) \Omega_0>1.
	\ea
The number of clones of the fundamental domain that can be observed in 
a small compact cosmos can be estimated by the number of copies that can
fit within the SLS: $V_{SLS}/{\cal V}_{\cal M}$.
We will often refer back to these scale comparisons in the
coming discussions.

\subsection{Fluctuations in the CMB}
\label{fluctcmb}

Regardless of the origin, there are small deviations from homogeneity
and isotropy which are treated by perturbing the metric
and matter tensors about the FRW solutions.  The equations of motion for these 
perturbations to linear order are determined by 
$\delta G_{\mu \nu}=8\pi G\delta T_{\mu \nu}$.  While the equations are
extremely complicated there is one quantity, the 
gauge invariant 
potential $\Phi$, which 
is a coordinate invariant combination of 
components of $\delta g_{\mu \nu}$ and 
is of central importance for our considerations.
It is so named since in the Newtonian limit it corresponds to the usual gravitational potential.

The hot and cold spots in the CMB are caused by these perturbations in the 
gravitational potential.   The fluctuations Doppler shift the photons
generating an anisotropy both at decoupling and as light traverses the 
time changing gravitational field.  
There are many carefully derived results for the temperature fluctuations in the
literature based on relativistic kinetic theory and relativistic
perturbation theory \cite{{peebk},{ll},{whu}}.  Since these topics comprise multiple review
papers on their own, we will not go through the detailed derivations
but will instead try to provide a cohesive, intuitive motivation for
each of the relevant concepts.  

To derive the famous Sachs-Wolfe
effect
\cite{sw}, we followed the pedagogical discussion of White and Hu
\cite{whu}.  Photons gain energy and are
therefore blueshifted as they fall into a potential well and lose
energy and are therefore redshifted as they climb out.  The Doppler
shifts will cancel in a static homogeneous and isotropic space.
However at the time that the photons last scatter, some photons
will be in
potential wells and some will be in potential peaks.  Therefore
this snapshot of the metric fluctuations becomes frozen into the CMB.
It is this snapshot or fossil record we observe.
Energy conservation gives a Doppler shift of 
	\be
	\left.{\delta T\over T}\right |_f -\left.{\delta T\over
T}\right |_i
	=\Phi_f-\Phi_i.
	\ee
The local gravitational potential 
$\Phi_f$ 
makes an isotropic contribution to ${\delta T/T}$ and
will be left
off hereafter.  
By adiabaticity, $aT=$constant, it follows that
	\be
	\left.{\delta T\over T}\right |_i=-{\delta a\over a}.
	\ee
Assuming flat space for simplicity, $a\propto t^{2/3}$ during matter
domination.  The shift in $a$
is then
	\be
	{\delta a\over a}={2\over 3}{\delta t\over t}.
	\ee
The shift $\delta t$ can be understood as a shift relative to the cosmic
time due to the perturbed gravitational potential.  In a gravitational
potential clocks appear to run slow by
	\be
	{\delta t\over t}\sim \Phi 
	\ee
from which it follows that 
	\be
	\left.{\delta T\over T}\right |_f={\Phi_i\over 3}.
	\ee
This term is the surface Sachs-Wolfe effect from the time of decoupling.
There is an additional contribution to the temperature fluctuation 
known as the 
integrated Sachs-Wolfe (ISW) effect which is accumulated as the photon
traverses space through a decaying potential.  Including this latter
contribution the full Sachs-Wolfe effect is
	\be
	{\delta T(\hat n)\over T}={1\over 3}\Phi(\eta_o \hat n)
	+
	2\int^{\eta_o}_{\eta_{\rm SLS}}d\eta
	\Phi^{\prime}(\eta,\vx)
	\label{sachswolfe}
	\ee
where a prime denotes differentiation with respect to conformal time
\cite{sw}.  In flat space with ordinary energy, the only contribution
to $\delta T/T$ is from the surface Sachs-Wolfe effect
(the first term in eqn.\ \ref{sachswolfe}).
The ISW
(the second term in eqn.\ \ref{sachswolfe}) makes a significant
contribution to large-scale fluctuations if space is curved and/or
when a cosmological constant dominates the energy density.
In the absence of a significant cosmological constant, the ISW is
important as perturbations decay along the line of sight during
the curvature dominated epoch at 
	\be
	1+z\sim (1-\Omega_0)/\Omega_0
	\label{curvdom}
	\ee
for underdense cosmologies.  This will prove to be important to 
the issue of topology since the ISW may camouflage the conspicuous
marks of topology imprinted in the surface Sachs-Wolfe effect.

Therefore the temperature fluctuations observed today can be 
predicted from the primordial fluctuations in the metric.  The shape of
$\Phi(\eta,\vx)$ will depend on the geometry of space as well as some initial
spectrum.
The perturbed Einstein equations 
give the equation of motion
	\be
	\Phi^{{\prime} {\prime}}+3{\cal H}(1+c_s^2)\Phi^{\prime} -c^2_s\nabla^2 \Phi
	+(2{\cal H}^{\prime} +(1+c_s^2){\cal H}^2)\Phi=0
	\label{feta}
	\ee
where ${\cal H}=a^{\prime}/a$ and $c_s$ is the speed of sound in the 
cosmological fluid.
In a matter dominated era
$c_s=0$.  The potential is separable and can be written
$\Phi=F(\eta)\Psi(\vec x)$ with $F(\eta)$ the solution to 
eqn (\ref{feta}).  The $\Psi(\vec x)$
can always be expanded in terms of the eigenmodes to the Laplacian;  that is,
	\be
	{\Phi(\eta,\vx)}=F(\eta) \int d^3\vec k
	\ \hat \Phi_{\vec k}\ \psi_{\vec k}(\vec x)
	\label{elem}
	\ee
with the eigenmodes $\psi_k$ satisfying
	\be
	\nabla^2\psi_{\vec k}=-k^2\psi_{\vec k}.
	\label{eigeneq}
	\ee
The Laplacian $\nabla^2=g_{\mu \nu}D^\mu D^\nu$ depends on the 
curvature through the covariant derivatives $D_\mu$.
For now we assume all the $k$'s are continuous eigenfunctions
although this will not be the case when the space is topologically 
identified.
The $\hat \Phi_k$ are initial amplitudes given by the statistical 
profile of the initial fluctuation spectrum.
All assumptions and/or predictions for the initial perturbations are contained
in the $\hat \Phi_{\vec k}$.

During inflation quantum fluctuations in the field are amplified by
the accelerated expansion.  These fluctuations about the ground
state of the field theory are known to be Gaussian distributed.  The amplitude of the fluctuations are related to the specific inflationary model
but for the most part are independent of the scale $k$.
So inflation predicts a Gaussian distribution of fluctuations independent
of scale.
Specifically this means the $\hat \Phi_{\vec k}$ are drawn from a random 
Gaussian ensemble consistent with 
	\begin{equation}
	\left\langle \hat \Phi _{\vec k}^{*}
	\hat \Phi _{\vec k}\right\rangle 
	\propto{\cal P}_\Phi (k)\delta^3 (\vec k-\vec k^{{\prime} })
	\ \ .   \label{assume}
	\end{equation}
The angular bracket $<>$ denote an ensemble average and ${\cal P}_\Phi$
is the predicted power spectrum.
De Sitter inflation delivers a flat, Harrison-Zeldovich spectrum
which corresponds to ${\cal P}_\Phi =$constant.
Other forms for ${\cal P}_\Phi\propto k^{n-1}$ have been
derived from specific inflationary models
with the spectral index $n\ne 1$ and give tilted as
opposed to flat spectra.

From the preceding we can fully predict $\Phi$ and
therefore $\delta T/T$ given the curvature of spacetime and the initial 
spectrum.  Armed with this prediction, fluctuations on a given manifold
can be compared to the data.  For data comparison it is 
customary to consider the correlation function.
Although $\delta T(\vx)/T$ permeates space, we only observe 
fluctuations from our SLS at location $\vx=\Delta \eta\hat n$
with $\hat n$ a unit directional vector.  
Since the fluctuations are taken to be Gaussian, the correlation function
contains all of the information about the temperature fluctuations.
The ensemble average
correlation function between any two points on the sky is 
	\be
	C(\hn,\hnp)\,  = \,
	\left\langle 
	{\delta T\over T}(\hn) {\delta T\over T}(\hnp) \right\rangle.
	\label{corrf}
	\ee
The theoretically predicted $C(\hat n,\hat n^{\prime})$
can then be statistically compared to the data to estimate the likelihood
a given model is responsible for the world we live in.

Because fluctuations are observed on a sphere it is customary to decompose
the data into spherical harmonics,
	\be
	{\delta T\over T}(\hat n)=\sum_{\ell m}a_{\ell m} Y_{\ell m}(\hat n)
	\label{sphereharm}
	\ee
where the $Y_{\ell m}$'s are the usual spherical harmonics.  The
spherical harmonics form a complete set of states on the sphere and
are orthogonal so that
	\be
	\int_0^{2\pi}d\phi \int_0^{\pi}
	\sin\theta d\theta Y_{\ell^\prime m^\prime}^*(\theta,\phi)Y_{\ell m}
	(\theta,\phi)=\delta_{\ell^\prime \ell}\delta_{m^\prime m}
	\ee
and unit normalized.
Using the orthogonality of the $Y_{\ell m}$,
invert (\ref{sphereharm}) to find
	\be
	a_{\ell m}=\int d\Omega Y_{\ell -m}(\hat n){\delta T\over T}
	(\hat n)
	\ \ .\label{invert}
	\ee
If the probability distribution is Gaussian, then the Fourier correlation
function 
	\be
	C_\ell=\left < \left |a_{\ell m}\right |^2 \right >
	\ee
contains
complete information about the fluctuations.  
The $<>$ again denotes an ensemble average.
Either non-Gaussianity or the breaking
of homogeneity and isotropy with topological identifications
will mean that the
$C_\ell$'s do not provide complete information.  We will
emphasize this in section \S \ref{obshyp}.  

Assuming for now a simply connected topology, homogeneity and isotropy
of the metric allows one to perform an angular average which is in
effect an average over $m$'s without loss of information.  
Such an average gives an estimator for
the ensemble average angular power spectrum $C_\ell$,
	\be
	C_{\ell}=\sum_{m=-\ell}^{m=\ell}|a_{\ell m}|^2/(2l+1).
	\label{est}	
	\ee
Using equation (\ref{invert}) gives
	\be
	C_{\ell}={1\over (2l+1)}\sum_m\left
	[\int d\Omega^{\prime} Y_{\ell -m}(\hat n^{\prime})
	\int d\Omega Y_{\ell m}(\hat n)
	\left <{\delta T\over T}(\hat n){\delta T\over T}(\hat n^\prime)
	\right >\right ]
	\ \ .
	\label{clong}
	\ee
The parameter $\ell $ can be associated with the angular size of a given
fluctuation, with $\ell\sim \pi /\theta$.  Large angle fluctuations are 
associated with low $\ell$ and small angle fluctuations are associated
with high $\ell $.
The lower the value of $\ell$, the fewer contributions there are to the sum
(\ref{est}).  
As a result, the estimator of the true ensemble average is poorer
for low $\ell$ than it is for high $\ell$ where there are many 
contributions to the average.  This is known as cosmic variance and can
be included as an error bar.  
For a homogeneous, isotropic space cosmic variance can be estimated as
	\be
	C_\ell \sqrt{2/({2 \ell +1})} .\ee
We only have one universe available to measure and we cannot be sure that
our universe is not just a randomly large deviation from average.
It is difficult to interpret the significance of low $\ell$ observations such 
as the lack of power in the quadrupole as observed by \cb.
The ambiguity of cosmic variance
is worsened with topological identifications as emphasized in \cite{bpsII}.

These are all the tools needed to determine $\delta T/T$ for a specific
theoretical model, the correlation function $C(\hat n,\hat n^\prime)$,
and the angular
average power spectrum $C_\ell$.  These predictions can then 
be compared with data.  
When we come to the influence of topology we will see the increasing importance
of using full sky maps of $\delta T/T$ and the full correlation function
$C(\hat n,\hat n^\prime)$.

We have described the scalar modes above.
There are also tensor modes which are gravitational waves and vector modes 
which are rotational perturbations and do not grow with time.  The
scalar modes are the only ones which couple to the energy density and
pressure and we restrict ourselves to scalar modes hereafter.
Topology may in fact influence the gravitational wave background but
to date this remains unexplored territory.

The primordial fluctuations catalyze the formation of galaxies and
galaxy clusters.  So the minute quantum fluctuations are amplified
into the gigantic structures we see today.  The primordial spectrum
can be tested by measuring the fluctuations in the CMB but there are 
other manifestations.  For instance, the eventual nonlinear growth of
perturbations can be simulated numerically to test theoretical
predictions for structure formation against astronomical
observations.  It would be interesting to know if topological features
could sculpt the distribution of structure as well \cite{lb_fractals}.

In the following subsections we find the eigenmode decomposition in 
simply connected spaces.
Perpetuating a cruel prejudice against $\es$ we consider only $\ah $ and
$\eu $.  However, the authors of Ref.\ \cite{newsphere} catalog all of 
the $\es$ topologies and discuss detection strategies based on the
crystallographic methods.
The motivation for neglecting $\es$ topologies in connection with 
the CMB is observational as
there is no outstanding evidence which supports an $\Omega >1$
universe \cite{open}.  
The current debate is whether $\Omega=1$ or whether $\Omega <1$.
The future satellites which aim to refine measurements of the CMB intend
to resolve this debate.
The overriding theoretical prejudice is for $\Omega=1$ consistent with
inflation.  More
recently the supernova data has drummed up enthusiasm for
$\Omega_\Lambda=0.7$ and $\Omega_m=0.3$ so that the total $\Omega=1$
\cite{saul}.  
The recent small angle experiments Boomerang and MAXIMA 
corroborate these values if certain prior assumptions constrain
the statistical analysis.  More quantitatively, the current numbers
are $\Omega =0.90\pm 0.15$, $\Omega_b h_0^2=0.025 \pm 0.010$
$\Omega_{CDM}h_0^2=0.13\pm 0.10$ and a spectral index of
$n=0.99\pm 0.09$ \cite{balbi} where
the subscript $b$ denotes the baryonic 
contribution and the subscript $CDM$ denotes the net contribution
from cold dark matter.
Still, the data are clearly underdetermined
and it is an unresolved issue how many different
models can match the data.

In Ref.\ \cite{addr} an argument is made for
a constraint on $\Omega_0$ with some model-independence.
They argue that a constraint is imposed if one
requires that a local maximum detected in
the correlation function of large scale structure 
($\sim 100-200h^{-1}$Mpc)
occurs at the same comoving positions at different redshifts.
The argument is independent of both CMB and supernova Ia data, and 
favors a hyperbolic universe.
Their result is a cosmology with $\Omega_0=0.9\pm0.15$ (95\% confidence)
(with $Omega_\Lambda$ in the vicinity of $\sim $ 0.65).

Instead of entering further into the parameter estimation debate, we will 
review the recent investigations in cosmic topology which have
focused on the flat and the hyperbolic manifolds.  
For an interesting catalog of topologies for more general
Bianchi classes see Ref.\ \cite{bark}

\subsection{Observing the CMB}
\label{obscmb}

Many billions of years after last scattering, we build the COsmic
Background Explorer (COBE) satellite to confirm the earlier detections
by Penzias and Wilson of the microwave background radiation.
The all sky map generated by the COBE satellite confirmed that 
the average temperature was homogeneous and isotropic at 
2.728 ${}^o$ K 
and that the spectrum was extremely thermal, a result
that drew spontaneous applause when presented in 1992.  The data also
confirmed the theoretical expectation that there are in fact minute
fluctuations as predicted by inflation at the $10^{-5}$ level that
appear to be scale invariant.
COBE measures fluctuations on large scales, $\ell \le 30$.  Low $\ell$s
correspond to 
fluctuations on scales far outside
the horizon at the time of decoupling and therefore can only be due to
fluctuations in the metric and density. 
So COBE
which measures $\ell \lta 30$ is a probe of the largest geometric
features and therefore is quite important as a probe of topology as
well.  Causal microphysics becomes important for $\ell \gta 100$.
At these large $\ell$ our calculation of 
$\delta T/T$ in eqn.\ (\ref{sachswolfe})
is insufficient as it neglects microphysical effects and 
a more detailed analysis is required.
Of particular importance are the high $\ell$ Doppler peaks.
Before recombination, sound waves in the baryon-photon fluid induce
additional Doppler shifts which produce a peak in power
on a scale related to the size of the horizon at the time of
decoupling.  The height of the peaks depends on the specific model
parameters.  The location of the first peak depends primarily on curvature, 
$\ell_{\rm peak} \simeq 220 \Omega_0^{-1/2}$, with some small $H_0$ dependence.
\cb does not measure high $\ell$ fluctuations but two important all-sky
satellites will, Microwave Anisotropy probe (MAP) and {\it Planck Surveyor}.
The recent balloon borne experiments locate
$\ell\sim 200$ and therefore contribute more evidence in favor of
$\Omega_0\sim 1$.

The task of the Differential Microwave Radiometer (DMR) experiment on 
the COBE satellite was to 
measure the large angle 
anisotropy of the entire sky \cite{{bennet},{smoot}}.
The temperature \cb measures in a given pixel can be written
	\be
	\left (\delta T\over T\right )_i=
	\sum_{\ell m} a_{\ell m} B_\ell Y_{\ell m}(\vec x_i)
	+n_i
	\ee
where $B_\ell$ is the experimental beam pattern and the noise in each pixel is 
$n_i$.  The DMR horns are characterized by an imperfect Gaussian beam pattern.
The noise is assumed to be uncorrelated and
Gaussian with mean 
$<n_i>=0$ and variance $ <n_in_j>=\sigma \delta_{ij}$.

\begin{figure}[tbp]
\centerline{
\psfig{file=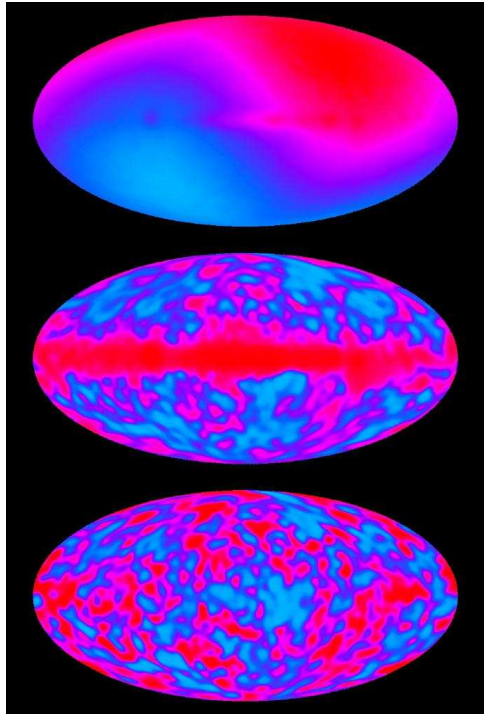,width=2.5in}
} \vskip 15truept
\caption{The three panels are, from top to bottom, the combined
data before the dipole is subtracted, the data after dipole subtraction but
before the galactic cut, the map minus both dipole and galactic emission.
Taken from the COBE webpage
at http://space.gsfc.nasa.gov/astro/cobe/
}
\label{cobepic}
\end{figure}

The experiment consists of three pairs of antennae.  Each pair measures
the temperature difference in two directions separated by $60^\circ$.
Each antenna has a $7^{\circ}$ beam and the data is smoothed on $10^\circ$.
A full sky scanned was performed several times over 4 years.
The \cb data is provided in the form of 6 maps in 3 frequencies, 31,
53
and 90 GHz \cite{{smoot},{bennet}}.
Each map has $N_p=6144$ of size $(2.6^o)^2$.  
Compressing to resolution 5 pixels, $N_p=1536$ pixels of 
size $(5.2^o)^2$, loses no information and
is sometimes implemented in data analysis.

The observed
$C_\ell$s have been determined from the \cb data by Gorski 
\cite{kris}, by
Tegmark \cite{teg}, and by Bond, Jaffe and Knox
\cite{jaffe}.
The monopole ($\ell =0 $) is just the average temperature
itself and can be discarded from the data.  
The largest scale anisotropy is the dipole ($\ell=1$) generated by our solar
system's peculiar velocity relative to the CMB.  Since this is not
cosmological in origin, the entire dipole is discarded from the data.
The first relevant cosmological observations begin with $\ell =2$,
the quadrupole, and it is curious to note that the measured
quadrupole is low.  This may just be cosmic variance in action but compact 
manifolds do happen to predict low power on large scales.  This is discussed
at length in section \S \ref{obsflat}.
The six maps can be 
compressed
into one weighted-sum map
with the monopole and dipole subtracted.  A galactic cut
of a region $\pm 20^o$
around the
Galactic
plane is also needed to eliminate Galactic emission.
This finally messaged data is ready for comparison to theoretical predictions.

The signal-to-noise of \cb is $\xi=2$.  By comparison, the projected
sensitivity of MAP is $\xi=15$ with a resolution of $0.5^\circ$,
a tenfold improvement in signal-to-noise and 30 in resolution.
Planck has even higher resolution but will launch much later.

\subsection{Topology and the CMB}
\label{obstop}

Several aspects of the fluctuation spectrum are altered when the 
manifold is multiconnected.  The most conspicuous and consistent
signatures of topological identifications are as follows:
(1)  Multiconnectedness destroys
global isotropy for all but the projective space and destroys global
homogeneity for all but the the projective space and the hypertorus.
(2)  The spectrum of fluctuations is discrete reflecting the natural 
harmonics of the finite space.
(3)  Since the fluctuations must fit within
the finite space, there is a cutoff in perturbations with wavelengths 
which exceeds the topological scale in a given direction.
(4)  Geometric patterns are encrypted in the spatial correlations.  The
patterns reflect the repeated occurrence of topologically 
lensed hot and cold spots.

The proposed methods of scanning the CMB for
evidence of topology can be split into direct
statistical methods and geometric methods.  The direct methods
begin with a theoretical model, compute the temperature fluctuations
as was done in \S \ref{simpflat} and \S \ref{simphyp} for the simply
connected spaces, and determine the likelihood of the model parameters
against the data.  There are shortcomings with this brute force approach.
All conclusions are model dependent and there are an infinite number of
models.  The model is not just the manifold but also the orientation and
the location of the Earth and the initial primordial spectrum.  While
some conclusions can still be drawn there is an appeal to a model independent
attempt at observing topology.  That is where the geometric methods come in.
The geometric methods treat topological lensing much like gravitational
lensing.  One just observes the lensed images and their distribution to 
reconstruct the geometry of the intervening lens.  The most amusing example
is the correlated circles of \S \ref{obscirc} although other patterns
can emerge as well as described in \S \ref{obspat}.

Before preceding to review the known approaches to observing topology 
it is worth expanding on point (1) above.
The global anisotropy and inhomogeneity means that the correlation function
depends on the location of the observer and the orientation of the manifold.
Additionally, $C(\hat n,\hat n^\prime)$ depends on both $\hat n$ and 
$\hat n^\prime$ and not just the angle between them.  As a result,
the angular average performed in the definition of the multipole moments
$C_\ell$ discards important topological information.
The information lost in the angular average 
can be quantified
by an enhanced cosmic variance $\left <C_\ell^2 \right>$ 
\cite{bpsII}.
The correlation function can be decomposed into isotropic and 
anisotropic pieces,
	\be
	C(\hat n,\hat n^\prime)=C^I(\hat n,\hat n^\prime)
	+C^A(\hat n, \hat n^\prime).
	\ee
By isotropy
	\be
	C^I(\hat n,\hat n^\prime)=\sum_{\ell} {\ell +1/2\over
	\ell(\ell +1)}C_\ell P_\ell(\hat n\cdot \hat n^\prime)
	\ee
and by anisotropy
	\be
	C^A(\hat n,\hat n^\prime)=\int d\Omega_{\hat n}
	\int d\Omega_{\hat n^\prime} C^A(\hat n, \hat
	n^\prime)P_\ell(\hat n \cdot \hat n^\prime)=0
	\ee
so that the anisotropic piece is orthogonal to the Legendre
polynomials.
The expectation value of the estimator for the $C_\ell$ depends
solely on the isotropic piece by construction 
	\be
	\left < \tilde C_\ell\right >={\ell(\ell+1)\over 8\pi^2}
	\int  d\Omega_{\hat n} \int  d\Omega_{\hat n^\prime} 
	C(\hat n, \hat n^\prime)P_\ell(\hat n\cdot \hat n^\prime),
	\ee
but the variance
	\be
	{\rm var}(\tilde C_\ell)=\left < \tilde C_\ell^2\right >-
	\left < \tilde C_\ell\right >^2 
	\ee
contains anisotropic pieces.
This can be interpreted as very large error bars due to cosmic 
variance.  As a result, conclusions based on the $C_\ell$'s alone
are weak \cite{{bpsI},{bpsII}}.  
Many of the statistical analyses described below do only
examine the $C_\ell$s and so can only rule out a topology but never 
really confirm topology.

However, others have argued that this increased error is not very
large for compact hyperbolic spaces.  An argument by Inoue for instance
expresses the variance as the sum of a geometric variance and the usual cosmic
variance.  The geometric variance contains the additional variance due
to topology and is intrinsically small \cite{inoueprog}.  The smallness of
the geometric variance can be traced to the randomness of the 
eigenmodes on compact hyperbolic models (explored further in \S
\ref{numersol}) which generates what Inoue refers to as
geometric Gaussianity \cite{ktinoue}.

\newpage
\section{Observing flat topologies in the CMB}
\label{obsflat}

Since the flat spaces are most easily constrained
using the \cb data we begin with them and discuss hyperbolic manifolds
in all their glory separately.
While small universes can be ruled out, it is rather fascinating to 
note that 
large cases are marginally consistent with the data.
After all, the observed quadrupole is low and 
so the infrared truncation in power seen in flat spaces and described
below can actually create a better fit to the data.
However, such a marginal flat space, just 
coinciding with the size of the observable universe is unaesthetic 
at best.  In fact, there is no natural scale for the size of a flat
universe and they are not the favored small universe candidates for this 
reason.  Still, they provide an important and accessible testing ground
for methods of observation and we discuss them next.

\subsection{Direct methods in flat space}
\label{directflat}

\subsubsection{Simply connected flat space}
\label{simpflat}

In flat, simply connected space, the spectrum of fluctuations is well known.
We need to find the elements of eqn. (\ref{elem}) with $F(\eta)$ 
the solution to eqn. (\ref{feta}).
In flat space ${\cal H}=2/\eta$ 
and the solution to eqn (\ref{feta}) decays as
$\Phi^\prime\propto \eta^{-6}$.
Consequently,
in flat space $F(\eta)$ is effectively constant and there is no ISW effect
during radiation or matter domination.  (There is however an ISW effect
if the universe is dominated by a cosmological constant.  This may turn
out to be important in salvaging finite flat models if the universe is 
accelerating as the recent supernovae observations indicate \cite{saul}.)
The Laplacian in flat space is simply
	\be
	\left ({\partial^2_x}+{\partial^2_y}+{\partial^2_z}\right )
	\psi_{\vec k}= -k^2 \psi_{\vec k}
	\ee
with solutions 
	\be
	\psi_{\vec k}=
	\exp\left (i{\vec k}\cdot \vx 
	\right ).
	\ee
The potential of eqn.\ (\ref{elem}) can then be expanded in terms of these
as 
	\be
	{\Phi(\vx)}=\int_{-\infty}^{\infty} d^3 k
	\hat \Phi_{\vec k}\exp\left (i{\vec k}\cdot \vx 
	\right )
	\ee
All of the assumptions about the statistics and shape of the spectrum are
contained in the $\hat \Phi_{\vec k}$.  All other quantities are 
determined by the geometry of the space.  
In accordance with the 
inflationary prediction the $\hat \Phi_{\vec k}$ are
assigned an independent Gaussian probability distribution 
consistent with the flat space normalization
	\begin{equation}
	\left\langle \hat \Phi _{\vec k}
	\hat \Phi^* _{\vec k^\prime}\right\rangle 
	={\frac{2\pi^2}
	{k^3}}{\cal P}_\Phi (k)\delta^3 (\vec k-\vec k^{\prime })
	\ \    \label{assumef}
	\end{equation}
\cite{mkb}.

From eqn. (\ref{sachswolfe}) the Sachs-Wolfe
effect is simply
	\be
	{\delta T(\hat n)\over T}={\Phi(\hat n)\over 3}.
	\ee

The correlation function between any two points on the SLS is then, from 
eqn (\ref{corrf}), 
	\be
	C(\hn,\hnp)\propto \int
	{d^3 k\over k^3}
	{\cal P}_\Phi
	       \exp\left(i\Delta \eta \vec k \cdot (\hn - \hnp ) 
	\right )
	\label{eq:cnn}
	\ee
up to a normalization.
For a homogeneous and isotropic space, 
the correlation function depends only on the angular
separation between $\hn$ and $\hnp$.  Consequently all the information on 
the theoretical sky is in the angular average.
Using the orthogonality relations of the
Legendre polynomials in eqn (\ref{clong}) gives
	\ba
	C_\ell 
	&= &
	\frac{1}{4 \pi} \int d \Omega \int d \Omega' C(\hn, \hnp) P_\ell(\mu)
	\nonumber \\
	&=& {4\pi\over 25}\int_0^{\infty}{dk\over k}j^2_l(\Delta \eta k)
	{\cal P}_\Phi.
	\ea
If we assume the initial fluctuation is a powerlaw
${\cal P}_\Phi\propto k^{n-1}$, this can be integrated to give
\ba
C_\ell \propto {\Gamma(\ell +(n-1)/2)\over\Gamma(\ell+(5-n)/2)}
{\Gamma((9-n)/2)\over \Gamma((3+n)/2)}
\label{clflat}
\ea
\cite{bondef}.

As described in 
\S \ref{obscmb}, 
the \cb data has been analyzed over the years by many groups to 
determine the $C_{\ell}$'s observed.  Statistical likelihood
analyses are then performed to compare the theoretical model
with the \cb data.  The infinite, flat model is generally found
to be consistent with the data for a nearly flat power spectrum
($n\sim 1$).
The question addressed in the next section, is how well finite
flat models match the data.  To flash forward, the answer is essentially
that finite flat models match the data well if they are comparable in size
to half the observable universe although some studies have put 
even stronger limits as described below.  All limits are subject
to caveats based on the
assumptions made about the perturbation
spectrum -- none of which have yet been tested observationally. 

\subsubsection{Compact, flat spaces}
\label{compflat}

We have already reviewed the standard decomposition of the temperature
fluctuations in a simply connected space.  When the manifold is compact and
the SLS exceeds the dimensions of the space, then the global topology 
is reflected in the CMB sky.
There are three notable alterations 
to the predicted fluctuations when
the manifold is compact:  (1)  The eigenvalue spectrum is discrete not
continuous.  (2)  There is a cutoff in the power of fluctuations on wavelengths
which exceed the natural size of the space.
(3)  The correlation function $C(\hn,\hnp)$ depends on orientation and
so depends on both $\hn$ and 
$\hnp$ explicitly.  It is not simply a function of the angular separation.

The exact eigenmodes can be found for all of the 6 compact orientable 
flat spaces \cite{{lss},{slsi}}.  This allows a direct statistical 
comparison of the theoretical predictions for compact, orientable spaces 
with the \cb data.  With any statistical comparison of a model with the 
data, the conclusions are model dependent.
The reader should bear in mind that the bounds quoted assume equal-sided 
spaces, $\Omega_\Lambda=0$, and a flat Gaussian distributed
power spectrum.  

The earliest bounds on the hypertorus constrained the topology scale to be
$\gta 0.8\Delta \eta$ \cite{sss} which is still less than the diameter of the
SLS.
There could still be as many as eight
copies of our universe within the observable horizon.
The analysis was later extended to the other 
compact, orientable flat spaces where similar bounds were obtained 
\cite{{lss},{slsi}}.  Stronger bounds were put on the hypertorus by comparing
the full covariance matrix against the 2-year \cb data \cite{deO1}.
The length of a side was set to be $\gta 1.2 \Delta \eta $.

The most conspicuous feature in the fluctuation
spectrum is a suppression of power
on large scales since large fluctuations cannot fit into the finite box.
Such an infrared cutoff can be deduced from Cheeger's inequality \cite{cheeger}
	\be
	k_{min}\ge {h_C\over 2},
	\quad \quad h_C=\inf_X{A(S)\over {\rm min}\left (V(M_1),V(M_2)\right )}
	,
	\ee
with the infimum taken over all possible surfaces $S$ that divide the space
into two subspaces ${\cal M}_1$ and ${\cal M}_2$ where 
${\cal M}={\cal M}_1\cap {\cal M}_2$.  $S$ is the boundary of the two subspaces,
$S=\partial {\cal M}_1=
\partial {\cal M}_2$. 
The isoperimetric constant $h_C$ depends on the geometry more than 
the topology.  Intuitively speaking, in very long thin manifolds
$h_C$ can be quite small leading to a lowered spectrum of eigenvalues.
However, this is highly correlated with thin bottleneck structures and is not
a common feature of spaces with a more regular shape \cite{{cheeger},{buser}}. 
From Cheeger's inequality it follows that
all flat hypertori have $k_{min}\ge 2/L$ with $L$ the longest side of the 
torus
as argued in \cite{bpsI}.  

Although the cutoff would seem a good indicator of topology, it just so happens
that
the longest wavelength
fluctuation observed, namely the quadrupole, is in fact low.
Some might even take this as evidence for topology \cite{workshop}.
Cosmic variance is also large on large scales.
Consequently, a fundamental domain the size of the observable universe
is actually consistent with the observed \cb $C_\ell$s \cite{lss}.  
However, since the fundamental domain has a particular orientation on the sky, 
the correlation is not simply a function of the angular separation
between $\hn$ and $\hnp$ as it is in the infinite case.  
Therefore conclusions based on the $C_\ell$s alone are weaker than a 
statistical analysis based on the full $C(\hat n, \hat n^\prime)$.
The strongest bounds on the hypertorus were placed
by comparing the full 
correlation function $C(\hat n, \hat n^\prime)$ to the data.
It was found that $L\gta 1.3 \Delta \eta$ \cite{{deO1},{bpsI},{bpsII}}, which 
is still less than the diameter of the observable universe, 
$\sim 2\Delta \eta$.
Asymmetric spaces were constrained by de Oliviera-Costa et. al. as described
in section \S \ref{geomflat}.


Despite the bounds placed on the equal-sided
compact, flat spaces, they still provide an excellent testing ground for 
geometric measures of topology.  They may also be saved by 
$\Omega_\Lambda\ne 0$ models.  We compile a list of
all the eigenmodes, eigenvalues and relations for the compact flat spaces
for completeness.  
These solutions are taken from Ref. \cite{slsi}.  The spectra were also
found in Ref. \cite{sss} for the other compact spaces, however, the critical
relations were overlooked leading to confusion about the cutoff in the
long wavelength modes.  It was mistakenly concluded because of the missing relations that in the twisted spaces longer
wavelengths could fit in the fundamental domain
since a wave needs to wrap more than once before coming back to 
a fully periodic identification \cite{css2}.
However it was shown in Ref. \cite{slsi}, that these long modes are 
forbidden 
by the relations among the $\hat \Phi_{\vec k}$, and all of the compact 
topologies have roughly the same long wavelength cutoff.

As a result of the global topology, 
all of these spaces are anisotropic and all except for the 
hypertorus are inhomogeneous.  Topology can be implemented by
imposing the boundary conditions
	\be
	\Phi_{\vec k}(\vx)=\Phi_{\vec k}(g\vx) \quad \forall g\in\Gamma
	.
	\ee
Compact topology 
always restricts the eigenvalues to a discrete
spectrum:
	\be
	\Phi(\vec x)=\sum_{\vec k}\hat \Phi_{\vec k} e^{i\vec k\cdot \vec x}
	\ \ .	
	\label{eigensum}
	\ee
In additional, relations
are imposed on the $\hat \Phi_{\vec k}$.

\begin{figure}
\centerline{\psfig{file=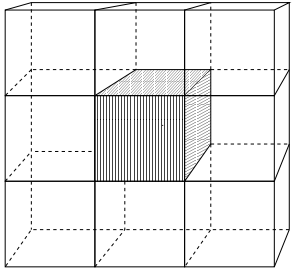,width=1.5in}}
\caption{Tiling flat space with parallelepipeds.}
\label{para}
\end{figure}

Demonstrating the discretization 
explicitly for the hypertorus, the identifications of the 
cube are expressed in the three boundary conditions 
$\Phi(x,y,z)=\Phi(x+L_x,y,z)=\Phi(x,y+L_y,z)=\Phi(x,y,z+L_z)$.
Imposing the first boundary condition on $\Phi(\vec x)$ of 
eqn.\ (\ref{eigensum}) gives
$e^{-ik_xx}=e^{-ik_x(x+h)}$ which requires $k_x={2\pi\over l_x}n_x$ with 
$n_x$ an integer.  The other two boundary conditions provide the discrete
spectrum
	\be
	k_x={2\pi \over l_x} n_x\quad \quad
	k_y={2\pi \over l_y} n_y\quad \quad
	k_z={2\pi \over l_z} n_z
	\ \ 
	\ee
with the $n_i$ running over all integers \cite{{zel73},{fanghoujun},
{sss}}.
It is clear that there is a minimum eigenvalue and hence a maximum wavelength
which can fit inside the fundamental domain defined by the parallelepiped
\cite{sss}:
	\[
	k_{\rm min}=2\pi \ {\rm min}\left ( {1\over L_i}
	\right )\quad 
	{\lambda}_{\rm max} ={\rm max}\left (L_i\right)\  .
	\nonumber
	\]
Although global anisotropy is broken by topology, we can still form
the $C_\ell$s to make a comparison, if incomplete, with the \cb 
$C_\ell$s.
Expanding the exponential and 
Legendre polynomials in terms of spherical harmonics, the 
$C_\ell$ eqn.\ (\ref{clong}) becomes
\be
C_\ell \propto
\sum_{\vec k} \sum_{\vec k'}
{{\cal P}_{\vec k}\over k^3}
 j_\ell^2(\Delta \eta k) 
\label{eq:cl}
\ee
Three other spacetimes are constructed from a
parallelepiped.
The first twisted parallelepiped 
has opposite faces identified with one pair of faces
twisted by $\pi $ before gluing.
The eigenmodes are 
$\vec k=2\pi(j / h, w/b,n/2c)$, with the additional relation 
$\hat \Phi_{jwn}=\hat \Phi_{-j -w n}\ e^{i {\pi}n}$.  The minimum 
mode consistent with the relations is $k_{\rm min}=2\pi/L$.  The
$C_\ell$s become
	\be
C_\ell \propto
\sum_{jwn} 
\frac{{\cal P}(k)}{k^3} 
\frac{j_\ell(\Delta \eta k)^2}
{2 \ell + 1} 
\sum_{m = -\ell}^{\ell}
|Y_{\ell,m}({\bf \hat k})|^2
( 1 + e^{i \pi (n+m)}).
	\ee

A
parallelepiped with 
one face rotated by $\pi/2$ has
discrete eigenmodes 
$\vec k=2\pi(j / h, w/b,n/4c)$, with the additional relations
$\hat \Phi_{jwn}=\hat \Phi_{w -j n}e^{in\pi/2}
=\hat \Phi_{-w -jn}e^{in\pi}=\hat \Phi_{-w j n}e^{i 3n\pi/2}$.
Again the minimum mode is $k_{\rm min}=2\pi/L$.
The $C_\ell$s are given by
\be
C_\ell  \propto
\sum_{jwn} 
\frac{{\cal P}(k)}{k^3} 
\frac{j_\ell(\Delta \eta k)^2}
{2 \ell + 1} \sum_{m = -\ell}^{\ell}
|Y_{\ell,m}({\bf \hat k})|^2 
( 1 + e^{i (n+m){\pi}/2} +
e^{i (n+m){\pi}} + e^{i 3(n+m){\pi}/2}).
\ee

The last parallelepiped 
has a fundamental domain of volume $2 h b c$ and is thus
a double parallelepiped.
The identification rules involve three rotations through
$\pi$ as shown in \cite{wolf}.
They are
so that $(x,y,z)\rightarrow (x+h,-y,-z)$, 
$(x,y,z)\rightarrow (-x,y+b,-(z+c))$, and 
$(x,y,z)\rightarrow (-(x+h),-(y+b),z+c)$. 
The resultant discrete spectrum is
$\vec k=\pi(j / h, w/b,n/4c)$ with relations
$\hat \Phi_{jwn}
=\hat \Phi_{j-w -n}\ e^{i\pi j}
=\hat \Phi_{-jw -n}\ e^{i\pi(w+n)}
=\hat \Phi_{-j-w n}\ e^{i\pi(j+w+n)}$.
The minimum mode is $k_{\rm min}=\sqrt{2}\pi/L$.
Since the volume of the space is $2L^3$, the
angular averaged long wavelength
is $\sqrt{2}/2^{1/3}\simeq 1.1 $ times the length of the fundamental domain.
The angular power spectrum is
\ba
C_\ell & \propto&
\sum_{jwn} 
\frac{{\cal P}(k)}{k^3} 
\frac{j_\ell(\Delta \eta k)^2}
{2 \ell + 1} \sum_{m = -\ell}^{\ell}
Y_{\ell,m}^*({\bf \hat k}) \times \\ \nonumber
&\big(&
Y_{\ell,m}({\bf \hat k}) (1 + e^{i (m+j+w+n)\pi}) 
+ Y_{\ell,-m}({\bf \hat k}) e^{i \ell \pi}
(e^{i (m+j)\pi} + e^{i (w+n)\pi}) \big).
\ea

\begin{figure}
\centerline{\psfig{file=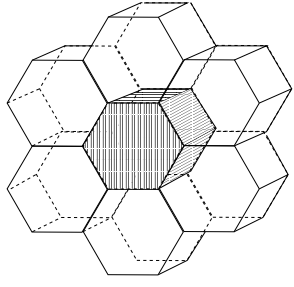,width=1.5in}}
\caption{Tiling flat space with hexagonal prisms.}
\label{hexag}
\end{figure}

The last two possible compact flat spaces are based on a
hexagonal tiling. With
the opposite sides of the hexagon identified and the prism faces
rotated relative to each other by 
$2\pi/3$, the potential can be written 
	\be
	\Phi =\sum_{n_2 n_3n_z} \hat \Phi_{n_2 n_3 n_z}
	e^{ik_z z}	 
	\exp {\left [
	i{2\pi \over h}\left [
	n_2\left ( x -{1\over \sqrt{3}} y \right )
	+n_3\left ( x +{1\over \sqrt{3}} y \right )\right ] \right ]}
	\label{hexmodes}
	\ee
with the eigenmodes
$\vec k=2\pi((n_2+n_3) / h, (-n_2+n_3)/b,n_z/3c)$.
The relations on this space are
$\hat \Phi_{n_2, n_3, n_z}
=\hat \Phi_{n_3, -(n_2+n_3), n_z} e^{i2\pi n_z/3}
=\hat \Phi_{-(n_2+n_3), n_2, n_z} e^{i4\pi n_z/3}$
and lead to
\be
C_\ell  \propto
\sum_{n_2n_3n_z} 
\frac{{\cal P}(k)}{k^3} 
\frac{j_\ell(\Delta \eta k)^2}
{2 \ell + 1} \sum_{m = -\ell}^{\ell}
|Y_{\ell,m}({\bf \hat k})|^2 
 ( 1 + e^{i 2(n_z+m){\pi}/3} + e^{i 4(n_z+m){\pi}/3}).
\ee

Lastly, the prism faces are glued
after rotation by $\pi/3$.
The potential can still be written as (\ref{hexmodes}), with eigenmodes
$\vec k=2\pi((n_2+n_3) / h, (-n_2+n_3)/b,n_z/6c)$
and a set of relations 
$\hat \Phi_{n_2, n_3, n_z}
=\hat \Phi_{(n_2+n_3), -n_2, n_z} e^{i\pi n_z/3}
= \hat \Phi_{n_3, -(n_2+n_3), n_z} e^{2i\pi n_z/3}
= \hat \Phi_{-n_2,-n_3, n_z} e^{i\pi n_z}
= \hat \Phi_{-(n_2+n_3), n_2, n_z} e^{i4\pi n_z/3}
= \hat \Phi_{-n_3, (n_2+n_3), n_z} e^{i5\pi n_z/3}$.
The $C_\ell$s are given by
\ba
C_\ell & \propto&
\sum_{n_2n_3n_z} 
\frac{{\cal P}(k)}{k^3} 
\frac{j_\ell(\Delta \eta k)^2}
{2 \ell + 1} \sum_{m = -\ell}^{\ell}
|Y_{\ell,m}({\bf \hat k})|^2 \\ \nonumber
&(& 1 + e^{i (n_z+m){\pi}/3} + e^{i 2(n_z+m){\pi}/3} 
+ e^{i (n_z+m){\pi}} + e^{i 4 (n_z+m){\pi}/3} 
+ e^{i 5 (n_z+m){\pi}/3}).
\ea
The volume of both of these topologies is $h^2 c \frac{\sqrt{3}}{2}$.

In summary, both the $\pi$-twisted and 
$\pi/2$-twisted tori have $C_\ell$s that are almost identical to that of the
torus.  
The harmonics do show distinctions 
between the parallelepiped topologies, but unfortunately
the variations fall well within cosmic variances.
In addition to the damping at low $\ell$, harmonics of the discrete 
spectrum create both dips and enhancements extending up to high values of 
$\ell$.  The dips are a natural consequence of the 
discretization and the resultant absence of certain modes in the 
spectrum.  The enhancements are due to the distribution 
of multiple images dictated by the geometry of the fundamental domain.

As a result of the low observed quadrupole, 
the predicted cutoff alone is not enough to rule out compact, flat models.
The likelihood of the $C_\ell$ in the full \cb range was 
compared to the relative likelihood of a flat, infinite cosmology in 
Ref. \cite{{sss},{lss},{slsi}}.  
Compact, flat spaces are 
tens of times less likely than their infinite counterparts if the topology scale exceeds about half
the diameter of the SLS.  
Since compact topologies do not give isotropic temperature
fluctuations, this lends ambiguity to any likelihood analysis.  The 
conclusions drawn were quite conservative, ruling out $\ell \gta 0.8 \Delta 
\eta$; that is, the length of a side must be $\gta 0.4$ the diameter
of the observable universe.  Again, it should be emphasized that these 
conclusions are contingent on the correctness of the
assumptions about the initial perturbation spectrum.

While the angular power spectrum
is sufficient to constrain symmetric, flat topology it is in general a poor
discriminant. The average over the sky fails to recognize the strong
inhomogeneity and anisotropy manifest in these cosmologies. 
Direct attacks on anisotropic spaces inspires more geometric approaches
as discussed next.

\subsection{Geometric methods in flat space}
\label{geomflat}


One of the first geometric, or pattern driven searches for topology was 
initiated in Ref. \cite{deO2}.  For an anisotropic hypertorus,
a symmetry plane or a symmetry axis can be identified in the CMB
\cite{{sokolov},{star},{fang}}.
For these asymmetric spaces, 
a search for patterns was emphasized in Ref. \cite{deO2}.
They develop a statistic that is independent of cell orientations but is
sensitive to the plane and axis symmetries of the rectangular spaces.
For
anisotropic models there is a stronger dependence on the orientation of the
manifold relative to the Galaxy cut.  
The statistic $S(\hat n_i)$ searches for reflection symmetries in a plane
perpendicular to $\hat n_i$ and is defined by
	\be
	S(\hat n_i)={1\over N_{pix}}
	\sum_{j=1}^{N_{pix}}{\left [{\delta T(\hat n_j)\over T}
	-{\delta T(\hat n_{ij})\over T}\right ]^2\over
	\sigma^2(\hat n_j)+\sigma^2(\hat n_{ij})}.
	\ee
$N_{pix}$ is the number of pixels after the Galaxy cut and 
$\sigma (\hat n)$ is the r.m.s. error associated with the pixels
in the direction $\hat n$.  The object
$\hat n_{ij}$
is the reflection of $\hat n_j$ in the plane with normal $\hat n_i$,
	\be
	\hat n_{ij}=\hat n_j-2(\hat n_i\cdot \hat n_j)\hat n_i.
	\ee
Lower values of $S(\hat n_i)$ corresponds to a higher degree of symmetry.
The temperature fluctuation is computed using the eigenmode expansion and 
a flat spectrum.  The resultant $\delta T/T$ depends on 
six parameters: the three spatial orientations fixing
the domain orientation and the three topology scales.

The smaller a given direction the more extreme the asymmetry.
In $T^1$ for instance, if the $z$ direction is extremely small,
then the size of fluctuations will be notably smaller in this direction
leading to a map where there is essentially less and less structure
at \cb resolution in this one direction and the structure will be drawn out
on the $(x,y)$ plane.  Similarly for $T^2$ there will essentially only be
structure along the large direction and so the pattern will appear to be rings
of fluctuations along the small directions.


Using this statistic, $T^1$ and $T^2$ models were constrained to be greater
than half the radius of the SLS
in their small dimensions.  Subsequently, the circle method was
used to study specific asymmetric models \cite{boudcirc}  where again
pessimistic conclusions were reached
as discussed in \S \ref{obscirc}.

\newpage
\section{Observing hyperbolic topologies in the CMB}
\label{obshyp}


The eigenmode decomposition is well known on simply connected $\ah$ 
and is given in \S \ref{fluctcmb}.  
However, when hyperbolic space is fully compact
the decomposition becomes intractable.  This is a stronger statement
than simply saying the eigenmodes and eigenvalues are {\it difficult}
to find.  It is actually formally impossible to write down the eigenmodes
analytically.  The boundary conditions are so intricate they resist
decomposition \cite{bv}.

The absence of an analytic solution to the eigenmode spectrum can be
directly related to the incipient chaos on compact manifolds.
On simply connected $\ah$ geodesics show the first critical ingredient
for the onset of chaos; that is, extreme sensitivity to initial conditions.
The geodesic deviation equation shows that two nearby trajectories 
diverge away from each other exponentially quickly:
	\be
	{D^2\zeta^\mu\over ds^2}=-R^\mu_{\alpha\beta \gamma}u^\alpha
	\zeta^\beta u^\gamma
	\ee
where $u^\mu=dx^\mu/ds$ and $\zeta^\mu$ is the separation of neighboring
geodesics.   Normal to the geodesic flow this becomes \cite{et}
	\be
	{d^2||\zeta_N^\mu\zeta_{N\mu}||\over ds^2}
	=-2\kappa||\zeta_N^\mu\zeta_{N\mu}||
	\ee
where $\kappa$ is the curvature.  If $\kappa<0$, these geodesics diverge
exponentially and a coordinate invariant Lyupanov exponent can be 
interpreted as $\lambda=\sqrt{|\kappa|}$.
Still, there is no chaotic motion on infinite $\ah$ since there
is no mixing and no folding of trajectories.  In other words, there is 
no loss of predictability as a result of the exponential deviation.
By contrast, when the space is made topologically compact, the mixing
and folding of trajectories is assured and the flows are well known to 
be fully chaotic.
An entropic measure of the chaotic flow
can be related to the volume of the manifold through 
the Kolmogorov entropy $S_K\propto {\cal V}^{-1/3}$
where ${\cal V}$ is the volume of the spacetime \cite{sinai}.  
Notice that if the space is infinite
this entropy vanishes.

Chaotic flows on compact 2-dimensional manifolds in particular 
have been studied at great length \cite{{bv},{aurichsteiner},{gutz},
{ott}}.  The cosmological implications of chaos have only been touched 
upon \cite{{css3},{lb_fractals}} and remain a largely unexplored terrain.
By and large, people have tried to obviate the chaotic
flows entirely when analyzing the CMB.  Methods include
brute force numerical determination of the modes, the method of 
images construction of CMB maps, and geometric methods.  We will 
discuss a catalog of such studies.

\subsection{Direct methods in hyperbolic space}
\label{directhyp}

\subsubsection{Simply connected hyperbolic space}
\label{simphyp}

Since all of the numerical methods will rely on the expansion of the
eigenmodes on the universal cover, we summarize those results here.
From the standard expansion (\ref{elem}).  We need to determine each 
of the factors in the decomposition.
In negatively curved space perturbations are time dependent according
to eqn (\ref{feta}) with solution
	\be
	F(\eta)={5\left (\sinh^2\eta-3\eta\sinh\eta+4\cosh\eta+4\right
	)\over \left (\cosh\eta-1\right )^3}.
	\ee

In a negatively curved space the Helmholtz eqn becomes
	\be
	{1\over \sinh^2 r}\left [
	\partial_r\left (\sinh^2r{\partial _r}\right )+
	{1\over \sin^2\theta} 
	{\partial_\theta}\left (\sin\theta \partial_\theta
	\right )+{1\over \sin^2\theta}\partial^2_\theta \right]\psi_{\vec k}
	=-k^2
	\psi_{\vec k}.
	\ee
As emphasized in Ref. \cite{lyth}, a complete orthonormal basis for 
the simply connected space is formed
by modes with real $k^2>1$ and so the eigenvalue range
is $k^2=[1,\infty]$.  These modes vary on a scale below the curvature
radius and are therefore subcurvature modes.  Standard causal theories 
such as inflation will not seed fluctuations on scales $k<1$ although some
other unforeseen mechanism may. 
It is customary to introduce another parameter 
	\be q^2=k^2-1\ee
with  the 
range $q^2=[0,\infty]$.
The eigenmodes on $\ah $ can be expressed as
	\be
	\psi_{q \ell m}=X_{q\ell}(r)Y_{\ell m}(\hat n)
	\ee
where the $Y_{\ell m}$'s are the usual spherical harmonics and
the $X_{q\ell}$ are the radial functions
	\be
	X_{q\ell}={\Gamma(\ell +1+iq)\over \Gamma(iq)}
	\sqrt{1\over \sinh r}P^{-\ell -{1\over 2}}_{iq-{1\over 2}}(\cosh r)
	\ee
where $\Gamma$ denotes the Gamma function and $P$ denotes the Legendre functions
\cite{lyth}.

The $\hat\Phi_q$ are drawn from a Gaussian probability distributions
with proper normalization for hyperbolic space
	\be
	\left <
	\hat \Phi_{q\ell m} \hat \Phi^*_{q^\prime\ell^\prime m^\prime}
	\right >
	={2\pi^2\over q(q^2+1)}{\cal P}_\Phi(q) \delta({ q- q^\prime})
	\delta_{\ell\ell^\prime}\delta_{m m^\prime}
	\ee

We can summarize the Sachs-Wolfe
effect as
	\be
	{\delta T(\hat n)\over T}=\int d^3{\vec k} \hat\Phi_{\vec k}(\hat n)
	L_{\vec k}
	\ee
with both the surface Sachs-Wolfe and the integrated Sachs-Wolfe accounted
for in 
	\be
	L_{\vec k}=
	\left [ {1\over 3}F(\eta_{sls})+2\int_{\eta_{sls}}^{\eta_0}
	d\eta F^\prime (\eta)\right ] \psi_{\vec k}(\vec x)
	\ \ .
	\ee

The Fourier multipoles are 
	\be
	C_\ell=2\pi^2\int_1^{\infty}{dk\over k(k^2-1)}{\cal P}_\Phi(k)
	\left [{1\over 5}X_{k\ell}(\Delta \eta)+{6\over 5}
	\int_{\eta_d}^{\eta_0}dr\ X_{k\ell}(r)F^\prime(\Delta \eta-r)
	\right ]^2.
	\ee
The \cb data alone tends to favor $\Omega\sim 0.3-0.4$
(see for instance \cite{krzy}) although both very low 
$\Omega $ and an $\Omega \sim 1$ are still compatible.
For a long time, the COBE results along with other astronomical
observations \cite{spergelcqg} put low $\Omega $ cosmologies
in the limelight.  However, more recent high $\ell $ observations \cite{exper}
in conjunction with the supernovae data \cite{saul} have 
pushed public opinion back towards $\Omega=1$ cosmologies.
The recent high $\ell $ observations only examine small patches of 
the sky.  A full sky map is required to advance topology observations
and so we reserve more commentary until the launch and hopeful
success of the future MAP and {\it Planck} missions.

\subsubsection{Cusps}
\label{obscusps}

There are multiconnected hyperbolic spaces which are not completely
compact (see Ref. \cite{lum} for some examples
such as those of Refs. \cite{{lobell},{seifertweber}}). 
For some of these the motion is not chaotic
and it is possible to find the
eigenmodes.  One topology of particular interest is the toroidal horn
\cite{{sos},{lbbs}}.  The horn is interesting since many
manifolds have cusped corners as stressed by \cite{css2}.  
Despite
their frequency in a set of generic manifolds,
it was argued in Ref. \cite{css2} that according to the thick-thin
decomposition,
it would very improbable for us to live
in the thin part of the manifold.  However, this argument is flawed.
Since the 
cusp narrows exponentially quickly, an observer can live in a fat
part of the manifold and still see photons coming from a constricted
part of the cusp.  
However, Olson and Starkman have developed an approximation scheme to study
cusps on full compact manifolds which shows the generic patterns do in 
fact persist.

\begin{figure}
\centerline{\psfig{file=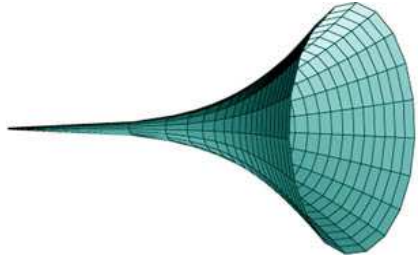,width=2.5in}}
\caption{Embedding of the cusp topology with one of the dimensions
suppressed.}
\label{horn}
\end{figure}

The cusp is most easily understood in the upper half space representation
of $\ah $ (see Appendix \ref{AppendixA})
	\be
	ds^2=-d\eta^2+dz^2+e^{-2z}\left (dx^2+dy^2\right )
	\label{coordsys}
	\ee
with the coordinate transformation
$e^{-z}=\cosh r-\sinh r\cos \theta, e^{-z}x=\sin\theta\cos\phi\sinh r,
e^{-z}y=\sin\theta\sin\phi\sinh r$.
The $(x,y)$ subspace looks like a conformally stretched flat space.
The topological identifications $x\rightarrow x+L_x$ and $y\rightarrow y+L+y$
make a 2-torus which is squeezed in area
by the exponential factor $e^{-2z}$ 
into a tightening cusp.
Since the subspace is conformally flat, there are no tangled geodesics and 
it is possible to decompose the eigenvalues analytically.

In the coordinate system (\ref{coordsys}),
the temperature fluctuations can be decomposed as
\begin{equation}
{\frac{\delta T}T}(\hat n_y)=\int_0^\infty dq\ \sum_{n_xn_y}\hat \Phi _{qn_xn_y}
(\hat n)N_{qn_xn_y}L_{qn_xn_y}  \label{it}
\end{equation}
with the normalization 
\begin{equation}
N_{qn_xn_y}=\left( {\frac{2q\sinh (\pi q)}{\pi ^2}}{\frac{(2-\delta
_{n_x0})(2-\delta _{n_y0})}{bh}}\right) ^{1/2}
\end{equation}
and
\be
L_{qn_xn_y}=\left[ {\frac 13}+ 2\int_{\eta _{{\rm i}}}^{\eta
_o}d\eta F^{\prime }(\eta )\right] e^zK_{iq}(Qe^z)\times   
 \pmatrix{ \sin\left({2\pi n_x\over b}x\right )\cr \cos\left({2\pi
n_x\over b}x\right )\cr}\pmatrix{ \sin\left({2\pi n_y\over h}y\right )\cr
\cos\left({2\pi n_y\over h}y\right )\cr}\ \ 
\ee
where $K_{iq}$ is a modified Bessel function with imaginary index. 
The
entire function is included here in the integration over $\eta $. The
argument of the Bessel function is
$Q^2=4\pi ^2(n_x^2/L_x^2+n_y^2/L_y^2)$.  Notice that $q$ is still a continuous
index since the $z$ direction is infinite.  There is no cutoff in 
the range of $q$ but there is a cutoff in the range of $(n_x,n_y)$.
This cutoff results in a severe suppression of power
as an observer lives nearer the cusp.  The suppression appears as a flat
spot in simulations of the \cb sky with concentric rings
of larger and larger structures as the cusp widens \cite{lbbs}.  

A standard likelihood analysis of the $C_\ell$s places the size of the 
torus at the location of the Earth to be $\gta \Delta \eta$.  This is 
not a particularly narrow part of the cusp relative to the SLS but nonetheless 
an observer can 
still see exponentially
deep down the throat putting large quiet regions in the CMB maps \cite{lbbs}.

The generic property of flat spots 
was tested in Ref. \cite{olsonstark}.  They showed that from a typical
location in a cusped manifold m003, flat spots of angular size $\sim 5^\circ$
would be visible for $\Omega_o=0.3$.  They conjecture that observable
flat spots on this
scale are typical of cusped manifolds.
To handle the fully compact case Olson and Starkman found a means to 
approximate the modes in the cusp without tackling the full eigenmode 
solutions.
They consider a horosphere, a sphere within the Poincar\'e representation
of $\ah$, which is tangent to the cusp and passes through the point
on the SLS.  On the horosphere, the transformation group simplifies and they
are able to isolate modes.

The specific space they consider is a cusped manifold, namely m003, constructed
from two tetrahedra with total volume
${\cal V}\approx 2.0299$.  The restriction to the horosphere results in 
a hexagonal tiling of $\eu$ and the eigenmodes are reminiscent of those
of eqn (\ref{hexmodes}).  Consequently the minimum eigenmode from
the hexagonal tiling provides an estimate for the extent of the flat spot.
Situating the observer at a symmetric point in the manifold they find  
that flat spots in the CMB from cusped regions subtend an average angle
of roughly $5^\circ$ for $\Omega_0=0.3$.  Larger values of $\Omega_0 $
will naturally diminish the angular scale.
They argue that in general cusped manifolds will show flat spots as 
characteristic features in the CMB maps.

\begin{figure}
\centerline{\psfig{file=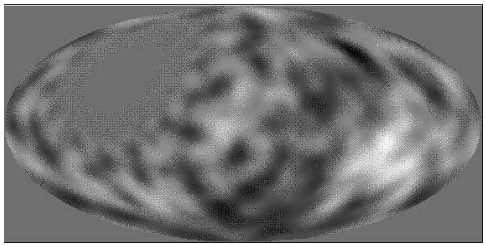,width=2.5in}}
\caption{Flat spot in a map of the sky in the cusp topology with 
scales
$b=h=1$ and $\Omega_0=0.3$}
\label{flatspot}
\end{figure}

\subsubsection{Numerical Eigenvalue Solutions}
\label{numersol}

The eigenmodes can always be obtained by brute force numerically.
Three groups have developed numerical methods to isolate a list of 
eigenmodes and eigenvalues for a small collection of manifolds
\cite{{inoue},{inouetomita},{inouethesis},{cornsperg},{aurich}}.  
A comparison of the simulated CMB sky with the \cb data shows the 
specific compact hyperbolic manifolds studied were consistent
with the data. However
all of the statistical analyses 
compare only the 
$C_\ell$'s and not the full correlation functions.
Using the method of images to simulate the CMB,
a full statistical analysis is performed 
by Bond, Pogosyan, and Souradeep as described in section \S \ref{obsmi}
where more negative 
conclusions are obtained; namely that the spaces studied were 
inconsistent with COBE except for maybe one orientation of the manifold.
It would be interesting if numerical eigenmode constructions 
would compare a full
statistical analysis to that performed using 
the method of images to check for consistency.

\centerline{\bf Eigenmodes of the Thurston Space}

Inoue was the first to find precise eigenmodes on a compact hyperbolic
3-space for the purposes
of simulating a cosmological model.  
(Aurich and Marklof were the first to compute the eigenmodes on an 
orbifold \cite{aurichmarklof}.)
His method was based on the direct 
boundary element method initially developed by Aurich and Steiner for the
study of $2$-dimensional compact hyperbolic spaces.
The numerical method is based on solving the Helmholtz equation
	\be
	(\nabla^2+k^2)\psi_{\vec k}(\vx)=0
	\label{helm}
	\ee
on a compact manifold, i.e.
with periodic boundary conditions on the universal
covering space.  Various methods for doing so
include the finite element method and the
finite difference method.
He uses the more precise but numerically more time consuming
method called the direct boundary element method
(DBEM) 
also used by Aurich and Steiner to study quantum chaos on 
a 2D compact space \cite{aurichsteiner}.
He isolated 
the first 36 eigenmodes of the 
manifold $m003(-2,3)$ from the SnapPea census, otherwise
known as the Thurston space in homage to the Fields Medalist 
W. Thurston.  
There are two particularly interesting results.
Firstly he finds a cutoff in the low $k$ eigenvalues and therefore
a maximum wavelength at a value
near the average diameter of the space.
Secondly, 
he discovered that 
the expansion coefficients of the eigenmodes ($\xi_{qlm}$ below)
are well described as 
pseudo-random in behavior
\cite{inoue}.
Despite the long-wavelength cutoff, 
the integrated Sachs-Wolfe effect is able to compensate for low 
$\Omega_0$, as expected.  Consequently, 
assuming an 
Harrison-Zeldovich power spectrum,  
he finds that the $C_\ell$'s are consistent 
with \cb for $0.1 \le \Omega_0 \le 0.6$.

\begin{figure}
\centerline{\psfig{file=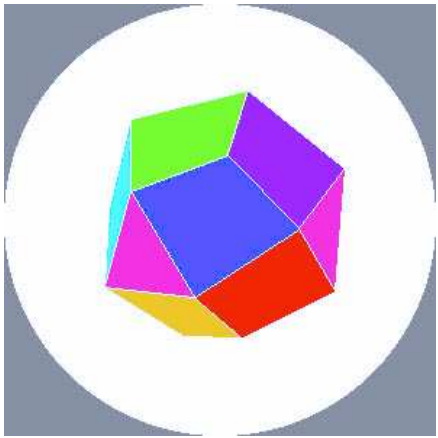,width=2.25in}}
\caption{
Dirichlet domain for the Thurston space}
\label{thurs}
\end{figure}

As in flat space, the eigenvalues are a discrete subset of the 
continuous eienvalues.  Since, the eigenmodes of a compact manifold
are a
subset of the eigenmodes on the universal cover (\S \ref{simphyp}), 
they can be expanded as
	\be
	\psi^{\cal M}_{q }=\sum_{\ell m}
	\xi_{q\ell m}X_{q\ell}(r)Y_{\ell m}(\hat n).
	\label{modexpan}
	\ee
The $\xi_{q\ell m}$ can be thought of us as containing the 
analogue of the relations found explicitly in flat space
in section \S \ref{compflat}.
The challenge of numerically isolating the modes can be reduced to the 
still difficult task of finding the $\xi_{q\ell m}$.
Inoue explicitly finds that the $\xi_{q \ell m}$ behave as random Gaussian 
numbers for ($\ell <19$ and $q>9.94$).  He extrapolates that this continues
to be true, even more so, for all higher modes and introduces this as 
an approximation to determine modes above $k=10$.
A distinction should be made between the random Gaussian
initial conditions
contained in the $\hat \Phi_{\vec q}$ and the random Gaussian behavior of
the $\xi_{\vec q}$.  The former is an 
assumption which may or may not be true depending on the history of 
the universe while the latter is a property of the manifold.

The DBEM was first used
to find the pseudo-Gaussian random numbers $\xi_{q\ell m}$
for $k<10$ \cite{inoue} and the analysis was then extended to include the 
first 36 eigenmodes with $k\le 13$ \cite{inouetomita}.  
(As discussed below, it is claimed in Ref. \cite{cornsperg}
that 2 modes were originally overlooked in this method).  
A numerical cutoff in the low $k$ spectrum was found at the value 
$k_1=5.41$.
The corresponding maximum wavelength 
$\lambda_{\rm max}=2\pi/k_1$ can be compared to an approximation for
the diameter of ${\cal M}$ which
he computes as the average of the in-radius and the out-radius.  
The manifold has
in-radius $r_{in}=0.535$, out-radius $r_{out}=0.7485$
(and volume ${\cal V}=0.98139$).
This
yields $k_{\rm min}=4\pi/(r_{+}+r_{-})=4.9$, an agreement to within 10\% of his
$k_1$.
There are no supercurvature modes in the Thurston space.

The pseudo-Gaussian behavior can be interpreted 
in a quantum mechanical context \cite{inoue}.
Let the Laplace-Beltrami operator be the Hamiltonian of a quantum system, then
the eigenmode is a wavefunction eigenstate.  
Since the underlying classical dynamics is chaotic, it has been conjectured
that the quantum mechanical system will be governed by the predictions
of random matrix theory (RMT) \cite{{berry},{ott}}.
One such prediction is that the square expansion coefficients are 
Gaussian distributed if expanded with respect to a generic bases.
In particular, the coefficients should be given by the
Gaussian Orthogonal Ensemble (GOE)
	\be
	P(x)={1\over \sqrt{2\pi x}}e^{-x/2}.
	\ee
To check this prediction for the modes (\ref{modexpan}), Inoue
first recasts the eigenmodes
in terms of real independent coefficients
$a_{q\ell m}$ and real functions $R_{q\ell m}$ as 
	\be
	\psi^{\cal M}_{q}=\sum_{\ell m}a_{q\ell m}R_{q\ell m}
	\ee
where the $a_{q\ell m}$ can be expressed in terms of the $\xi_{q\ell m}$
and the $R_{q\ell m}$ can be expressed in terms of the $X_{q\ell}Y_{\ell m}$
\cite{inoue}.
Following \cite{aurichsteiner}, he examines the statistical behavior of
	\be
	x={|a_{q\ell m}-\bar a_q|^2\over \sigma_q^2}
	\ee
where $\bar a_q$ is the average and $\sigma_q^2$ the variance.  
The probability is singular
at $x=0$ and so it is customary to
compare the numerics to the cumulative distributions
	\be
	I(x)=\int^x_0P(x)dx={\rm erf}(\sqrt{x/2}).
	\ee
To test the RMT 
prediction, the numerically determined cumulative distribution are compared
for a goodness of fit.  Very good agreement with the GOE prediction is found
confirming Gaussian behavior for the low lying states.  The Gaussian behavior
is expected for the highly excited states as a consequence of the classical
chaos but Gaussian behavior for the low lying states is less obvious.
Yet this is in fact what his observations confirm.
He also tests the randomness of the $a_{q\ell m}$ and finds that they do 
behave as random variables.
The randomness is not a property of the eigenmodes but 
is due rather to
the almost random distribution of images in the
universal cover.
The number of copies of the fundamental domain inside a sphere with 
radius $\eta_0$ is
	\be
	n_1={\pi(\sinh(2\eta_0)-2\eta_0\over {\cal V}}
	\ee
giving $n_1\sim 29$ for $\eta_0=1.6$.

The number of eigenmodes increases as $k^3$ and the number of boundary 
elements increases as $k^2$.  The task of computing high $k$ modes becomes
unmanageable with this method.  
For higher $k$, Ref \cite{inoue} suggests taking the clue
from the behavior of low $k$ modes and approximating the highly-excited states
as random Gaussian numbers with a variance proportional to $q^{-2}$.
Using Weyl's asymptotic formula,
	\be
	N[q]={{\cal V}q^3\over 6\pi^3}
	\quad \quad q\equiv \sqrt{k^2-1}\quad \quad q>>1 
	\ee
with $N$ an integer.  
Since for a CH space the volume is fixed, Weyl's formula effectively
relates the number of states at a given $k$ with a topological feature
of the space.  It also allows an estimate of the $k_j$.
The spacing between discrete eigenvalues decreases as the inverse of 
$k$ for large $k$ and so approaches a continuous spectrum in the large
limit, as expected.
Weyl's formula 
is well obeyed and 30 of the 36 modes show random Gaussian behavior.
Interestingly,  
six degenerate states are found
which correspond to a nearly symmetric mode 
reflecting the global symmetry of the fundamental domain.
While a linear combination of the 
degenerate modes shows Gaussian
behavior again.
This is typical of classically chaotic systems where the classical chaos
leads to a Gaussian behavior in the quantized eigenmodes although 
occasionally the global symmetry of the space can surface in the eigenmodes
as non-Gaussian behavior. 
(The connection with quantum chaos led
Inoue to use the Selberg Trace Formula to compute eigenvalues for a large
number of CH manifolds \cite{inouetrace}.  This method proved to be quicker
and easier to implement numerically.)

The 
$C_\ell$'s are calculated
by sewing together the numerically obtained 
eigenmodes for the eigenvalues in the range $5.4 \le k\le 13$ with 
the above approximation for the expansion coefficients for 
$13\le k<20$. 
With the assumption that ${\cal P}_\Phi=$constant,
the resultant $C_{\ell }$ is compared with the \cb data
to conclude that the spectrum is consistent with \cb for
$0.2 \le \Omega_0\le 0.6$.  
The cutoff in the spectrum due to the minimum
mode $k=5.4$ is buried under the contributions to low $\ell $ power from 
the integrated Sachs-Wolfe effect.  The ISW becomes important during the
curvature dominated epoch at about $1+z\sim (1-\Omega_o)/\Omega_o$.
They find the contribution from $k \le 13$ modes
to $C_\ell $ for $2\le \ell \le 20$ is 7 \%
for $\Omega=0.2$ and 10\% for $\Omega_o=0.4$.  The rest is due to the ISW.
For this reason, the spectra appear to have a gradual peak near the 
long-wavelength cutoff or are nearly
flat down to low $k$ and are less severely constrained 
for $\Omega\ge 0.1$ than the flat
models of section \S \ref{obsflat} \cite{{inoue},{inouetomita}}.
In fact they find very good agreement with
\cb for $\Omega_0=0.6$, better than other FRW models.  In particular,
an alignment of the peak in the \cb data at $\ell\sim 4$ with the peak
due to the cutoff accounts for the better fit.

The conclusion cannot be stated as proving that the Thurston space is 
consistent with the observations, only that it does not appear inconsistent.
A comparison of the full correlation function to the data may not survive
consistency and in fact does not according the method of images analysis 
of \cite{{bpsI},{bpsII}}.

\centerline{\bf Eigenmodes of an Orbifold}

Aurich was able to compute a huge number of eigenmodes, 
the first 749, in an orbifold with negative curvature.  
Orbifolds can have compact volume but possess points which are
not locally $R^3$.  They can also have rotation group elements among
their isometries, unlike manifolds which have no groups with fixed points
in their isometries.  In the absence of a predictive theory for the topology
of the cosmos, there is
no convincing reason to omit them from the catalog.
Interestingly, they can have even smaller volumes than the minimum bound
on compact hyperbolic spaces.

The fundamental cell is a pentahedron which is symmetric along a plane
dividing the fundamental domain into two identical tetrahedra.
This allows a
desymmetrizing of the pentahedron useful for deducing the eigenmodes
following early work \cite{aurichmarklof}.  
There are 9 compact hyperbolic tetrahedra and
the one built into the pentahedral orbifold
is known as $T_8$.  The volume of $T_8$ is
${\cal V}_{\rm tetrahedron}\simeq 0.3586524$ which is smaller than
the Weeks space.
The smallest compact hyperbolic tetrahedron $T_3$ has volume 
${\cal V}=0.03588506 $, ten times smaller than $T_8$ 
and smaller even than the existing bound on CH manifolds
($V_{\cal M}=5/(2\sqrt{3}){\rm arcsinh}^2(\sqrt{3}/5)\sim 0.167$).
Using the boundary element method
749 eigenmodes on $T_8$ are obtained.

Both radiation and matter are included in his analysis which evolves
the metric
perturbations according to the usual adiabatic, linear perturbation theory
and solves $F(\eta)$ numerically.
In his analysis, the eigenmodes are expanded 
as
	\be
	\Phi(\eta,\vx)=\sum F_q(\eta)\psi_q(\vx)
	\label{exp1}
	\ee
where the $\psi_q$ are the eigenmodes and all of the time dependence is
in $F_{\vec q}(\eta)$.  
Notice that in comparison to the standard
expansion \ref{fluctcmb}, he does not
include the usual randomly seeded fluctuation amplitude $\hat \Phi_k$.  Instead
all of the initial conditions are absorbed into $F(\eta)$ with
	\be
	F_q(\eta_i) =\alpha/q^{3/2}
	\quad \quad F^\prime_q(\eta_i)=0.
	\label{exp2}
	\ee
There is however some element of randomness in the sign of the fluctuation.
The eigenfunctions have a freedom in the phase.  Since the phase is always
chosen to be real, this freedom results in a $\pm $ ambiguity in the 
eigenfunction which is chosen randomly.
The constant $\alpha $ is normalized against the \cb data.
This is consistent with the Harrison-Zeldovich flat spectrum in terms of 
normalization but lacks any of the randomness that real fluctuations
would have.  
Because the randomness of the temperature fluctuations about the mean is
neglected, it would not be quite right to
interpret the simulated maps of Ref \cite{aurich} in terms of
an actual universe.  The advantage however is that the maps
show a random nature 
which can only be due to the random character of the modes
themselves.  This is important for the reasons discussed at length in the 
previous
subsection \ref{directhyp}.

Since the $C_\ell$ produces a global averaging, 
an ensemble average is generated by the expansion 
(\ref{exp1})-(\ref{exp2}) even if the standard
random initial fluctuations were ignored.  Although
the simulations depend on the location of the observer and the orientation
of the manifold,
this dependency is not examined and the observer
is fixed, offset from the origin.
The $C_\ell $'s are compared to 
the $\ell <30$ \cb data, the Saskatoon data around $\ell \sim 100$
\cite{saskatoon} and the QMAP data \cite{qmap}
above $\ell\sim 80$.
All modes are computed up to $k_{max}=55$.  
Aurich varies $\Omega_0$ between $0.2$ and $0.6$ and finds reasonable
agreement for $\Omega_0\simeq 0.3-0.4$.
For smaller $\Omega_0$ he
finds that the numerical $C_\ell$ increases too quickly with $\ell$.
We suggest that this is evidence of a long wavelength cutoff in the spectrum.
Although for a lower $\Omega_0$, the ISW is important in terms of building
up the low $\ell$ power, it is also true that the suppression is more
severe.  To put it another way,  if the constant $\alpha $ were normalized
to \cb at high $\ell$, as it should be in order
to minimize the effects of cosmic variance, 
then the steep slope would be seen as a suppression of large
angle power due to the finite extent of the space.

\begin{figure}
\centerline{\psfig{file=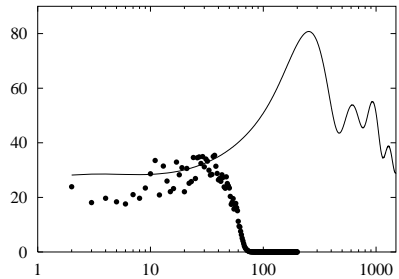,width=2.25in}}
\caption{
This figure was supplied courtesy of R. Aurich.  
The figure shows $T_\ell$ measured in micro-Kelvin
versus $\ell $ as computed 
by Aurich and Steiner for an $\Omega_0=0.9$ cosmology with 
$\Omega_\Lambda=0.6$.  The angular power spectrum $C_\ell $ for the
orbifold is shown in comparison to an infinite topology as obtained
by CMBFAST.  The suppression for $\ell > 30$ is only due to the truncation of
the sum over the eigenmodes since only the first 749 eigenmodes have
been taken into account.}
\label{aurcl}
\end{figure}

The same orbifold was studied again in Ref.\ \cite{as} where both 
a cosmological constant and a smooth dark energy component were 
added to the radiation and matter
energy density.  The total energy is taken to be nearly
but not quite flat to address the recent results from the small-angle
CMB experiments \cite{{exper},{balbi}}.
They still find a suppression in power
for $\ell \lta 10$, as illustrated in figure
\ref{aurcl}.

Although the results of Ref. \cite{aurich}
to be at variance with the results of 
\cite{{bpsI},{bpsII}} we emphasize again the limitations of comparing only the 
$C_\ell$'s.  This is expanded upon in section \S \ref{obsmi}.

\centerline{\bf Fast Method for Isolating Eigenmodes}

A fast numerical method for obtaining the eigenvalues on compact hyperbolic
manifolds was developed by Cornish and Spergel \cite{cornsperg}.  In comparison
to the boundary element method developed by Aurich and Steiner
\cite{aurichsteiner}, it is technically inferior but far quicker and easier to
implement numerically.
The DBEM has only been applied to a few
cosmological spaces, the tetrahedral orbifold of
\cite{aurich}, the Thurston manifold \cite{inoue}, and the 
Weeks manifold to name a few \cite{inouetrace}, to name a few,
while the method of
Ref. \cite{cornsperg} allows them to obtain the lowest eigenvalues and
eigenmodes for 12 manifolds, requiring
only the generators as input.

The method is to simply solve
	\be
	\Psi_k(x)=\Psi_k(gx)
	\ee
using a singular value decomposition.
The eigenmodes are again
expanded in terms of the eigenmodes on the universal
cover $\ah$
as in eqn (\ref{modexpan}).
Random points are selected within the fundamental domain and 
all of the images out to some distance are located in the covering space.
For each point $p_j$ there are $n_j$ such images located with the
generators $g_\alpha$.  The
$n_j(n_j+1)/2$ boundary conditions are
	\be
	\Psi_k(g_\alpha p_j)=\Psi_k(g_\beta p_j)
	\quad \quad \alpha \ne \beta
	\label{bclist}.
	\ee
Using the expansion of the eigenmodes in terms of the eigenmodes of the
universal cover \ref{simphyp}, they write (\ref{bclist}) 
as a collection of difference
equations which they then solve using a standard 
singular value decomposition
method for handling over constrained systems of equations.

The example they studied in detail is the Weeks space m003(3,-1).
  They also looked at the Thurston space and
found agreement with the DBEM method applied by Inoue except they found
2 new modes missed previously but later confirmed \cite{ktinoue}.
Again, for higher $k$, the number of modes obeyed Weyl's formula well.
Also, the modes were well described by random matrix theory which predicts
the expansion coefficients obey a Gaussian
Orthogonal Ensemble.

\begin{figure}
\centerline{\psfig{file=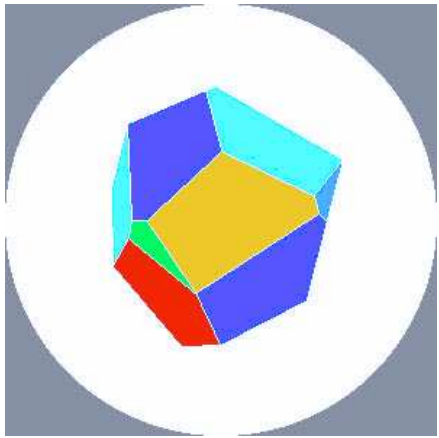,width=2.25in}}
\caption{
Dirichlet domain for the Weeks space}
\label{weeks}
\end{figure}

\begin{figure}
\centerline{\psfig{file=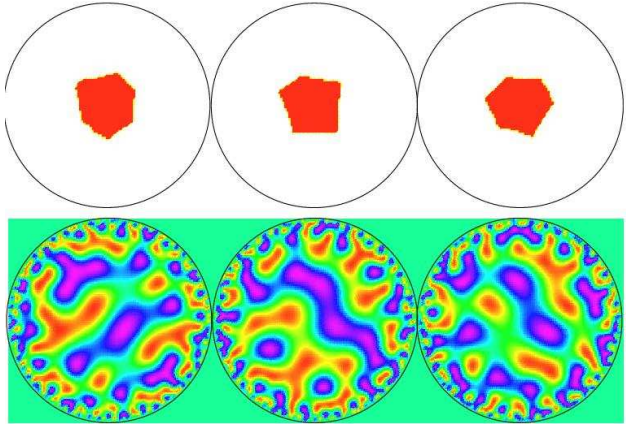,angle=-90,width=3.25in}}
\caption{This figure was supplied courtesy of N. Cornish.
The lowest eigenmode in the Weeks space is shown.  The space
is drawn in the Poincar\'e representation.  The three panels
represent different slices through the Poincar\'e ball.
The upper figure shows the slice through the fundamental domain
while the lower figure shows the slice through the eigenmode.
Specifically,
the leftmost panel is the $x=0$ slice, the middle is the 
$y=0$ slice, and the rightmost panel is the $z=0$ slice.
}
\label{cornspergfig}
\end{figure}

The Weeks space has a large symmetry group which ensures a large degeneracy of
eigenmodes.
The symmetry group is the Dihedral group of order 6.
They found the first 74 eigenmodes, many of which are degenerate.  The higher
the mode, the more degeneracies in agreement with Weyl's asymptotic formula.
A view of the lowest eigenmode in the Weeks space can be found in 
figure \ref{cornspergfig}.

\begin{figure}
\centerline{\psfig{file=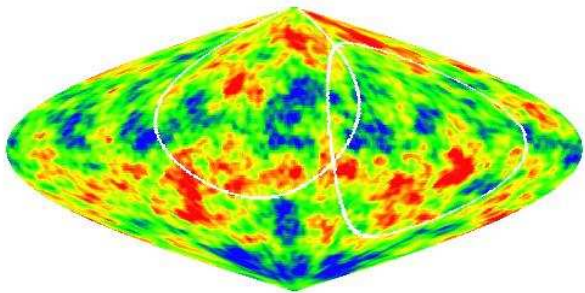,width=3.25in}}
\vspace{8mm}
\caption{This figure was supplied courtesy of N. Cornish.
A numerical realization of the CMB in the Weeks topology with
$\Omega_o=0.3$. One pair of matched circles is indicated by the
white lines in this projection.}
\label{weekscirc}
\end{figure}

Implementing
their eigenmodes in a simulation of a cosmological
model, they compare the numerically generated $C_\ell$'s to the \cb
data \cite{cornsperg}. 
Their numerically modeled cosmologies are based on the first 100+ modes of
the following small spaces:
The Weeks space m003(3,-1) with ${\cal V}=0.9427$, the Thurston space
m003(-2,3)
with ${\cal V}=0.9814$,
s718(1,1) with ${\cal V}=2.2726$ and v3509(4,3) with ${\cal V}=6.2392$.  
One realization of the Weeks space is shown in figure \ref{weekscirc}.
Generically,
a kind of statistical isotropy prevails even though the spaces are globally
anisotropic.  Since the expansion coefficients are pseudo-random, Gaussian
distributed numbers which are statistically independent of $\ell $
and $m$, they generate nearly isotropic eigenmodes and lead
to a kind of isotropy across the microwave sky.
The authors suggest inflation may not be need to explain why the universe 
is nearly isotropic.
However inflation is still needed 
to explain why local values are so marginally near flat.

A standard likelihood analysis based on 100
realizations of each topology was performed.
A mode cutoff in the spectrum is again evident and
gets worse as $\Omega_0$ is lowered.  
For all of the manifolds studied, they find that 
	\be
	\lambda^q_1=(1.3\rightarrow 1.6)D
	\label{lq}
	\ee
where $\lambda^q_1=2\pi/q_1$ (although it is customary 
to define the wavelength corresponding to a given mode as $\lambda^k=2\pi/k$)
which obeys the bound
	\be
	{4\tilde D\over D^2(\sinh\tilde D+\tilde D)^2}\le k_1^2\le 1+\left ({2\pi\over
D}\right )^2
	\ee
with $\tilde D$ the square root of the smallest integer $\ge D^2$.
They do caution that they cannot find modes with $q=[0,1/4]$ using their method,
and so perhaps supercurvature modes lurk.  
However the ISW becomes increasingly more
important as $\Omega_0$ is lowered.  There is an optimal value where the 
two effects compete to create a spectrum with a slight tilt at low $\ell$
and give a better match to the \cb data.  For the 
Weeks' space this occurs at $\Omega_0=0.3$.  

Additionally, once the spectrum is normalized to \cb they
find that the size of fluctuations on 
$8h^{-1}$ Mpc is naturally increased above the value of $\sigma_8=0.6$ 
for a simply connected universe with $\Omega=0.3$ to a value of 
$\sigma_8=0.75$ for the Weeks manifold with $\Omega=0.3$.  
The present day cluster abundance seems to imply 
$\sigma_8=0.9\pm 0.1$ for $\Omega_0=0.3$ \cite{pen} and so higher values
are desirable.

They find the $C_\ell$'s have
a better fit to the \cb data with a relative likelihood of 
$\sim 20 $ and a better fit to 
the large-scale structure date ($\sigma_8$ increases by 
$\sim 25\%$).  However, as with all of the other statistical analyses they
have applied a weak test by analyzing the angular averaged
$C_\ell$s and not the full
correlation function.  Their conclusions can be interpreted as finding the 
manifolds are not ruled out by the $C_\ell$ alone.

\subsubsection{The Method of Images}
\label{obsmi}

The fluctuations can be simulated 
using the method of images.  The correlation
functions can be calculated without the explicit eigenmodes and eigenvalues.
The procedure is to sum the correlation function on the universal cover 
over all images out to some large radius.  The
more distant images are handled in a continuous approximation.
The eigenspectrum can be calculated but the
correlation
functions are obtained to better accuracy for a given order in the sum.
Bond, Pogosyan and Souradeep implemented a detailed method of images
\cite{{bpsI},{bpsII},{bps_texas},{bps_moriond},{bps_cwru}}.
They emphasize that the full
correlation function $C(\hat n, \hat n^\prime)$ must be compared to the
data for a meaningful statistical analysis.  The philosophy is to perform
a statistical search for patterns.
They find
two principle  effects
(1) anisotropic patterns and (2) a long wavelength 
cutoff in the power spectrum.
The patterns appear as spikes of positive correlation when a point on 
the SLS and one of its images is correlated.  The larger the SLS relative
to the out-radius, the more statistically significant will the patterns 
be.  If the SLS is smaller than the in-radius then naturally the correlation
function
is very close to that for the simply connected space.  

The angular correlation function $\cq $ is 
computed from the spatial two-point correlation function
$\spc\equiv \left <\Phi(\vec x, \tau_{LS})\Phi(\vec x^\prime,
\tau_{LS})\right >$ which can be expressed as
	\begin{equation}
	\spc=\sum_i\pow(k_i)\sum_{j=1}^{m_i}\Psi_{ij}(\vx)\Psi^*_{ij}(\vxp)
	\end{equation}
with the sum over a discrete ordered set with multiplicities $m_i$.
Notice that the $\Psi_{ij}$ obey
	\begin{equation}
	\left (\nabla^2 + k_i^2\right )\Psi_{ij}=0
	\end{equation}
as always, however with the slightly different two-index notation.  The 
$j$ index accounts for any degeneracies.
The spatial correlation function $\xi^c$ 
on the compact
manifold ${\cal M}^c$, can be expressed in terms of the spatial 
correlation function $\xi^u$ on the universal cover.  A derivation
of the relation between $\xi^c$ and $\xi^u$
exploits orthonormality and completeness with the following equations:
	\begin{equation}
	\int_{{\cal M}}d\vxp \spc^c (\vx,\vxp) \Psi^c_{ij}(\vxp)=\pow(k_i)
	\Psi^c_{ij}(\vxp)
	\label{spcc}
	\end{equation}
	\begin{equation}
	\int_{{\cal M}^u}d\vxp \spc^u(\vx,\vxp) 
	\Psi^u_{j}( k,\vxp)=\pow(k_i)
	\Psi^u_{j}( k,\vxp).
	\label{spcu}
	\end{equation}
Since eigenfunction on the compact space must also be eigenfunctions on
the universal cover (although the converse is not true) it follows that
	\bea
	\int_{{\cal M}^u}d\vxp \spc^u(\vx,\vxp) 
	\Psi^c_{j}( k,\vxp) &=& \pow(k_i)
	\Psi^c_{j}( k,\vxp)\nonumber\\
	&=& \int_{{\cal M}}d\vxp \spc^c(\vx,\vxp) 
	\Psi^c_{ij}( \vxp) \label{e3}.
	\eea
Combining (\ref{e3}) with (\ref{spcu}) gives
	\be
	 \int_{{\cal M}}d\vxp \spc^c(\vx,\vxp) 
	\Psi^c_{j}( k,\vxp) = 
	\int_{{\cal M}^u}d\vxp \spc^u(\vx,\vxp) 
	\Psi^c_{ij}( \vxp) 
	\label{e4}.
	\ee
Since 
${\cal M}$ tessellates
${\cal M}^u$ we can re-express the left-hand-side of (\ref{e4})
as
	\begin{equation}
	\sum_{g \in \Gamma}
	 \int_{{\cal M}}d\vxp \spc^u(\vx,g \vxp) 
	\Psi^c_{ij}( \vxp) 
	=
		 \int_{{\cal M}}d\vxp
	\left [\tilde{ \sum_{g \in \Gamma}}\spc^u(\vx,g
\vxp) 
	\right ]
	\Psi^c_{ij}( \vxp) 
	\end{equation}
where in the last step a regularization is needed and denoted
by the tilde above the summation.
Identifying integrands
gives
	\begin{equation}
	\spc^c(\vx,\vxp)=\tilde{ \sum_{g \in \Gamma}}\spc^u(\vx,g
\vxp) 
	\end{equation}
and the $\spc^c$ can be calculated as the sum over images in the universal
cover with only a knowledge of the group elements.
The spatial correlation function on the universal cover is known to be
	\be
	\spc^u(\vx,\vxp)\equiv \spc^u(r)=\int^\infty_0{dq q\over (q^2+1)}
	{\sin(qr)\over q\sinh r}{\cal P}_\Phi(q)
	\ee
and ${\cal P}_\Phi$ is taken to be a Harrison-Zeldovich spectrum.
This is the core of the method of images.

The regularizer requires some additional effort.
The correlation function on the universal cover does not have compact support:
	\begin{equation}
	\int_{{\cal M}^u}d\vxp \spc^u(\vx,\vxp) 
	=\infty.
	\end{equation}
The need for regularization is not a result of the CH space having a large
number of periodic orbits as incorrectly claimed in \cite{css_cqg}
nor is it dues to the chaotic nature of the trajectories but rather is 
a result of the correlation function not having compact support.
Even flat spaces require regularization.
The regularized spatial correlation function can be written	
	\begin{equation} \tilde \xi^u_\Phi(\vx,\vxp)
	\equiv \spc^u(\vx,\vxp)-{1\over V_{{\cal M}}}
	\int_{g{\cal M}}d{\bf{\vec x}^{\prime\prime}} 
	\spc^u(\vxp,{\bf{\vec x}^{\prime\prime}}) 
	\end{equation}
for $g $ such that $\vxp$ lies in $g {\cal M}$.
The regularization ensures
	\begin{equation}
	\int_{{\cal M}^u}d\vxp \tilde \xi^u_\Phi(\vx,\vxp)
	=0.
	\end{equation}
Finally then,
	\begin{equation} 
	\spc^c(\vx,\vxp)=\sum_{g \in \Gamma}
	\tilde \xi^u_\Phi(\vx,\vxp)
	=\sum_{g \in \Gamma} \spc^u(\vx,\vxp)-{1\over V_{{\cal M}}}
	\int_{{\cal M}^u}d{\bf{\vec x}^{\prime\prime}} 
	\spc^u(\vxp,{\bf{\vec x}^{\prime\prime}}) .
	\end{equation}

The regularization prescription is not unique.  
The actual limiting procedure 
utilized in Ref. \cite{bpsI} sums images up to a radius $r_*$ and then 
regularizes by subtracting the
integral of $\spc^u(r)$ over a spherical ball of radius $r_*$:
	\be
	\spc^c(\vx,\vxp)=\lim_{r_*\rightarrow \infty}
	\left [\sum_{r_j<r_*}\spc^u(r_j)-{4\pi\over V_{\cal M}}
	\int^{r_*}_0 dr \sinh^2 r \spc^u(r)\right ]
	\ee
with $r_j=d(\vx, g_j \vxp) \le r_{j+1}$.  The value of 
$r_*$ is numerically pushed out to 4 or 5 times the out-radius in order to 
get a convergent result.
This procedure
is simpler and does well for large $r_*$.
It also does not require a detailed dependence on the complicated shape
of the fundamental domain.

\begin{figure}
\centerline{\psfig{file=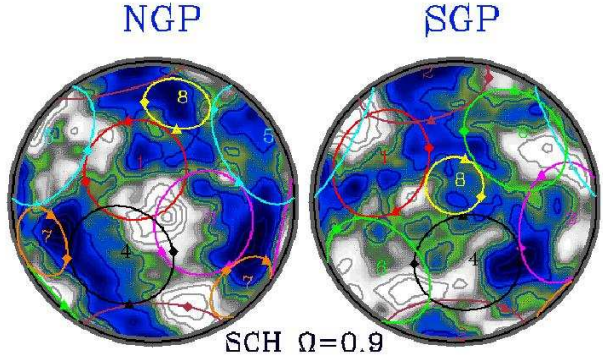,width=3.25in}}
\vskip 5truept
\caption{This figure was supplied courtesy of T. Souradeep.
The full CMB sky is represented by the two hemispherical caps -- one 
in the direction of the South Galactic Pole (SGP) and the other in the
direction of the North Galactic Pole (NGP).  The label SHC refers
to a small compact hyperbolic model, namely m004(-5,1).
As well as showing the fluctuations, this figure shows correlated circle
pairs explained in more detain in a following section.}
\label{bpsfig}
\end{figure}

An example of the maps they generate using the method of images is shown
in figure \ref{bpsfig}.

They are also able to estimate the density of states and therefore obtain
a rough estimate of the power spectrum using the method of images.  This is
in addition to their primary result of having computed the correlation 
function.
When they do estimate the density of states they always find an infrared cutoff
as expected for long wavelength modes.

This is consistent with the expectations.
For a 3-dimensional CH space there is a bound on Cheeger's isoperimetric 
constant of \cite{{chavel},{berard}}
	\be
	k_{min}\ge h_C/2\ge {1\over d_{\cal M}}\left [ 2\int^{1/2}_0
	dt \cosh^2(t)\right ]^{-1} =0.92/d_{\cal M}
	\ee
and so there are no supercurvature modes for $d_{\cal M}<0.92 $.
The suppression of power is covered in part by the ISW and so is less prominent
than in flat models.

Given their numerical calculation,
they do a Bayseian probability analysis comparing a few
compact hyperbolic cosmologies to the \cb data.
Their analysis represents by far the most complete of the statistical 
tests that has been performed and in principle is the most complete
test that can be performed.
They find the compact hyperbolic models they study are inconsistent
with the \cb data for most orientations although for a 
small set of special orientations they do find a better
fit than the standard infinite models.
Their results are qualitatively independent of the matter content.  
The analysis magnifies the inadequacy of using 
the $C_{\ell}$ alone.  In particular they point to the 
large cosmic variance that results from 
the break down associated with isotropy.

With the assumption of Gaussianity both in the noise and in the raw
$\delta T/T$, the probabilistic comparison used is
	\be
	{\cal P}(\delta |C_T)={1\over (2\pi)^{N_p/2}\|C_T\|^{1/2}}
	e^{{1\over 2}\delta\dagger C_T^{-1}\delta}
	\ee
where 	$ C_T=C(\hat n,\hat n^\prime)$ and $\delta $ is the data.
To numerically
obtain the full correlation function requires an evaluation of the
spatial correlation function at $N_p(N_p+1)N_L^2/2$ pairs of points
where $N_p$ is the number of pixels 
and $N_L$ is is the number of points along the line of sight used to
integrate the ISW.  They find an $N_L\sim 10$ is satisfactory.
They estimate the likelihood function 
	\be
	{\cal L}(C_T)\equiv 
	{\cal P}(\bar \delta |C_T)=\int d\delta 
	{\cal P}(\bar \delta |\delta)
	{\cal P}( \delta |C_T)
	\ee
with $\bar \delta $ the map which maximizes the conditional
probability and the integration is
performed over all realizations of the simulated sky $\delta $.
The resultant likelihood function they use is
	\be
	{\cal L}(C_T)=
	{1\over (2\pi)^{N_p/2}\|C_N+C_T\|^{1/2}}
	e^{{1\over 2}\bar \delta\dagger(C_N+C_T)^{-1}\bar \delta}
	\ee
with $C_N$ the noise covariance matrix.
What is actually obtained is model dependent relative likelihoods.
The model-dependent parameters are
${\cal M}$, the orientation, the location of the observer,
$\Omega_m$, $\Omega_0$ and $\Omega_\Lambda$, and the initial assumptions
regarding the spectral shape and character.
The manifolds studied were m004(-5,1), which is relatively small,
and v3543(2,3), which is comparatively large.  ($\Omega_\Lambda=0$
always.  Adding $\Omega_\Lambda$ relaxes the constraints.)
They varied $\Omega_m=\Omega_0$ over three values arranged so that 
the SLS was comparable to the out-radius and varied over 24 different
orientations.
They do leave the observer at the maximum of the injectivity radius.
This location gives the most symmetric perceived shape to the
Dirichlet
domain which one might expect to lead to the most conservative bounds.
It seems fair to assume that
moving the observer to a thinner region in the manifold
for instance will only amplify asymmetries and the constraint on 
the correlation function should only be more severe.
However, others have argued that locating the Earth away from the local
maximum of the injectivity radius can increase the likelihood fit of
a CH model to the data.  For instance, 
to check the effect of the inhomogeneity,
Inoue moved the observing point for the same model m004(-5,1)=Thurston 
manifold and performed the Baysean analysis as Bond, Pogosyan, and
Souradeep did.  He found
that there are a few choices of the 
position and orientation for which the likelihood is much 
larger than that of the infinite counterpart \cite{inoueprog}
(as did \cite{bpsII}).
The best-fit positions are scattered in the
manifold and far from the local maximum of the injectivity
radius. 

In any case, the results of \cite{{bpsI},{bpsII}}
are presented as the relative likelihood of a given
model
in comparison with a simply connected hyperbolic CDM model with 
the same $\Omega_m=\Omega_0$.
They uniformly find that the compact hyperbolic models have very
small relative likelihoods as compared with the simply connected models.
The rare exception occurred near $\Delta\eta\approx r_{+}$ with a
particular
orientation.  The statistical significance is unclear since a
fortuitous
alignment of the measured fluctuations with simulated
topological images could enhance the likelihood when taken over 
all realizations which reinforce this correlation.
The authors
defer conclusions to future tests such as the circle method.
Although again, Inoue argues the statistical significance {\it is}
clear.  If the $C_\ell$s fit the data poorly, there will be no 
orientation of the manifold that leads to an alignment of measured
fluctuations with topological images.  Put another way, it may be
unfair to suggest that a good fit to the data is a fortuitous
alignment instead of acknowledging this good fit to be a good fit.

Bond, Pogosyan, and Souradeep 
draw the general conclusion that $d_{\cal M}/2 > 0.7 \Delta\eta$.
The compact models they tested are excluded at the $3\sigma$ level
with the exception of those with very special orientations.
By contrast, if they were to only analyze the statistical likelihood
of the $C_\ell$ they would mistakenly conclude that the compact models
were preferred at the $1\sigma$ level
over the simply connected cosmology.  They emphasize
that
the error bars on the $C_\ell$'s are huge because of the exaggerated
cosmic variance in the topologically connected models.
It may also be worth noting here that there is some argument over
the interpretation of these error bars (see Ref. \cite{inoueprog}).

Due to the global breaking of homogeneity, they find the variance is 
spatially dependent.  This is another reason to be weary of
conclusions based on $C_\ell$ alone.  They found characteristic loud
spots,
that is regions in the sky with larger variance than others.
Loud regions correspond to intersections of the SLS through smaller
regions of the fundamental domain.  The regions are geodesicaly small,
that is to say, there are shorter geodesics relative to other regions in
the volume.  Unless the region is very very small, the ISW can obscure
this particular feature.
These loud and quiet regions are familiar from the cusp topology
studied in Ref. \cite{lbbs}.  
There, even the ISW cannot compensate for the deepest regions
of the cusp.

\begin{figure}
\centerline{\psfig{file=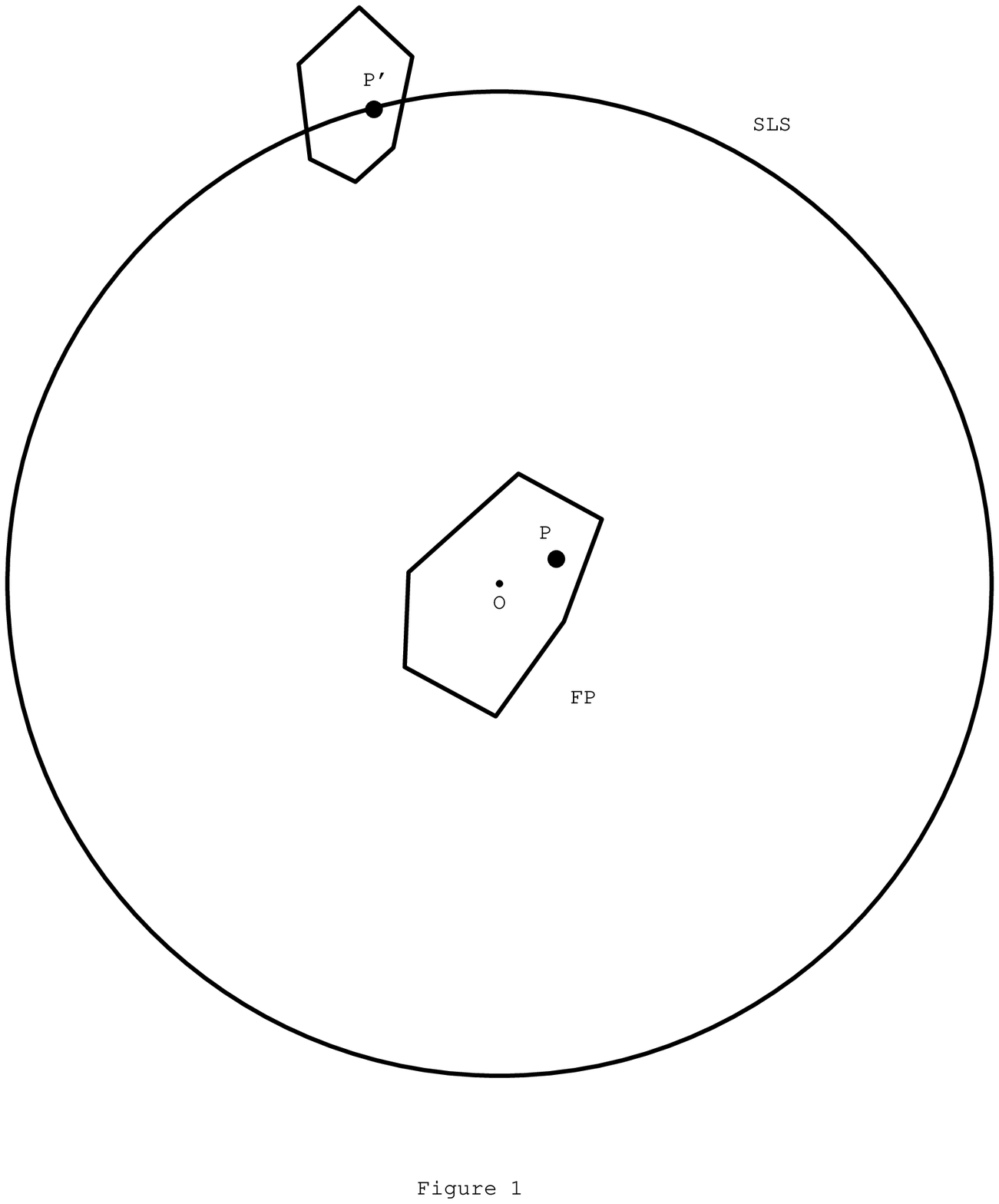,width=2.in}}
\vskip 5truept
\caption{
This figure was supplied courtesy of H. Fagundes.  It shows the
fundamental polyhedron as well as a copy of the fundamental polyhedron
that intersects the spherical surface of last scattering.  The point of
intersection identifies the original location of a CMB spot in the universe.}
\label{scheme}
\end{figure}

Another attitude to take would be to argue that if we do live in 
a compact hyperbolic manifold, some orientation is necessarily going
to be much more likely, namely the right one.
The special status of the `correct' orientation for our manifold,
assuming it is compact hyperbolic, was taken seriously
by one of the pioneers in cosmic
topology, Helio Fagundes \cite{{morehf},{morehf2}}.  He reasoned that the
particularly significant hot and cold spots in the COBE maps 
found by Cay\'on and Smoot \cite{cas} may shed light on the orientation 
of the manifold. These loud spots may be patches of high and low regions
in the physical density, as opposed to being just statistical fluctuations.
Since the density fluctuations eventually evolve into large-scale structure,
then the hot and cold spots should correspond to physical 
superclusters
and voids respectively.  Fagundes used this idea to try to match sources
in catalogs of galaxy superclusters and voids and thereby fix the orientation
and location of the earth in the manifold.  His algorithm 
begins with a point $P^\prime$ centered on one of the spots
isolated in Ref.\ \cite{cas} as illustrated in the schematic figure
\ref{scheme}.  The point $P^\prime \in SLS$. He then 
maps ghost images of $P^\prime=\gamma P$ for $\gamma \in \Gamma$
so the $\gamma$ are composite words built from the
face-pairing matrices.  He tries to alien the image points with 
voids and superclusters for different orientations of the space and 
different basepoints, i.e. different locations of the earth.
He performed these scans for
the ten smallest compact hyperbolic manifolds known
\cite{hw}.

\subsection{Geometric methods in hyperbolic space}
\label{geomhyp}

Given the likelihood analysis of the previous section, should we generically
conclude that all compact hyperbolic spaces must be large relative to the
observable universe?  The answer is really ``no'' 
since there is so much model
dependence in any of the direct methods of the previous section.  We 
cannot be confident about our assumptions for the initial fluctuation
spectrum, the choice of manifold, the orientation, location of observer,
local parameter values or even the statistics themselves.
Remembering that 
there are an infinite number of manifolds to consider and 
the ambiguities in the model parameters, statistical conclusions 
are limited to the specific.
These attributes beg for a template independent method of searching
for topology.  Much like gravitational lensing, we might hope to simply
look at the sky and see evidence of topological lensing without prior
assumptions about the shape of the lens.  There is hope for such model
independent observations as exemplified 
in the circles in the sky described below.
Generically, we distinguish these
geometric methods which search for
patterns from statistical methods which rely on a specific model.

\subsubsection{Circles in the Sky}
\label{obscirc}

\begin{figure}
\centerline{\psfig{file=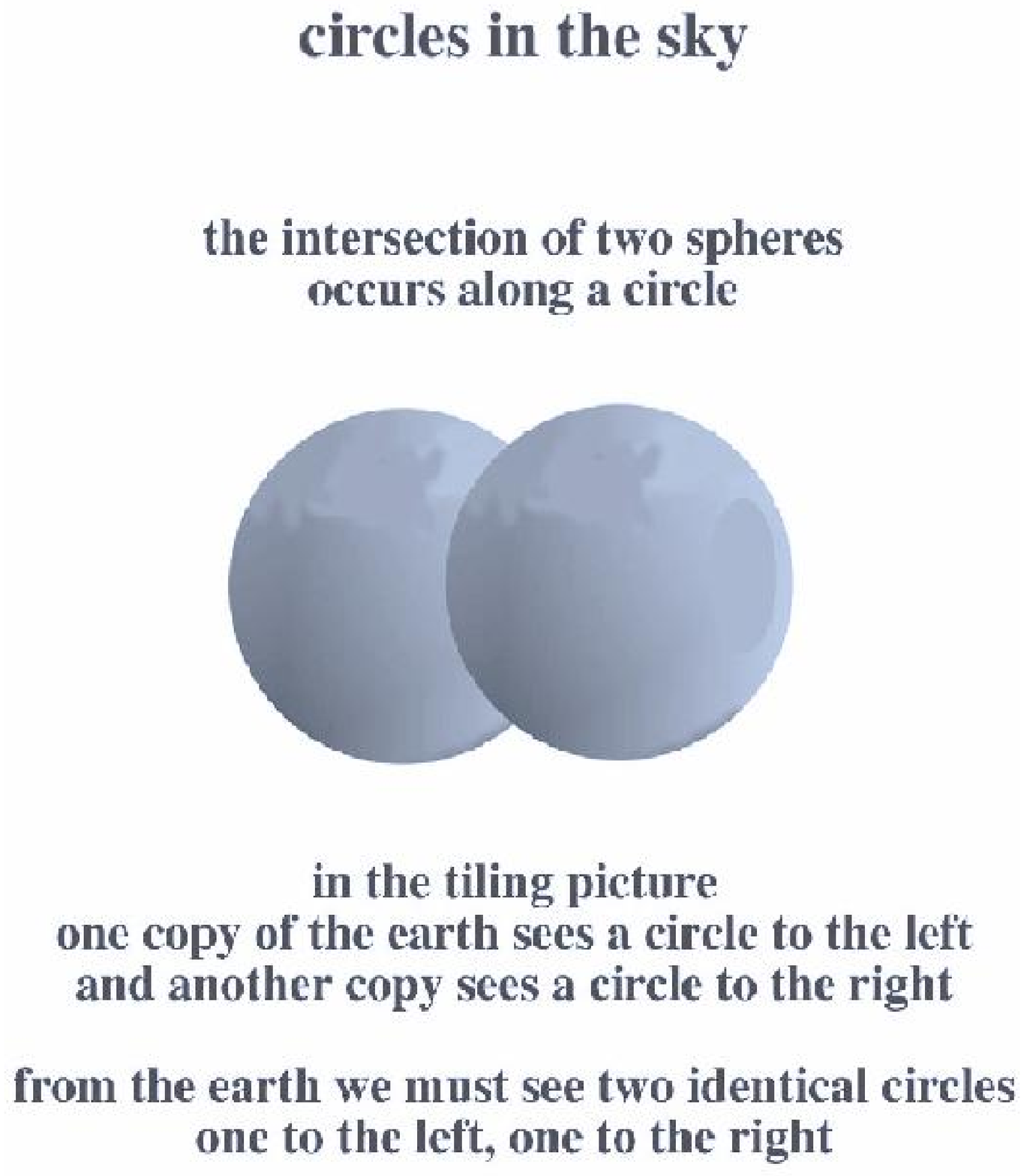,width=2.in}}
\vskip 5truept
\caption{}
\label{circles}
\end{figure}

\begin{figure}
\centerline{\psfig{file=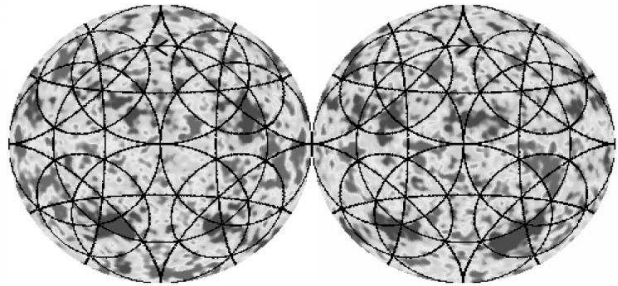,angle=0,width=3.25in}}
\voffset -0.5truein
\vskip 5truept
\caption{This figure was supplied courtesy of N. Cornish
and shows the northern and southern hemispheres of the microwave
sky in flat hypertorus.  There are 13 matched circle pairs indicated
by black lines.
}
\label{toruscirc}
\end{figure}

Possibly the nicest geometrical observation made thus far has been 
the prediction of
circles in the sky of Cornish, Spergel, and Starkman 
\cite{{css1},{css2},{css_cqg}}.
It has quickly become a popular topic in 
conversations on
the topology of the universe.  
The circles are most easily seen in the tiling representation of a
compact
space.  Each copy of the Earth will 
come complete
with its own surface of last scatter.  If the two images of the Earth are near
enough,
these identical copies of the SLS will intersect.
The intersection of two 
spheres occurs along a circle.  
Since the observers at the center of the intersecting
spheres are actually just copies of one observer,  the circle of
intersection must always come in pairs as
illustrated in fig.\ \ref{circles}.  To emphasize, 
we will not look up in the sky
and see intrinsic circles in the microwave; that is,
the circle pairs do not have 
identical temperatures along a circle but rather the temperature
varies identically when taken along correlated circles
\cite{css_cqg}.  An illustration of the location of circle pairs
in a finite flat torus is shown in figure \ref{toruscirc} from 
Ref.\ \cite{cornweeks}.

\begin{figure}
\vspace{60mm}

\includegraphics{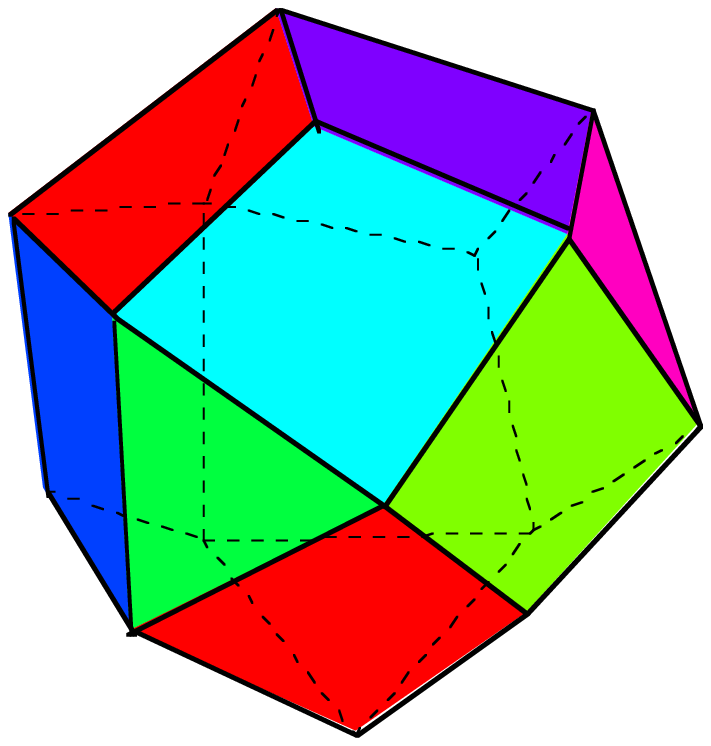}
\includegraphics{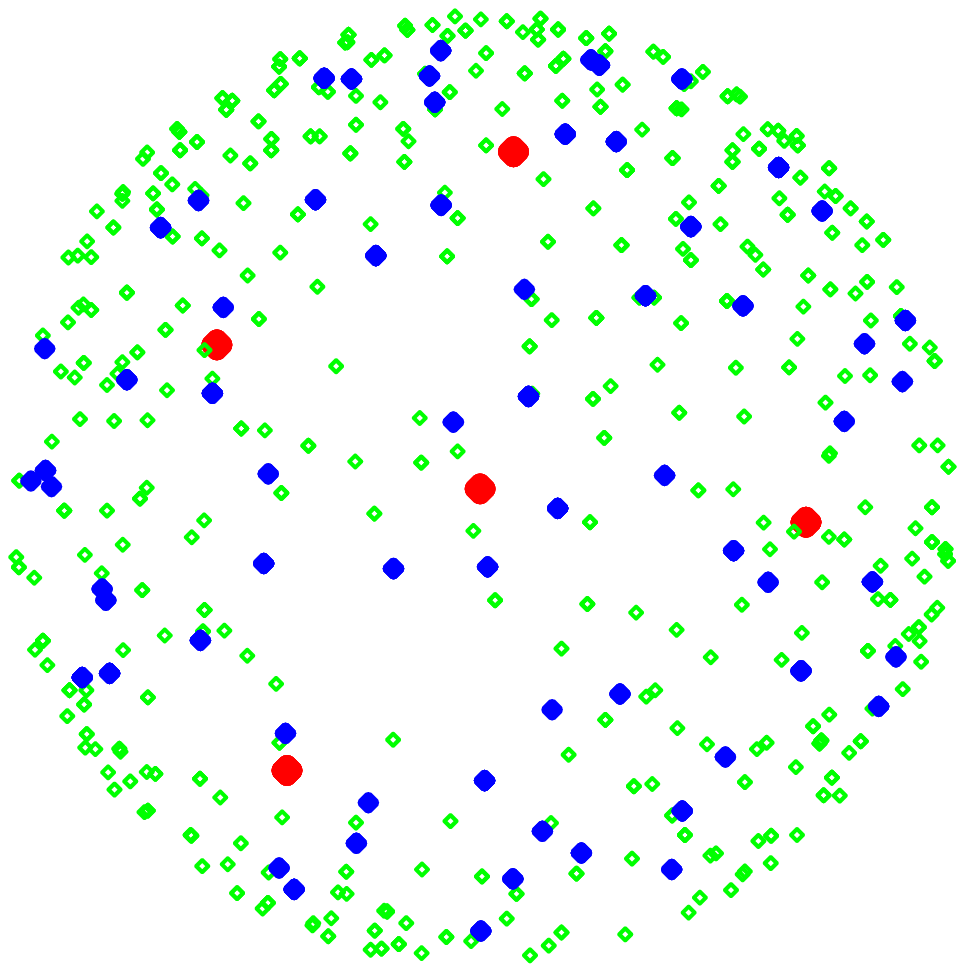}

\vspace{2mm}

\caption{This figure was supplied courtesy of N. Cornish
and illustrates
just how many circles are to be expected.
A fundamental domain for the Thurston space is shown.  The distribution
of points mark our clones out to a distance of three times the curvature
radius.
The large points are within one
curvature radius, the medium sized points are within two
curvature radii and the small points are within three curvature
radii.  Each of these clones will have a clone surface of last scatter.
The clones which generate circle pairs will
depending on the size of surface of last scatter compared to the
clone distance and therefore depends on the value of $\Omega_0$.}
\label{thurcirc}
\end{figure}

The most important aspect of the circles approach is that it applies to all
multiconnected topologies and does not require a template nor any
a priori assumptions
about a model.
All compact spaces will have circles if any part of the geometry
is smaller than the SLS.  The radius, number, and
distribution of the circles will vary from space to space and hence
the topology can be reconstructed from the circle pairs.
A distribution of clone images in the Thurston space can be seen in 
figure \ref{thurcirc}.
Since the shape of the Dirichlet domain is not a topological invariant
but rather depends on the location and orientation of the observer,
even the Earth's location in the universe can be deduced.

\

A statistical scan of the sky must be performed to draw the
correlated circles out of the maps.  The circle pairs will be
completely hidden.  To pull them out, consider two rings
each with angular radius $\alpha$ and with relative phase $\phi_*$
centered on arbitrary points 
$\vx $ and ${\bf \vec y}$.  To test whether these arbitrary rings are 
in fact correlated circles of intersection,
the comparison statistic
	\be
	S(\phi_*)={\left < 2T_1(\pm \phi)T_2(\phi+\phi_*)\right > \over
	\left < T_1(\phi)^2+T_2(\phi+\phi_*)^2\right >}
	\label{circstat}
	\ee
has been proposed
where $<>=\int_0^{2\pi} d\phi $ 
with S range $[-1,1]$
\cite{css_cqg}.  Perfectly matched circles have $S=1$ while an ensemble
of uncorrelated circles will have a mean value of $S=0$.
(Roukema includes the small scale Doppler effect in a slight alteration of the
statistic \cite{boudcirc}.)
Orientable topologies will have clockwise-anticlockwise correlations while
non-orientable topologies will have a mixture of clockwise-clockwise and
clockwise-anticlockwise correlations.
In flat space, matched circle pairs have angular radius
	\be
	\alpha={\rm arccos}\left ({X\over 2\Delta \eta}\right )
	\ee
with $X$ the distance between the Earth and its image.
In hyperbolic geometry
	\be
	\alpha={\rm arccos}\left({\cosh X-1\over \sinh X\tanh \Delta \eta}
	\right ).
	\ee
There are no circle pairs if the image is too far for the spheres
to intersect and the
expressions are invalid for $X>\Delta \eta$.
The number of images grows exponentially with $X$ in hyperbolic space
so that
most circles have small
radii, although the statistic works best for large circles. 

Noise will degrade the circle
statistic so that $S\ne 1$.  The experimental noise
in each pixel can be approximated as random Gaussian noise 
	\be
	P(n)={1\over \sigma_n
	\sqrt{2\pi}}e^{-n^2/2\sigma_n^2}
	\ee
with variance $\sigma_n$.  The true temperature fluctuations
are also taken to have a Gaussian distribution with variance $\sigma_s$.
The probability for the matched circle is then
	\be
	P^m(S)dS={\Gamma(N)2^{-N+1}\over \Gamma(N/2)^2}
	{(1+2\xi^2)^{N/2}\over
	(1+(1-S)\xi^2)^N}(1-S^2)^{N/2-1}dS
	\ee
with $\Gamma(N)$ the gamma function and $\xi=\sigma_s/\sigma_n$ is
the signal-to-noise of the detector while $N$ is the number of pixels
[?].  For large $N$, the distribution has a maximum
	\be
	S^m_{\rm max}={\xi^2\over 1+\xi^2}+{\cal O}(N^{-1}).
	\ee
The higher the resolution and the signal-to-noise, the better this
statistic will fare.
If the experimental angular resolution is $\delta \theta $, then there are 
$N\simeq 2\pi\sin\alpha/\delta \theta$ data points around each circle of
angular radius $\alpha$.  For
\cb, the signal-to-noise ratio is $\xi=2$, $\delta \theta =10^o$ and 
$N\simeq 36\sin\alpha $ pixels while for MAP, $\xi\simeq 15$, 
$\delta \theta=0.2^o$ 
and $N\simeq 1800\sin\alpha $
in its highest frequency channel.
Using MAP
parameters, only circles with $\alpha >4^o$ are detectable.

To make a relative comparison with a probability distribution 
for unmatched circles they advocate
	\be
	P^u(S)dS=
	{\Gamma(N)2^{-N+1}\over \Gamma(N/2)^2}
	(1-S^2)^{N/2-1}dS
	\ee
which is centered at $S=0$ with a FWHM$\simeq(8\ln 2/n)^{1/2}$.
The unmatched probability distribution $P^u$ was derived
by assuming the temperature at each point is an independent Gaussian
random variable. There are some weaknesses in this assumption since
there are known correlations in the CMB.

Cornish and Spergel applied
the circles test to maps generated with their numerically
isolated eigenmodes of section \ref{directhyp}.
For a realization of the Weeks space they found the uncorrelated ISW 
clouded the statistic resulting in a poor match
for circle pairs.  To combat this pollution they remove all power in
modes below $\ell=21$ and find a substantial improvement in the match
with values of $S\approx 0.9$.  Their expectations for the future
satellite missions are high.

The circles method was first applied to simulated maps by Bond,
Pogosyan
and Souradeep using the method of images of section \S \ref{obsmi}.
From
their correlation function they are able to make full sky maps.
Doing so they search for circles using the cross-correlation
coefficient
between fluctuations along two circles $C_1$ and $C_2=g C_1$,
	\be
	\rho_{12}\equiv
	{\left < \delta T(\vx)\delta T(g \vx)\right >
	\over \left [\left <\delta T(\vx)^2 \right >
	\left <\delta T(g \vx)^2 \right >\right ]^{1/2}
	}
	\ee
with $\vx\in  C_1$ and $g \vx\in C_2$.
A statistical average is implied by the angular brackets.  For one
given realization an integration over $\vx $ along the circles is
performed
instead.  They find $\rho_{12}$ is in the range $0.6-0.95$ for
$\Omega_0=0.9$ and m004(-5,1)
and $0.2-0.6$ for $\Omega_0=0.6$ for v3543(2,3).  
Notice that $\Omega_0$ is selected for each manifold so that 
${\cal V}$ is comparable to the volume of the SLS.
The matches they find are good even at \cb resolution.
The correlations
along circles gets worse as the ISW contribution is enhanced at 
low $\Omega_0$.

In addition to the circular intersections of the boundary of the SLS,
there are
intersections of the volume.
Let S represent the collection of points contained within the SLS.  Then the
intersection
of $g S  \cup S$ defines a lens-shaped region
(see their figure) in the volume
and must be identical
to the intersection of $S\cup g^{-1} S$.
As long as $r_{-}<\Delta \eta$, then there will be lens-shaped
intersections 
of the volumes and
circular intersections of the copies of the SLS.
Since the anisotropy occurs along the entire line of sight,
Bond, Pogosyan and Souradeep 
consider the correlations in the full lens-shaped volume.  
Based on 
the \cb constraints obtained via the method of images
as described in section \S \ref{obsmi} they took
the topology scale
comparable
to the SLS.  They argue
that correlations must then be very near the faces of
the Dirichlet domain and so the correlations provide 
a quick sketch of the
shape
of the universe \cite{bpsII}.

The volume intersections are relevant since, contrary to intuition, 
they found the ISW is not entirely uncorrelated.
While there are an infinite number of lines of sight which share
at least one common point, there are also pairs of lines which have
segments in common and so an enhanced correlation.  Every pair of 
lines of sight which are directed toward the center of matched
circles,
necessarily contain segments of identical points
(see fig.) and leads to correlated patterns.
These substantial
anti-correlated features tend to lie at the centers of matched 
circles.
The anticorrelation comes from the interference term between the 
surface Sachs-Wolfe effect and the ISW.
As described in section \S \ref{obsmi}, their likelihood
comparison of a handful 
of spaces to the existing data led to pessimistic conclusions.

Another
application of the circles method using \cb
sought to identify an asymmetric flat 3-space
\cite{boudcirc}. \cb is not ideal for detecting circles in hyperbolic
space since the ISW accounts for most of the power in the
\cb range of detection.  However, for flat spaces 
with no cosmological constant the sky is determined by the
surface Sachs-Wolfe effect alone without the obscuring effects of the ISW.
Despite the low resolution, \cb might still see circles if we live
in a compact flat space.

Roukema considered a specific asymmetric torus
which was put forth as a candidate
model to match cluster observations
\cite{{roukemaedge},{roukemablanloeil}}.  
Although he used the recent circles prediction to test his hypothesis,
the symmetries of the space and the geometric approach are reminiscent
of the earlier work of \cite{deO2} described in section \S \ref{geomflat}.
Hot X-ray bright gas in large galaxy clusters were used to search for
topological images \cite{roukemaedge}.
Two clusters at redshifts
$z\sim 0.4$ very nearly form a right angle with the Coma cluster
with very nearly equal arms.  
On the basis of this geometric
relation, they take a toroidal geometry for the universe to explain 
these clusters as topological
images of the Coma cluster.  
The distance from the Coma cluster to CL 09104+4109
and the distance from Coma to the cluster
RX J1347.5-1145 are both $\approx 960h^{-1}$ Mpc for 
zero cosmological constant.  
The third dimension 
is taken to be larger than the diameter of the SLS and hence topological
effects from this direction are essentially unobservable.
The size of the small dimension was taken to be roughly 
$\Delta \eta/13.2$.  Using the circles statistic,
the candidate was ruled out at the 94 \% confidence level, provided
that the cosmological constant is zero.
Specifically, the statistic Roukema used was
	\be
d\equiv \left<
         {  \left({\delta T \over T}\right)_i 
                 - \left({\delta T \over T}\right)_j 
     \over { 
\left\{ \left[\Delta \left({\delta T \over T}\right)\right]_i^2 +
      \left[\Delta \left({\delta T \over T}\right)\right]_j^2 
\right\}^{1/2}  
} } \right>
	\label{bstat}
	\ee
where $[\Delta(\delta T/T)]_i$ are the observational error estimates 
on the temperature fluctuations 
$(\delta T/T)_i$ as estimated by the COBE team.
This statistic directly tests the consistency 
of temperature values within observational error bars \cite{boudcirc}.
If the ISW effect is treated as noise as in Ref.\ \cite{boudcirc}, 
the application of the
circles principle using the statistic \ref{bstat}
can still enable rejection of a specific partly compact model  
or
show that a model one tenth of the horizon diameter is consistent
with the COBE data.

In general, there are obstacles to implementing the circle method in practice.
For one, the much emphasized ISW effect is not correlated
in this way and so can obscure the circle pairs. Other problematic
effects include the velocity and thickness of the SLS.  
In Ref. \cite{boudcirc}, it was noted that 
many circles are partially lost along with 
$20^\circ$ galactic cut and that
both detector noise and foreground contamination posed 
difficulties for the circle detections.
Realistically it may not be possible to observe circles even if they 
are there and this will be the challenge faced in realistic
analyses of the future satellite data.

\subsubsection{Pattern Formation}
\label{obspat}

Developmental biology and condensed matter physics have long exploited
pattern formation induced by periodic boundary conditions.  Since compact
topologies can be understood as a set of intricate boundary conditions,
pattern formation has a cosmological analogue.  The 
emergence of patterns in the sky are not so clear since many modes
are competing for attention.  The superposition of many otherwise
distinct geometric patterns can lead to something apparently random.
The task of geometric methods is to separate the patterns out of the sky.
The statistic of eqn.\ (\ref{circstat}) manages to draw out circles.
Other patterns 
in addition to the circles can emerge as described in this section.

The patterns are best siphoned off a CMB map by scanning for correlations
\cite{pat}.  As an example consider a map of the antipodal correlation
	\be
	A(\hat n)=C_{\cal M}(\hat n,\hat n^\prime)
	\ee
which measures the correlation of fluctuations received from opposite 
points on the SLS \cite{pat}.
In a simply connected space opposite sides of the SLS should have no 
communication between them and a
map of antipody would generate nothing
more than a monopole.  It is possible that accidental correlations appear
at random but no geometric structure would emerge.
By contrast, if the universe is topologically connected, then 
points on the manifold which seem to be far apart in the tiling picture
may actually be quite close together in the fundamental domain.  Therefore
opposite points on the SLS may be strongly correlated, may in fact be
the same point.  This is another example of ghost images but in an 
antipodal map collections of ghosts are caught and a picture of the 
symmetries of the space emerges \cite{pat}.

\begin{figure}
\centerline{\psfig{file=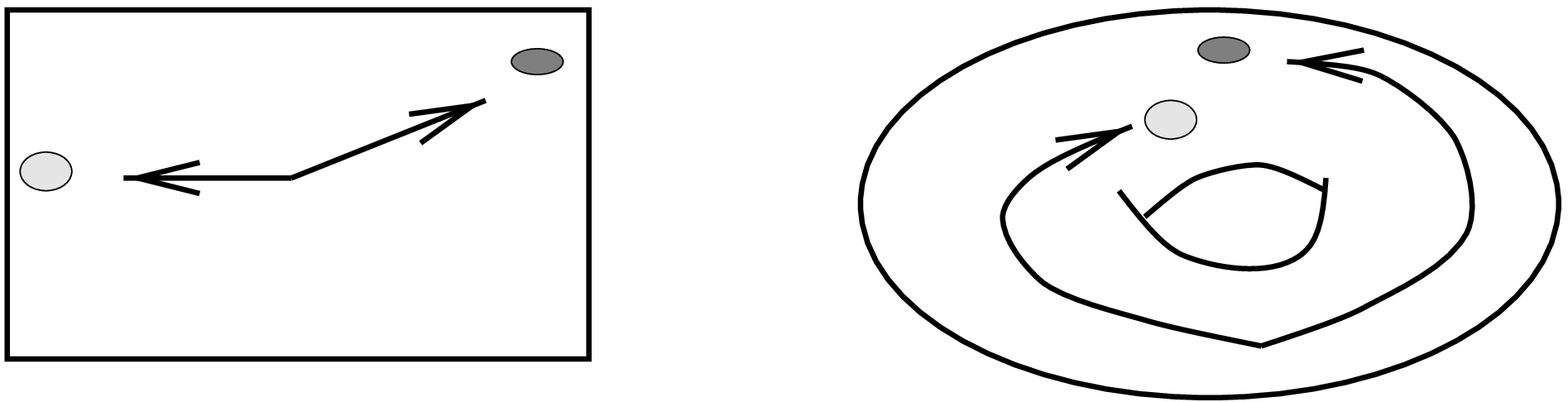,width=2.in}}
\vskip 5truept
\caption{}
\label{close}
\end{figure}

The size of a spot can be estimated at the Silk damping 
scale below which 
fluctuations have smoothed some.
The angular size of these spots are too small for \cb to have detected
but will be visible to the high resolution MAP and {\it Planck
Surveyor}.

Although the search for pattern formation in correlated maps is model
independent, a zoo illustrating the variety of structures compact
manifolds produce can be built with some simple approximations.
The correlation between two points 
on a compact manifold 
can be estimated as the correlation they would have on the universal
cover given their minimum separation:
\begin{equation}
C_{{\cal M}}({\hat n},{\hat n}^{\prime })\approx C^U\left[ d_{{\rm min}}( 
\vec x({\hat n}),\vec x^{\prime }({\hat n}^{\prime }))\right] \ \ ,
\label{eq:approx}
\end{equation}
where $C^U$ is the correlation function on the universal cover
and $d_{{\rm min}}$ is the minimum distance between the two points in the
topological space.
The estimate is 
effectively the lowest order term in the method of images.  It is inadequate
for use in a likelihood analysis but is sufficient for predicting the types
of patterns which emerge from topological lensing.
The images of a given point out 
to order $m$ are found with the generators of the identifications as
\begin{equation}
\vec y_{k_m,..,k_i}=\prod_i^mg_{k_i}\vec x^{\prime }(\hat n^{\prime })\ \ .
\end{equation}
The image point which lands closest to $\vec x(\hat n)$ determines $d_{{\rm  
min}}$. 

\begin{figure}[tbp]
\centerline{
\psfig{file=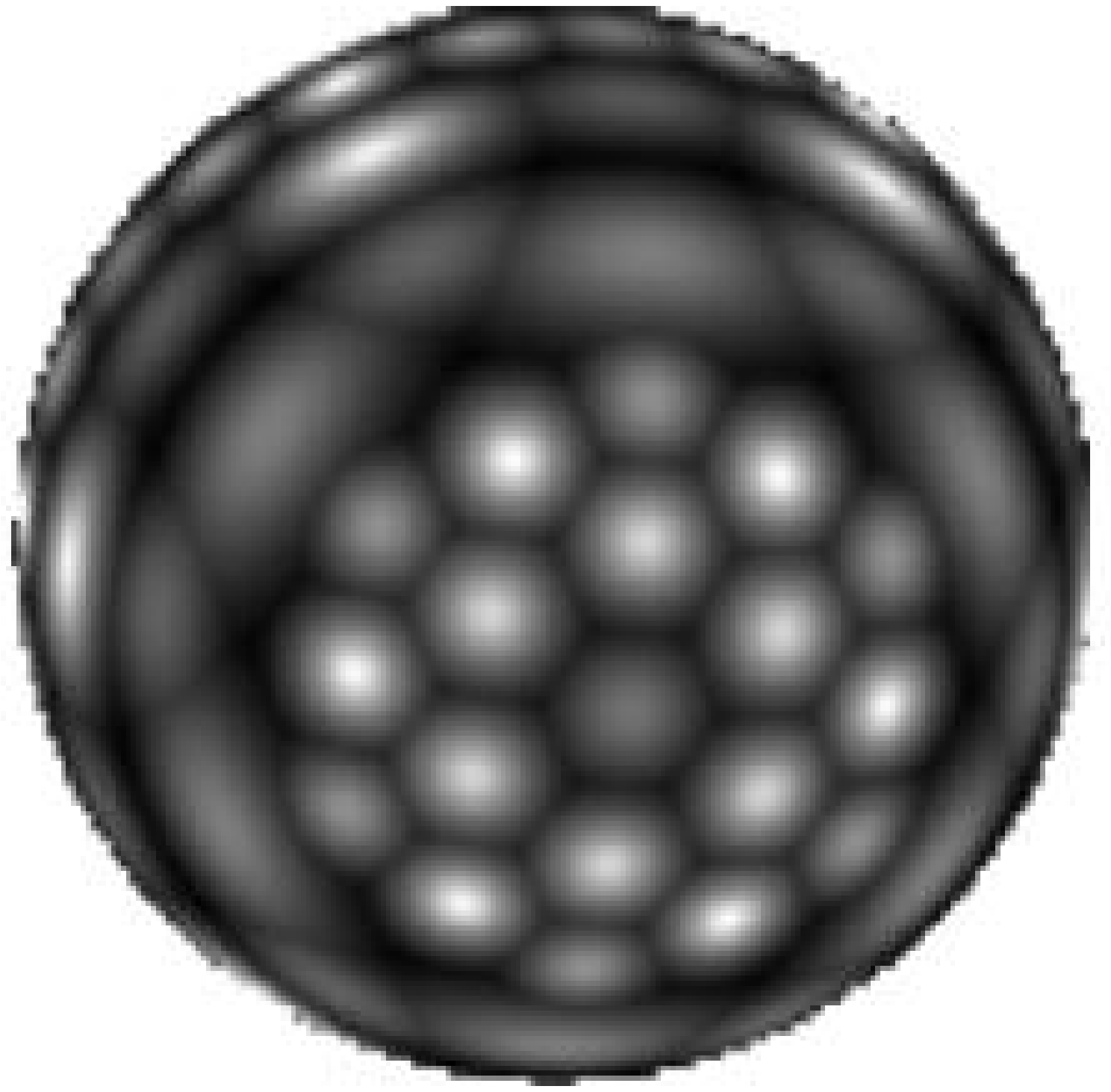,width=2.2in}
\quad\quad
\psfig{file=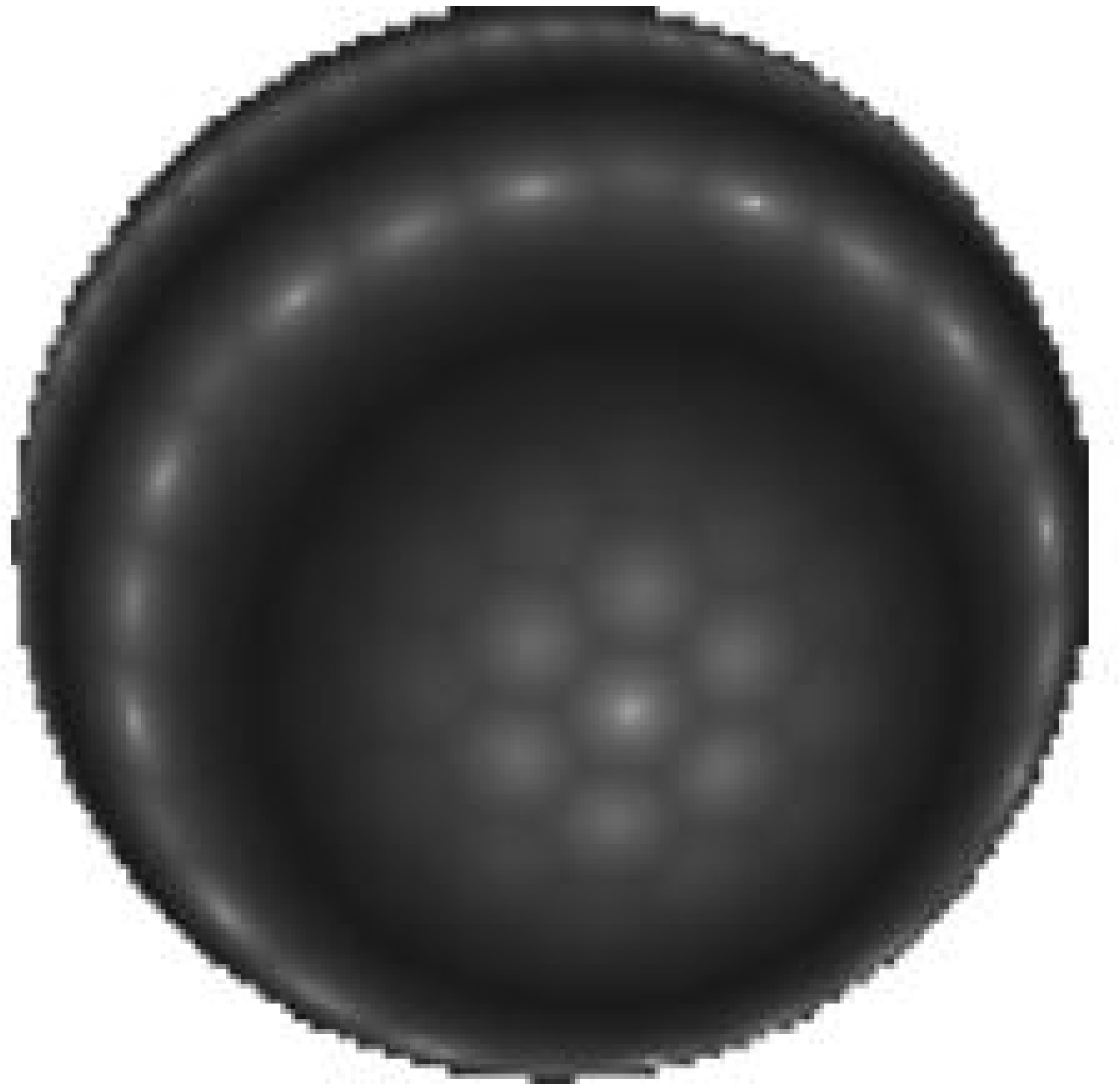,width=2.2in}
} \vskip 15truept
\caption{Left:  Orthographic projection of ${\rm A({\hat n})}$ for a hexagonal
prism with $L= 0.6 \Delta \eta$. 
Right: ${\rm A({\hat n})} $ for $\pi/3$-twisted hexagonal prism.
The length of the prism
direction
is $.24$ while $L=1$. There are circles }
\label{hexpat2}
\end{figure}

As an example, we show the antipodal map for a $2\pi/3$-twisted hexagonal
prism in fig.\ \ref{hexpat2}.
For the 
antipodal map we prefer the orthographic projection which shows the genuine
shape of the surface of last scattering
instead of the Aitoff projection customary in 
$\delta T({\hat n})/T$ maps.
Notice the clear hexagonal face drawn out by the correlated map.
Since antipody is symmetric under $\pi$, the back is a copy of the front.
For the $\pi/3$-twisted hexagon, if the space is small enough,
pairs of circles appear
in $A(\hat n)$, as
shown in the right-most panel of fig.\ \ref{hexpat2}.  These are the circles
in the sky of section \S \ref{obscirc}.

\begin{figure}[tbp]
\centerline{{\psfig{file=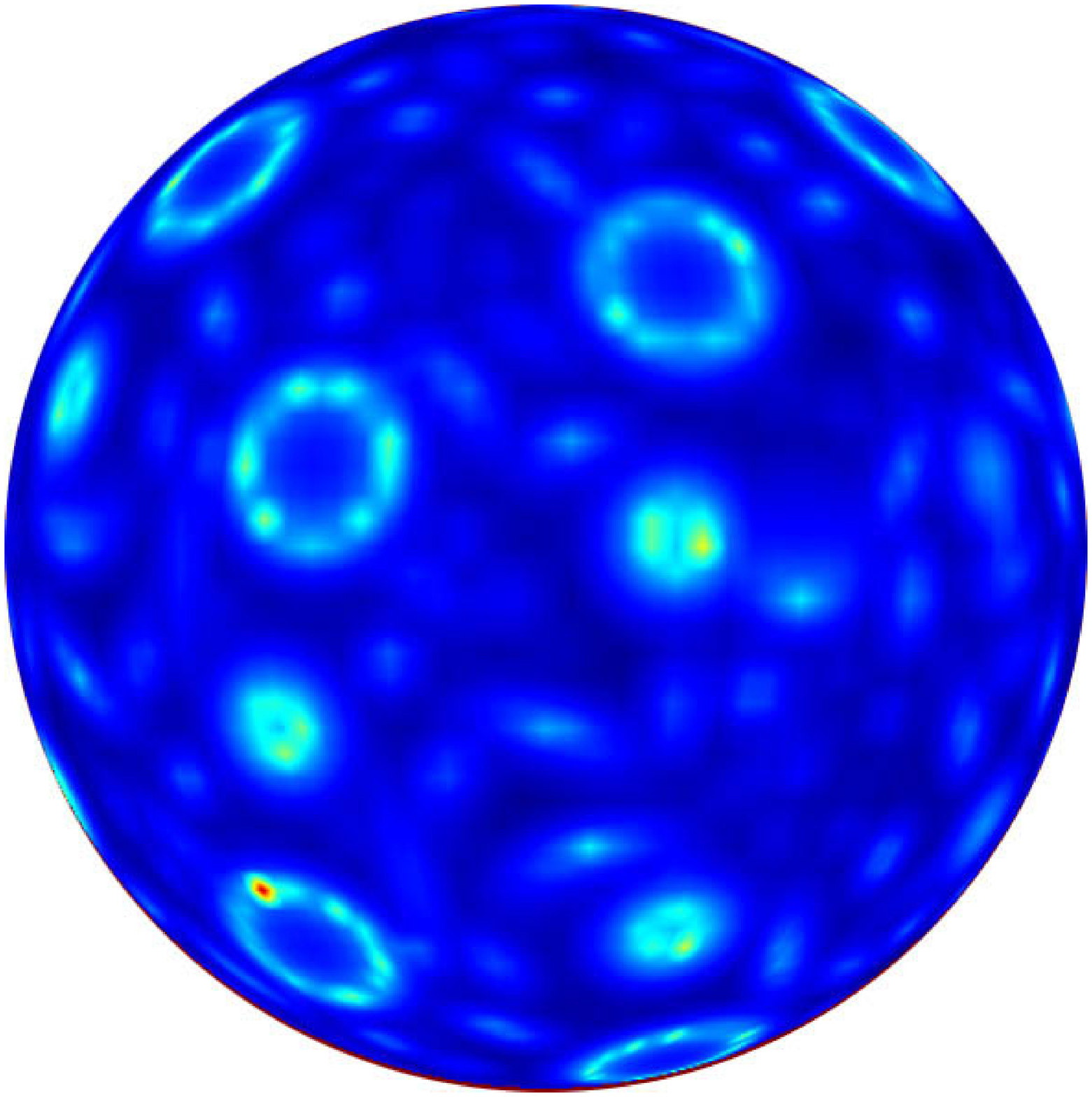,width=1.8in}}
\quad\quad
\psfig{file=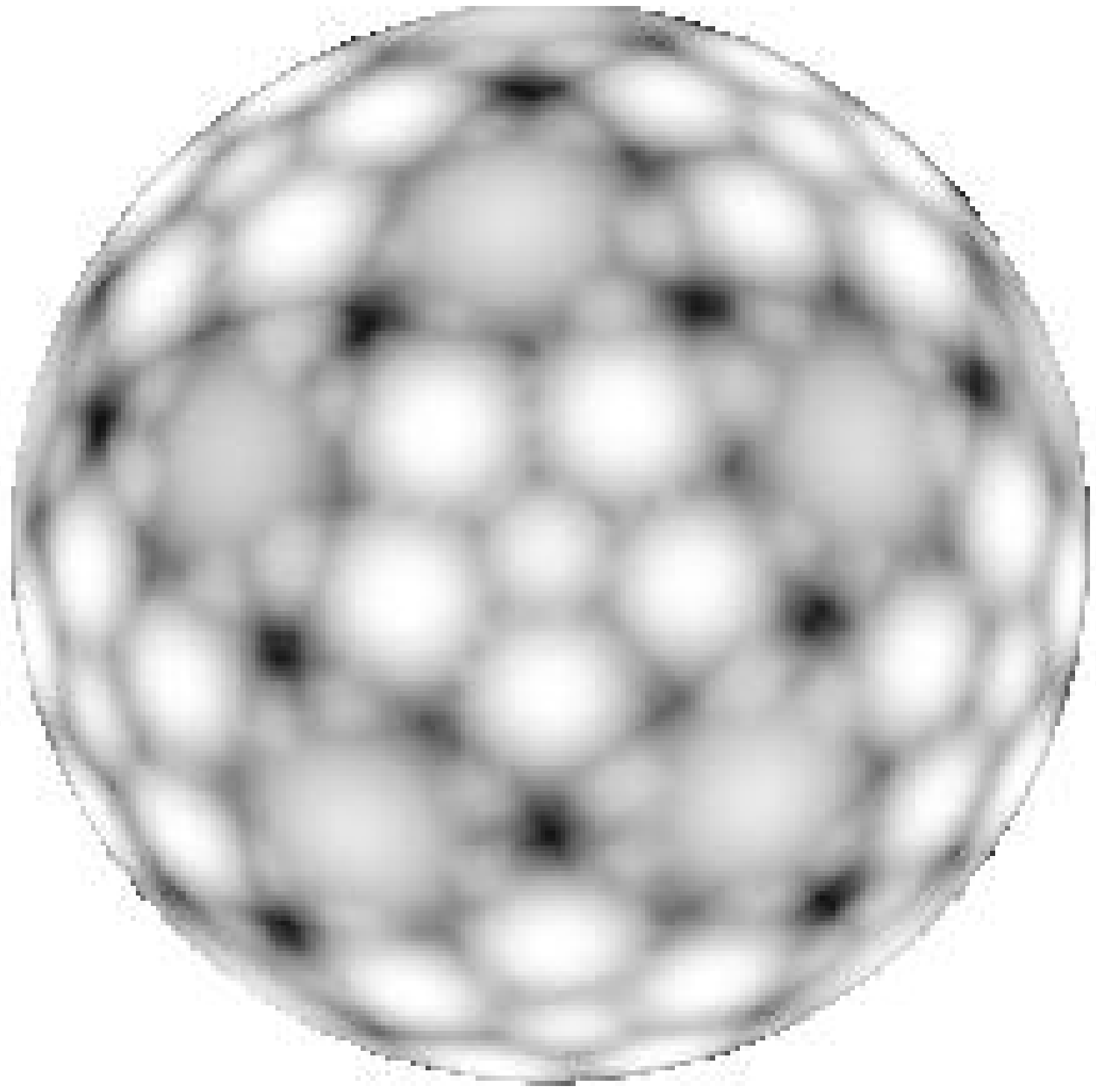,width=2.5in}}\vskip 15truept
\caption{Left: $A(\hat n)$ for the Best space with 
$\Omega=0.3$.  Right: The 
point-to-sphere correlation.}
\label{bestant}
\end{figure}

\begin{figure}[tbp]
\centerline{{\psfig{file=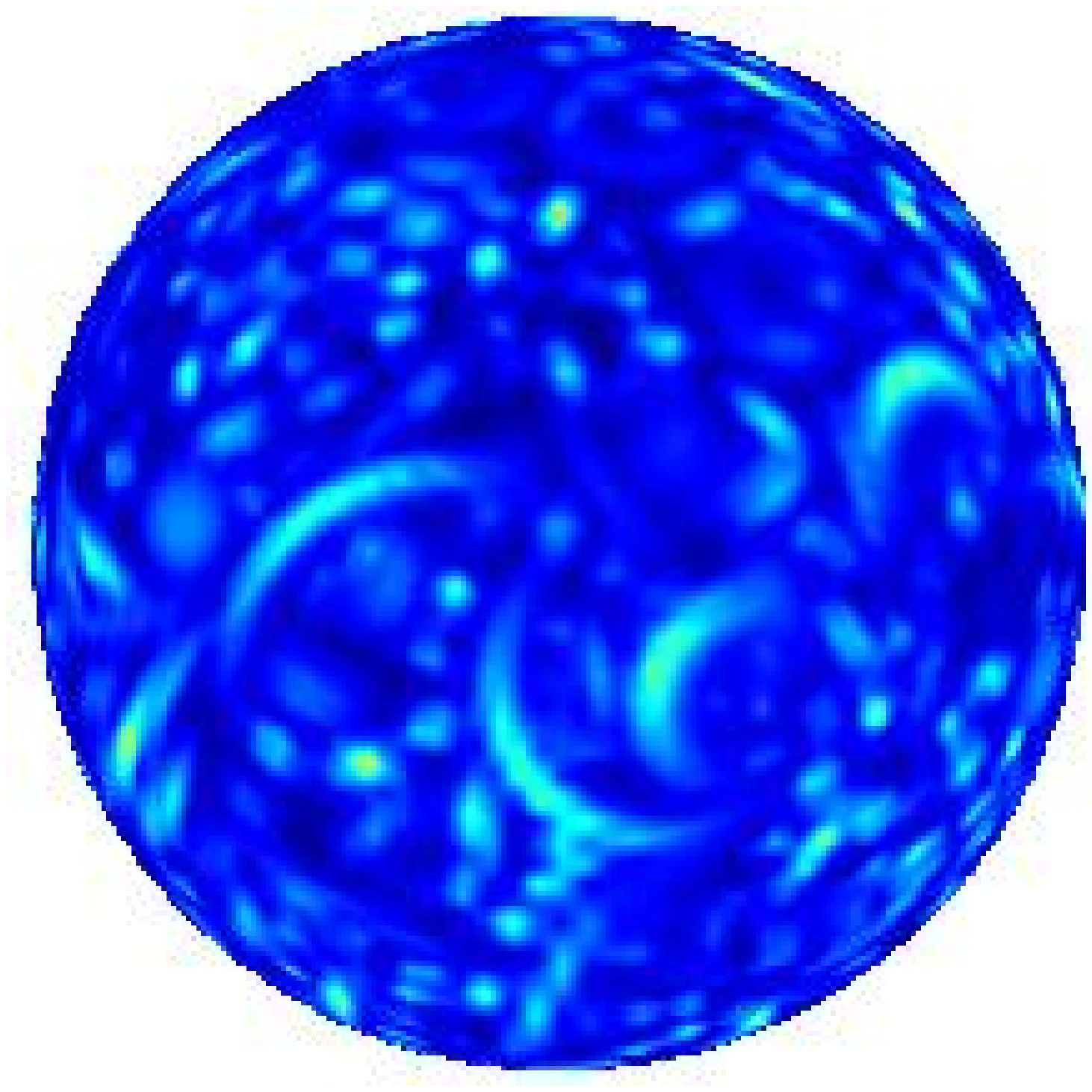,width=2.5in}}}
\vskip 15truept
\caption{Antipody in the Thurston manifold.}
\label{thursant}
\end{figure}

A compact hyperbolic space shows distinct patterns.  
The compact icosahedron known as
the Best space after the mathematician who identified the manifold
provides the best testing ground \cite{best}.
The map of $A(\hat n)$ is shown in the left of
fig.\ \ref{bestant}.  Antipody outlines
pairs of identified triangular faces and also locates
circles.  Clearly a symmetry group for the Best space is located in this
map.
Another correlation function is also shown which compares one point on the
SLS to the rest of the sphere.  The point selected is a copy of the origin
and so reflects the most symmetric observation of the fundamental domain.
Another example is given by the antipodal map for the Thurston space in 
fig.\ \ref{thursant}.
Some of the correlated features such as the arcs 
in fig.\ \ref{thursant} may be secondary correlations
and it is not clear they will ever be bright enough to be observed.

\begin{figure}[tbp]
\centerline{\psfig{file=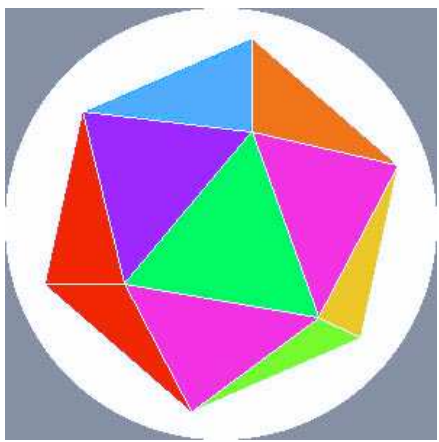,width=2.25in}}\vskip 15truept
\caption{Dirichlet domain for the Best space.
}
\label{best}
\end{figure}

In fairness, it is difficult to know if any of these patterns will really
be measurable in a realistic experiment with physical complications such
as the thickness of the surface of last scatter, additional Doppler effects,
noise, fictitious correlations etc..  
To read the correlations from the future data, real space statistics will need
to be developed which handle smoothings, subtractions of low order multipoles,
noise and fictitious correlations.  While some statistics have been promoted,
a realistic approach will likely develop only when the data is actually
available.

\newpage
\section{Beyond Standard Cosmology}
\label{extradsect}

\def\theequation{\thesection.\arabic{equation}}

\setcounter{equation}{0}

\subsection{The twin paradox and compact time}

So far we have ignored the issue of time.  We have explicitly considered
spacelike hypersurfaces $\Sigma$
of constant curvature foliated by a natural
conformal time.  Compactification of these surfaces leads to the
pictures we have described without compactifying time.  However,
an observer moving on an inertial worldline which is not at rest
with respect to the cosmic expansion will perceive an identification
which mixes spacetime coordinates 
as dictated by the Lorentz transformations $\Lambda $ so that
$\bar x=\Lambda x$ with $x=(\eta,\vec x)$ comoving coordinates.
As a result, a compactification of $\Sigma$ in the comoving coordinates
of the form $(\eta,\vec x)\rightarrow (\eta,\vec x+\vec L)$ for instance
will
result is an identification of 
a time shift as measured by the non-comoving observer of
$(1-\beta^2)^{-1/2}\left (\eta -\vec \beta \cdot \vec x \right )
\rightarrow (1-\beta^2)^{-1/2}\left (\eta -\vec \beta \cdot (\vec x
+\vec L)\right )$ where $\beta$ is the velocity relative to 
$\Sigma$.

\begin{figure}
\centerline{\psfig{file=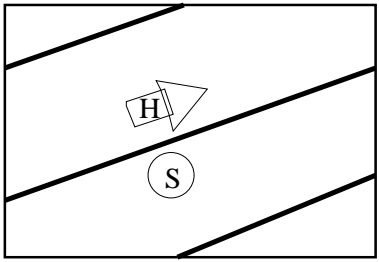,width=2.in}}
\vskip 5truept
\caption{An observer $O$ at rest with respect to the compact spacelike
hypersurface $\Sigma$ and an observer $\bar O$ moving at constant
velocity with respect to $\Sigma$ on a
periodic orbit of the torus.  The orbit
corresponds to 
$x_{\rm end}=T_yT_x^2x_{\rm start}$.
}
\label{fund2}
\end{figure}

As an explicit demonstration, consider the twin paradox
\cite{{jdbi},{jefftwin},{uzan}}.  Let $O$ be at
rest in a comoving frame where a flat spacelike hypersurface is compactified
into a torus.  Let $\bar O$ be an inertial observer on a periodic
orbit with respect to $O$
as in fig.\ \ref{fund2}.  
The periodic orbit will obey the boundary
condition
$\bar x=\Lambda x \rightarrow \Lambda \gamma x$ where $\gamma $ is 
the corresponding word.
From \S \ref{tools} we have $\gamma=T_y T_x^2$
(to include the physical time in the embedded coordinates
and in $T_x, T_y$ see Ref.\ \cite{jdbi}).
According to $\bar O$, both space and time points have been identified.
As a result it becomes impossible for $\bar O$ to synchronize her clocks
\cite{peters}.
The lack of synchronicity
will be given by the time component of $\Lambda(1-\gamma)x$ \cite{jdbi}.

Since both $O$ and $\bar O$ are inertial, by the principles of 
relativity, each should believe the other's clocks run slower and therefore
each could expect the other to be younger at their reunion 
leading to a paradox.
However, the topological identification breaks the general invariance
and selects a preferred frame, namely the frame in which the topological
identification is purely on $\Sigma$.  In that frame clocks can be
synchronized and the volume of space looks smallest.  The 
observer on the periodic orbit, though inertial, will discover that
their clocks cannot be synchronized and this additional
shift leads them to compute 
an older age for their twin.
Both agree the twin at rest with respect to $\Sigma $ is older.

This simple thought experiment emphasizes that 
compact topology selects a
preferred frame, namely the frame in which the 
universe looks smallest and in which observers can synchronize 
their clocks
\cite{peters} (see also Refs. \cite{lh}).
To generalize to curved space,
$\Lambda$ can be replaced by an appropriate diffeomorphism and 
the spacetime topology generalizes to
${\cal M}_c=R\otimes G/\Gamma$.

Finite 
spaces reverse some of Copernicus' philosophical advances by selecting
a preferred location at the center of the space, a preferred
observer at rest with respect to the compactification and a preferred
time.  While Copernicus may have removed us from the center of the universe, 
topology 
puts some observers back there.

This also raises the question of compactifying time outright.
Probably time could be compactified so that there were closed
timelike curves that were not causality violating.  Very 
restrictive possibilities would result since only events which could
repeat ad infinitum would obey the boundary conditions.
It would be hard to envision a compact time model being consistent
with the laws of thermodynamics, except perhaps on a cosmological
timescale.  The compact time scale would have to be much shorter or
much larger than biological timescales or
no children could sensibly be 
born since they would somehow have to grow young again.  
A universe could go through a big
bang and eventual big crunch only to repeat the history of the
cosmos with another big bang.  The same galaxies, stars and the same
people would be born, live and die.  Fated to repeat their
paths ad infinitum.  Quantum gravity may reset the initial conditions
at each big bang allowing new galaxies, new stars and new organisms
to form, sparing us from a relentless cosmic boredom.
The fanciful possibility of a compact time magnifies the already strange
and distinct nature of time.

\subsection{Extra dimensions}
\label{stringsection}


Topology it has been suggested is a discrete feature of space
\cite{lum}.  As such, it may be better integrated in a quantized
theory of gravity while there is no prediction for topology in 
classical relativity.  
The earliest attempts at creating a finite universe grew out
of semiclassical quantum cosmology \cite{{carlip},{gibbons},{hfq},
{hfq2}}.
It may seem intuitive that a smaller universe should be easier to 
create from nothing than a larger universe.  Therefore the probability
for creating a small, finite cosmos may be relatively high, if only
we could compute the wave function.  However, technical and conceptual
difficulties dog the semiclassical approach such as defining a measure
on the space of states, normalizing the wave function etc.
A convincing statement about the topology of the universe will very likely
require a fully quantized theory of gravity.

Interestingly, additional dimensions
have featured prominently in attempts to 
quantize gravity.  These extra dimensions are always topologically
compact and very small.
While no quantum gravity theory is yet able to predict the topology 
of space, the possibility that compact
internal dimensions will have topologically compact external
(that is, large) dimensions is certainly alluring.
Ultimately, a fundamental theory should predict the global topology
of the entire manifold whether it be $3D$ or $11D$.  In the meantime
the hierarchy between small and large dimensions remains mysterious
although some recent suggestions have created a bit of a stir
\cite{randallco}.

Kaluza-Klein theories introduced extra compact dimensions in an
attempt to unify fundamental theories of physics \cite{kk}.
Upon compactification, the radii of the small dimensions behave as scalar
and tensor field theories.  More modern string theories naturally
invoke extra dimensions in a manner reminiscent of these early
Kaluza-Klein models.
Recent fervor in string phenomenology has involved compact extra dimensions
of moderate to large size
in an attempt to explain the hierarchy problem in standard particle
physics.  The hierarchy problem questions the 
disparity in scales from the Planck mass of $10^{19}$ GeV to the 
mass of the electron at a few
eV.  A unification of fundamental theories has to 
naturally justify this span over $25$ decades of energy scales.
As in Ref.\ \cite{glennco}  we consider a spacetime with 
$4+{\cal N}$ dimensions of the form ${\cal M}=R\otimes M^3\otimes
{\cal M}^{\cal N}$ where $R$ represents time,
$M^3$ is a constant curvature Friedman-Robertson-Walker metric and 
${\cal M}^{\cal N}=G^{\cal N}/\Gamma$ is an 
${\cal N}$-dimensional compact internal space.  The 
$(4+{\cal N})$-dimensional gravitational action becomes
	\be
	A=\int d^{(4+{\cal N})}x\sqrt{-g^{(4+{\cal N})}}
	{\cal R}^{(4+{\cal N})} M_{*}^2.
	\ee
In the simplest Kaluza-Klein picture, integration over the compact
extra dimensions leads to the $(3+1)$-dimensional action we experience
	\be
	A\sim \int d^4x \sqrt{-g^{(4)}} {\cal R}^{(4)}M_{pl}^2
	+ ...
	\ee
plus additional dynamical terms
where $M_{pl}^2=M_*^{2+{\cal N}}R^{{\cal N}}$ and $R$ is the radius
of the internal dimensions.  The hierarchy between 
$M_{pl}$ and $M_*$ is large if 
$RM_*$ is large \cite{glennco}.  The disparity in energy scales 
then becomes a dynamical and geometric question.

The modern ideas inspired by string 
theory involve the localization of matter to a $3$-brane nested
in the full $(4+{\cal N})$-dimensional space.
The scale $M_*$ is expected to be $\sim $TeV as predicted
by supersymmetry.  Since ordinary matter is confined to a $3$-brane, we would
be unaware of these extra dimensions regardless of their size unless we 
try very hard to look for them.  
Some laboratory experiments are underway to probe any additional
dimensions which may be lurking there.

The topology of the compact extra dimensions is not well understood.
There are only a few requirements the internal dimensions must 
satisfy in order to break supersymmetry at a physically sensible scale.
The internal spaces fall loosely under the
broad category of Calabi-Yau manifolds.
Nearly all topologies investigated so far are 
modeled on flat geometries.  An outgrowth of cosmic topology has been the
suggestion that these internal dimensions be compact and negatively 
curved \cite{glennco}.  
The hyperbolic internal spaces have some advantages over flat space 
including a less demanding tuning of geometric parameters
and a suppression of 
astrophysically harmful graviton modes from the extra dimensions
\cite{glennco}.
Another advantage of the compact hyperbolic extra dimensions
over compact flat extra dimensions may be that chaotic mixing on finite
hyperbolic manifolds could explain the smoothness and flatness of the
large dimensions as argued in Ref.\ \cite{sst}.  We briefly discuss
chaos in the next section.

The profound connection between small dimensions and large
have only begun to be forged.  As many times in the past,
cosmology provides a unique terrain in which to test fundamental 
theories \cite{hormar}.

\subsection{Chaos}
\label{chaossect}

\def\theequation{\thesection.\arabic{equation}}

\setcounter{equation}{0}

We have touched upon the chaotic motions of particles on a compact 
hyperbolic space.  Chaos refers to the thorough mixing of orbits 
which show an extreme sensitivity to initial conditions.
Geodesics deviate on a surface of negative curvature as mentioned
in \S \ref{obshyp} so that they are extremely sensitive to 
initial conditions.  The motion becomes fully ergodic upon compactification
of the surface.  Formally, chaos in these Hamiltonian systems
means the geodesics equations are nonintegrable; there is no smooth
analytic function which can interpolate between orbits with different
initial conditions.  

Despite the resistance of chaotic systems to conventional 
integration methods a great deal about the structure of phase
space can be determined.  Much like thermodynamics, chaotic systems
can be understood in terms of a set of states dense in the phase space.
For chaos this set is provided by the collection of unstable
periodic orbits which grow exponentially with length.
All aperiodic orbits can be understood
in terms of this special subset.
The periodic orbits pack themselves
into a fractal set in order to fit within the finite phase space.
Recall the fractional dimension maintains the set
at a finite volume but allows for an infinite area.  In this way
fractals try to maximize the information content, so to speak,
while minimizing the volume.  The explicit fractal structure
on a compact octagon was isolated in Ref.\ \cite{jdbi}

Pursuing the analogy with thermodynamics, 
entropy can be defined as

	\be
	h(\mu)=\lim_{k\rightarrow \infty}{1\over k}
	\sum_i^{N(k)}\mu\ln(1/\mu(k))
	\label{genen}
	\ee
The calculation goes like this.  Draw a fundamental domain.  Now use
the generators $g_1...g_n$ to tile $\ah$
with copies.  The measure at order $k$ is then defined as 
the fraction of the total volume where each tile has
the same volume:
	\be
	\mu(k)={1\over N(k)}
	\ee
and $N(k)$ is the number of unique tiles.
The sum 
in eqn.\ (\ref{genen}) reduces
to 
	\be
	h(\mu)= \lim_{k\rightarrow \infty}{1\over k} \ln(N(k))
	=H_T.
	\ee
This is equivalently the topological
entropy, a symbolic entropy which counts the 
number of accessible states; that is, closed loops or equivalently 
tiles in the tessellation.

Recall that
periodic orbits can then be counted symbolically with the homotopy group.
We do not distinguish between loops
of varying lengths if they are homotopic.
The spectrum of periodic orbits so defined are therefore a 
topological feature.
The number of unique words that can be built of length 
$k$ out of a $n$-letter alphabet with $r$ relations 
is equivalent to the number of neighbors at order $k$ in the 
tiling. 
Not all
of these neighbors will be unique.  The repeats are accounted for by
the relations and we know that the number of neighbors at order $k$
is bounded by
	\be 
	(2n)k\ge	N(k)\le (2n)^k
	\label{Nk}
	\ee
the lower bound being simple spaces such as the flat nonchaotic torus and the
upper bound the unpruned maximally chaotic case.
The topological entropy of the torus is zero while the entropy of an 
unpruned hyperbolic space is $H_T=\ln(2n)$.  The compact octagon was
shown to have a topological entropy of $H_T=\ln 7$ \cite{jdbi}.

Another kind of entropy, the metric entropy introduced in 
\S \ref{obshyp}, 
describes how quickly mixing takes place in a chaotic
system and can be expressed as the sum of the positive Lyapunov
exponents: $h=\sum \lambda $.

This is just a taste of a rich area in dynamical systems theory 
and the reader is referred to the many excellent texts on the 
subject \cite{{ott},{gutz}}.

Although cosmologists gingerly avoid this complex feature of the 
dynamics, chaos does have profound and unavoidable consequences.
One of the original motivations for a compact topology exploited the chaotic
motions.  The chaotic mixing could lead to a dilution of initial
anisotropies leading to the symmetric universe we observe today.
These initial attempts failed since the space could not 
be made small enough to allow for sufficient mixing and still be the
large cosmos we observe today \cite{ellis}.
Variants on this idea fuse the chaotic mixing with an 
$\Omega<1$ inflation model where an early episode of chaotic mixing 
provides the moderate initial conditions needed to 
permit a subsequent inflationary phase to succeed \cite{css3}.

We have already encountered the 
nonintegrability in trying to find eigenmodes on compact hyperbolic
cosmologies.  For the initial spectrum of fluctuations, researchers
continue to use the assumption of a flat Gaussian spectrum as motivated by 
inflation.  Still, since the consistency of inflation and an observably small
topology is shaky, a more consistent approach would be to 
examine a distribution of initial fluctuations on a compact space
in the absence of inflation.
The numerical results of \S \ref{directhyp} do suggest
that a Gaussian flat spectrum is in fact natural on
compact hyperbolic spaces, even in the absence of inflation.
While this work is very suggestive, a primordial quantum system has
never been very thoroughly thought through.
Because of the importance of this issue we take a moment to discuss
quantum chaos.
The quantum system is 
relevant to our discussion regardless since the expansion of fluctuations 
will be analogous
to the expansion of the semiclassical wave function.  The stationary 
Schr\"odinger equation is the usual Helmholtz eqn (\ref{helm}).

The quantization
of the chaotic system is still not fully understood but interesting 
features have been conjectured and confirmed.  
Reminiscent of the Feynman path integral approach, the wave function can 
be thought of as the sum of classical trajectories.  Since the classical 
trajectories chaotically mix in phase space, it has been conjectured that
the wave function would be a random Gaussian function of the eigenvalue
$q$ with a spectrum given by the Wigner distribution function.  
Wigner's function is an attempt to generalize 
Boltzmann's formula for the 
classical distribution of a statistical system 
to a quantum system \cite{wigner}.  
It provides a description
of general attributes of wave functions \cite{gutz} and shows how the
function tends to distribute itself over classical regions of phase space.
Notice
the similarity between this suggestion and the method-of-images as well
as the numerical results of Ref \cite{{inoue},{cornsperg}}.

The nature of the discrete spectrum of energy levels can also be related
to the classical chaos.
As already argued in the numerical work on the eigenmodes of a compact space,
the number of eigenstates can be related to the volume of the space through
Weyl's asymptotic formula.  Additional features, such as the spacing between
energy levels, can be directly associated with the underlying chaos.

Despite this conjecture of randomness, 
distinct remnants of the underlying chaotic
dynamics have been found in the form of scars \cite{heller}.  Scars are 
regions of enhanced intensity in high $k$ states along periodic orbits.
It is hard to imagine the very high $k$ modes having cosmological significance
since it is the low $k$ modes which probe the largest cosmological distances.
Scaring is expected to be less prominent in low $k$ modes since the width
of the enhancement will be correspondingly more diffuse.  
Although diffuse, scars on large scales could provide the small
catalyst needed to order structure on large-scale.
Scarring on a $2D$ double doughnut was investigated in 
references \cite{aurichsteiner}.  The underlying fractal
structure of periodic orbits on the double
doughnut was studied in detail in Ref.\ \cite{jdbi}.  There it was suggested
that the filamentary structure we observe in the distribution of galaxies
may be a consequence of this slight enhancement of the seeds of 
structure formation.

For other important chaos articles see Refs.\ \cite{{ellistavakol},  
{detchaos},{gott},{lockhart}}


\newpage
\section{Summary}

The creation of the universe is still not well understood.
Clarity on the earliest moments will likely come only with a 
fully functional quantum theory of gravity.  In the meantime,
we know that space is curved and evolving and most also possess
{\it some} topology. If the curvature of the universe falls within
the observable horizon, then topology may also.
The CMB provides the deepest probe of the universe on the largest-scales
and we have reviewed the many ideas on how to extract the topology
of space from maps of the
microwave sky.  The methods fall into two primary categories:
direct statistical methods and generic geometric methods.
The salient features in CMB maps which reveal topology are
(1) a discretization of the sizes of hot and cold spots,
(2) a cutoff in the spectrum for wavelengths too big to fit within
the finite space
and (3) an anisotropic and inhomogeneous distribution of 
correlations corresponding to repeated ghost images of the same
spots.
Data from the future satellite missions MAP and 
{\it Planck Surveyor} are needed to determine if the CMB does in fact
encode such features.  When the new high resolution maps are in hand,
we will have to face the potentially prohibitive difficulty of foreground
contaminations, the thickness of the surface last scattering, and fortuitous
correlations.
If we are lucky enough to surmount these observational trials, we 
may be able to see the entire shape of space.

Still, as many fear, the topology scale may naturally be far beyond 
the observable universe.  If this is the case, we can turn to  
physics on the smallest scales to learn something about what we will never
see on the largest scales.  If an ultimate theory of gravity beyond
Einstein's is able to predict the geometry and topology of small
extra dimensions, there is every reason to hope we will learn, if 
only indirectly, the geometry and topology of the large dimensions.

\section*{Acknowledgements}
I appreciate the generous contributions of suggestions, ideas, 
and figures from the topology/cosmology
community.
I am grateful to R. Aurich, J.D. Barrow,
J.R. Bond, N.J. Cornish, H. Fagundes, K.T. Inoue,
J-P. Luminet, B. Roukema, D. Spergel and G. Starkman
for useful discussions.
I am especially grateful to Jeff Weeks for his mathematical insight
and his direct contributions to this review.  I thank
Theoretical Physics Group at 
Imperial College for their hospitality.
This work is supported by PPARC.

\newpage
\appendix
\section{Appendix:  Representations of Hyperbolic space}
\label{AppendixA}

Because of the recent emphasis on hyperbolic models, we include a
section here on other useful coordinate models for manipulating
$\ah $.
We have already seen two different coordinate systems
for the metric eqns. (\ref{stanmet}) and (\ref{cosmet}).
Another important coordinate system comes from embedding
$\ah $ in the $(3+1)$-Minkowski space as discusssed in \S \ref{ovgeom}.
The Minkowski metric is
	\be
	d\sigma^2=-dx_0^2+dx_1^2+dx_2^2+dx_3^2
	\ee
The $3D$-hypersurface is constrained to have pseudoradius
	\be
	-x_0^2+x_1^2+x_2^2+x_3^2=1
	\ee
and the curvature radius has been taken to $1$.  In Minkowski coordinates
the generators $\gamma \in \Gamma$ take the convenient
form of $O(4)$, orthogonal
$4\times 4$ matrices.  
These coordinates can be related to those of eqn.\ (\ref{cosmet})
with the transformation
	\ba
	x^\mu &=& \pmatrix{x^0\cr x^1\cr x^2\cr x^3}
	=\pmatrix{\cosh \chi \cr
	\sinh \chi \sin\theta \cos \phi \cr
	\sinh \chi \sin\theta \sin \phi \cr
	 \sinh\chi \cos \theta }.
	\ea

There are several other useful representations of the hyperbolic plane
including the Poincar\'e 
ball with
	\be
	\vec x=\tanh(r/2)\hat n
	\ee
with $\hat n$ the usual unit vector in spherical coordinates.
The Poincar\'e ball model maps $\ah$ to the 
open ball $\{\vx  \in \eu | \vec x\cdot \vec x < 1\}$
with 	
	\be
	d\sigma^2={4d\vx\cdot d\vx\over \left(1-{\vx\cdot \vx}\right )}.
	\ee
It is useful to know the geodesic distance is
	\be
	d(\vx,\vxp)={\rm arccosh}\left [1+{2\left |\vx-\vxp\right |
	\over (1-|\vx|^2)(1-|\vxp|^2)}\right ].
	\ee
All geodesics intersect the boundary orthogonally and are therefore semicircles
or straight lines which are the diameters of the ball.

There is also the upper-half space representation
 $\{\vx  \in \eu | x_3>0\}$,
	\ba
	d\sigma^2 &=& {\left (dx^2+dy^2 dz^2\right )\over z^2}\nonumber \\
	&\equiv &	d\rho^2+e^{-2\rho}\left (dx^2+dy^2\right )
	\label{upperhalf}
	\ea
with $e^\rho=z$.  This coordinate system is particularly useful 
for cusped manifolds which are discussed in \S \ref{obscusps}.

Finally there is the three-dimensional Klein model with coordinates
	\be
	\vec x=\tanh\chi\hat n
	\ee
which is used often in constructing the Dirichlet domain in three-dimensions
and mapping the periodic geodesics.  Geodesics are mapped into 
straight lines in this model.


\addcontentsline{toc}{section}{Acknowledgements}


\newpage

\addcontentsline{toc}{section}{References}

\begin{thebibliography}{999}

\bibitem{lum} M. Lachieze-Rey and J. -P. Luminet, Phys. Rep.
{\bf 254}, 136, (1995).

\bibitem{css_cqg} N.J.Cornish, D.N.Spergel and G.D.Starkman,
Class. Quant. Grav. {\bf 15} 2657 (1998).

\bibitem{galreviews1} 
J.-P. Uzan, R. Lehoucq, and J.-P. Luminet,
Proceedings of the XIXth Texas meeting, Paris 14-18 December 1998,
Eds. E. Aubourg, T. Montmerle, J. Paul and P. Peter, article-no: 04.25.

\bibitem{galreviews2} J.-P. Luminet and B.F. Roukema,
Proceedings of Cosmology School held at Cargese, Corsica, August 1998
astro-ph/9901364.

\bibitem{roukreview}
B.F. Roukema, Bull.Astr.Soc.India, 28, 483 (astro-ph/0010185) (2000);
B.F. Roukema,  Marcel Grossmann IX Conference, eds Ruffini  et al.  (2001)
(astro-ph/0010189).

\bibitem{ghosts1} L. Z. Fang and H. Sato, Comm. Theor. Phys,
(China) {\bf 2}, 1055, (1983); H. V. Fagundes, {\em Astrophys. J.}
{\bf 291}, 450, (1985); {\it ibid.}, Astrophys. J. {\bf 338 },
618, (1989); H. V. Fagundes  and U. F. Wichoski,  Astrophys. J. Lett.
{\bf 322}, L5, (1987).

\bibitem{ghosts2} H. V. Fagundes, Phys. Rev. Lett. {\bf 70}, 1579,
(1993); R. Lehoucq, M. Lachiese-Rey and J. -P. Luminet,
Astron. Astrophys. {\bf 313}, 339, (1996); B. F. Roukema and
A. C. Edge, Mon. Not. Roy. Astron. Soc. {\bf 292}, 105, (1997);
B. F. Roukema and V. Blanloeil, Class. Quantum Grav. {\bf 15}, 2645,
(1998);  B.F. Roukema,
MNRAS, 283, 1147 (1996);
B.F. Roukema \& S. Bajtlik, MNRAS, 308, 309 (1999).



\bibitem{thursweeks} W.P.Thurston and J.R.Weeks, Sci. Am. July 94
(1984).

\bibitem{glennpop} J.P.Luminet, G.D.Starkman and J.R. Weeks, Sci. Am., April
(1999).

\bibitem{cqg}  The entire issue of
Class. Quant. Grav. {\bf 15} 2589 (1998).

\bibitem{ctp98}   V. Blanl{\oe}il, B.F. Roukema, editors, 
 Proceedings of the Cosmological Topology in Paris 1998 meeting,
 astro-ph/0010170 (2000). 


\bibitem{css1} N.J.Cornish, D.N.Spergel and G.Starkman, gr-qc/9602039.

\bibitem{css2} N.J.Cornish, D.N.Spergel and G.Starkman,
Phys. Rev. D {\bf 57} (1998) 5982.

\bibitem{conf}   J. Levin,
E. Scannapieco and J. Silk, Class. Quant. Grav. {\bf 15}, 2689,
(1998).

\bibitem{pat}  J. Levin, E. Scannapieco, G. Gasperis, J. Silk and
J. D. Barrow,
{\em Phys. Rev. } { D} {\bf 58} (1998) article 123006.

\bibitem{imogen}  I. Heard and J. Levin,
Proceedings for ``Cosmological Topology'' in Paris 
(CTP98), astro-ph/9907166.

\bibitem{bpsI} J. R. Bond, D. Pogosyan and T. Souradeep,
Phys. Rev. D. {\bf 62} (2000) article 043005.

\bibitem{bpsII} J. R. Bond, D. Pogosyan and T. Souradeep,
Phys. Rev. D. {\bf 62} (2000) article 042006.

\bibitem{wolf} J. A. Wolf, {\it Space of Constant Curvature (5th
ed.)}, (Publish or Perish, Inc., 1994).

\bibitem{ellis} G.F.Ellis, Q.J.R.Astron. Soc. {\bf 16} 245 (1975);
G. F. R. Ellis, Gen. Rel. Grav. {\bf 2} 7(1971).

\bibitem{thurclass} 
W.P. Thurston, Bull. Am. Math. Soc. {\bf 6} (1982) 357.

\bibitem{Thurston}
W.P. Thurston, ``Three-dimensional geometry and topology''
(Ed: Silvio Levy, Princeton University Press, Princeton, N.J. 1997).

\bibitem{Bianchi} L. Bianchi, Mem. Soc. It. Della. Sc. (Dei. XL)
11, 267 (1897).

\bibitem{snappea} J. R. Weeks, {\it SnapPea: A computer program for
creating and studying hyperbolic 3-manifolds}, Univ. of Minnesota
Geometry Center (freely available at http://www.northnet.org/weeks).

\bibitem{mp} G.D. Mostow, Ann. Math.,  Studies {\bf 78}
Princeton University Press, Princeton, 1973);
G. Prasad, Invent. Math. {\bf 21} 255 (1973).

\bibitem{weekspace} J.R. Weeks, PhD thesis, Princeton University (1985).

\bibitem{bv}  N.L. Balazs and A. Voros, {\em Phys. Rep.} {\bf 143}
(1986)
109.

\bibitem{jwp} The mathematical section was contributed largely by 
Jeff Weeks who unfortunately was unable to coauthor this paper;
J. Weeks, private communication.

\bibitem{Francis-Weeks} 
G. Francis and J. Weeks, American Mathematical Monthly  106 (1999) 393-399.

\bibitem{Shape} 
J. Weeks, The Shape of Space, Marcel Dekker Inc., 1985.

\bibitem{Tbull} W.P. Thurston, Bull. Am. Math. Soc. {\bf 6}
357 (1982).

\bibitem{pscott} 
P. Scott, London Math. Soc. {\bf 15} 401 (1983).

\bibitem{Friedmann1924}
A. Friedmann, Zeitschrift f\"ur Physik {\bf 21} 326 (1924).

\bibitem{Löbell1929} 
 F. L\"obell, Ber. d. S\"achs. Akad. d. Wiss. {\bf 83} 167 (1931).


\bibitem{SW1932?}
C. Weber and H. Seifert,  
Math. Zeitschrift {\bf 37} 237 (1933).


\bibitem{Kojima??}.
S. Kojima, Isometry Transformations of Hyperbolic 3-Manifolds,
Topology Appl., {\bf 29} 297 (1988).

\bibitem{toco}  Miller et al. (TOCO experiment), ApJ 524, L1 (1999).

\bibitem{exper}  P. de Bernardis, et. al, astro-ph/0011469;
J.R. Bond, et. al., astro-ph/0011378; A.H. Jaffe, et.al. astro-ph/0007333;
S. Hanany, et. al. astro-ph/0005123.

\bibitem{vilenkin}
A. Vilenkin, Phys. Rev. D {\bf 27}{2848}(1983).

\bibitem{linde} 
A.D. Linde, Mod. Phys. Lett.{\bf A1}{81} (1986);
A.D. Linde, Physics Letters {\bf 175B}{395}(1986).

\bibitem{guth0} A.H. Guth, Phys. Rev. D. {\bf 23} 347 (1981).

\bibitem{guth1} A.H. Guth, Phys. Rep. 333 (2000) 555.

\bibitem{topstrings} C. Contaldi, astro-ph/0005115.

\bibitem{sokolov} I. Y. Sokolov, JETP Lett. {\bf 57}, 617, (1993).

\bibitem{fagundes1} D. Muller, H.V. Fagundes, and R. Opher 
{\it in press} Phys. Rev. D
gr-qc/0103014.

\bibitem{css3}   N.J.Cornish, D.N.Spergel and G.Starkman,
Phys. Rev. Lett. {\bf 77} (1996) 215.

\bibitem{sst} G.D. Starkman, D. Stojkovic, and M. Trodden,
hep-th/0106143 (2001).

\bibitem{bark} J.D. Barrow and H. Kodama, 
Class. Quant. Grav. {\bf 18} 1753 (2001); J.D. Barrow and H. Kodama,
gr-qc/0105049 (2001).

\bibitem{mkb} V.F. Mukhanov, H.A. Feldman, R.H. Brandenberger, Phys. Re;. {\bf 215}, 203 (1992).

\bibitem{hu} W. Hu, in ``The Universe at High-z, Large Scale Structure and 
the Cosmic Microwave Background'', eds. E. Martinez-Gonzalez and J.L. Sanz
(Springer Verlag) 1995.

\bibitem{ll} A.R. Liddle and D.H. Lyth, Phys. Rep. {\bf 231}, 1 (1993).

\bibitem{whu} M. White and W. Hu, Astronomy and Astrophysics .


\bibitem{waldbk} R. Wald, {\it General Relativity}.

\bibitem{mtw} Misner, Thorne, and Wheeler, {\it Gravitation}.

\bibitem{swein} S. Weinberg, {\it Gravitation and Cosmology}.

\bibitem{kolbturner} E.W. Kolb and M.S. Turner,
{\it The Early Universe} (Addison-Wesley, Reading, 1994).

\bibitem{peebk} P.J.E. Peebles, {\it The Large-Scale Structure of the 
Universe} (Princeton University Press, 1980).

\bibitem{sw} R.K. Sachs and A.M. Wolfe, Apj {\bf 147}, 73 (1967).

\bibitem{lb_fractals} J.Levin and J.D.Barrow, Class. Quantum Grav.
{\bf 17} L61
(2000).

\bibitem{newsphere}  E. Gausmann, R. Lehoucq, J.-P. Luminet, J.-P. Uzan,
and J. Weeks, gr-qc/0106033.

\bibitem{open} A. Dekel, D. Burstein and S. D. M. White, 1996, in
{\it Critical Dialogues in Cosmology}, ed. N.Turok, World Scientific;
N. Bahcall
and X. Fan, {\it preprint} (astro-ph/9804082).

\bibitem{inoueprog} K.T.Inoue, 
Progress of Theoretical Physics, {\bf 106} 39 (2001). 

\bibitem{ktinoue} K.T.Inoue, Phys. Rev. {\bf D} 103001, 1-15 (2000).

\bibitem{saul} S. Perlmutter, et. al., Ap. J. {\bf 517} 565 (1999).

\bibitem{balbi} A. Balbi, et. al., astro-ph/0005124.

\bibitem{addr} B.F. Roukema, G.A. Mamon, S. Bajtlik, submitted Astron.\&
Astroph., astro-ph/0106135 (2001).

\bibitem{bennet} C.L. Bennett, et al. 
ApJ {\bf 391} (1992) 466.

\bibitem{smoot}  G.F. Smoot et al., Ap. J. {\bf 360} 685 (1990);
G.F. Smoot et al., Ap. J. {\bf 396} L1 (1992).

\bibitem{kris} K. Gorski, 
Proc. Moriond XVI, 
ed. F.R. Bouchet et. al. (Gif-Sur-Yvette:  Editions Fronti\`ers) (1997).

\bibitem{teg} M. Tegmark {\em Phys. Rev.}
{\bf D 55} (1997) 5895.

\bibitem{jaffe} J.R. Bond, A. Jaffe and L. Knox,
preprint CfPA-97-TH-11; {\it ibid.} in preparation;
J.R. Bond and A. Jaffe,
Proc. Moriond XVI, 
ed. F.R. Bouchet et. al. (Gif-Sur-Yvette:  Editions Fronti\`ers) (1997).

\bibitem{bondef}  J.R. Bond and G.P. Efstathiou, MNRAS {\bf 226} 655 (1987).

\bibitem{lss}J. Levin, E. Scannapieco, and J. Silk, Phys. Rev.
D {\bf 58} (1998) article 103516.

\bibitem{slsi}  E. Scannapieco, J. Levin, and J. Silk, MNRAS
 303 (1999) 797.

\bibitem{sss}  D. Stevens, D. Scott and J. Silk, {\em Phys. Rev. Lett.} {\bf 
71} (1993) 20.


\bibitem{deO1}  A. de Oliveira-Costa and G.F. Smoot, Ap. J. {\bf 448}
447 (1995).

\bibitem{cheeger} J. Cheeger, in {\it Problems in Analysis, (A
Symposium in honor of S. Bochner)}, (Princeton University Press,
1970).

\bibitem{buser} P. Buser, Proc. Symp. Pure Math. {\bf 36}, 29, (1980).

\bibitem{workshop} Entire Volume of
Class. Quant. Grav. {\bf 15} (1998).


\bibitem{zel73} Ya B. Zel'dovich, Comm. astrophys. Space Sci.
{\bf 5} 169 (1973).

\bibitem{fanghoujun} L.Z. Fang and M. Houjun, Mod. Phys. Lett. A
{\bf 2} 229 (1987).

\bibitem{deO2}  A. de Oliveira Costa, G. F. Smoot and A. A. Starobinsky,
Astrophys. J. {\bf 468}, 457 (1996).

\bibitem{star} A. A. Starobinsky, JETP Lett. {\bf 57}, 622 (1993).

\bibitem{fang} L.Z. Fang, Mod. Phys. Lett. A {\bf 8} 2615 (1993).

\bibitem{boudcirc} B.F. Roukema, MNRAS, 312, 712 (2000);
B.F. Roukema, Class. Quant. Grav {\bf 17} 3951 (2000).

\bibitem{et} G.F.R. Ellis and R. Tavakol, in {\it Deterministic
Chaos in General Relativity}, Eds. D. Hobbill, A. Burd, and A. Coley,
(Plenum Press, New York, 1994).

\bibitem{sinai} Y.G. Sinai, Sov. Math. Dok., {\bf 1} 335 (1960).

\bibitem{aurichsteiner}  
R. Aurich and F. Steiner, Physica D {\bf 39}, 169,
(1989); R.Aurich and F.Steiner, Physica D {\bf 64} 185 (1993).

\bibitem{inouetrace}
K.T.Inoue,
Classical and Quantum Gravity, {\bf 18} 629 (2001),
math-ph/0011012.

\bibitem{gutz}  M.C. Gutzwiller, {\it Chaos in Classical and Quantum
Mechanics} (Springer-Verlag New York Inc, New York, 1990);
M.C. Gutzwiller, {\it J. Math. Phys.}, {\bf 12}
(1971) 343.

\bibitem{ott} E. Ott, {\em Chaos in dynamical systems}, (CUP, 
Cambridge, 1993).

\bibitem{lyth}  D.H.Lyth and A. Woszczyna, Phys. Rev. D {\bf 52} 3338 (1995).

\bibitem{krzy} K.M. Gorski, B. Ratra, N. Sugiyama, and A.J. Banday, Ap. J.
{\bf 444} L65 (1995).


\bibitem{spergelcqg}  
D. N. Spergel, Class. Quant. Grav. {\bf 15}, 2589, (1998).


\bibitem{lobell} F. L\"obell, Ber, S\"achs. Akad. Wiss. Leipzig {\bf 83} (1931).

\bibitem{seifertweber}  H.Seifert and C.Weber, Math. Z. {\bf 37} 237 (1933).


\bibitem{sos}  D. D. Sokolov and A. A. Starobinskii, {\em Sov. Astron. }{\bf 
19} (1976) 629.

\bibitem{lbbs}  J. Levin, J.D. Barrow, E.F. Bunn and J. Silk,
{\em Phys. Rev. Lett.} {\bf 79} (1997) 974 .



\bibitem{olsonstark} D. Olson and G.D. Starkman,
Class. Quant. Grav. {\bf 17} 3093 (2001).

\bibitem{inoue} K.T.Inoue, Class. Quantum Grav. {\bf 16} 3071 (1999).


\bibitem{inouetomita} K.T.Inoue, K. Tomita, and N.Sugiyama,
MNRAS {\bf 314} L21 (2000).

\bibitem{inouethesis} K.T.Inoue, Ph.D. Thesis, astro-ph 0103158.

\bibitem{cornsperg} N.J.Cornish and D.N.Spergel, math.DG/9906017;
N.J.Cornish and D.N.Spergel, Phys. Rev. D. {\bf 62}
(2000) article 087304.

\bibitem{aurich}
R. Aurich, Ap. J. {\bf 524} 497 (1999).

\bibitem{aurichmarklof} R. Aurich and J. Marklof, Physica {\bf D 92},
101 (1996).

\bibitem{berry}  M.V. Berry, {\it J. Phys. A} {\bf 12} (1977) 2083;
M.V. Berry in {\it Chaotic Behavior of Deterministic
Systems}, eds. G. Iooss, R. Hellman, and R. Stora, Les Houches
Proc. 36 (North-Holland, NY, 1981).

\bibitem{saskatoon} C.B.Netterfield, M.J.Devlin, N.Jarosik, L.A.Page
and E.J.Wollack, Astrophys. J. {\bf 474}, 47 (1997).

\bibitem{qmap} A. de Oliveira-Costa, M.J.Devlin, T.Herbit, A.D.Miller,
C.B.Netterfield, L.A.Page, and M.Tegmark, Astrophys. J. Lett.
{\bf 509}, L77, (1998).

\bibitem{as} R. Aurich and F. Steiner, MNRAS {\bf 323} 1016 (2001).

\bibitem{pen} U.-L. Pen, Astrophys. J. {\bf 498} 60 (1998).

\bibitem{bps_texas} J. R. Bond, D. Pogosyan and T. Souradeep, in: {\it
Proceedings of the XVIIIth Texas Symposium on Relativistic
Astrophysics}, ed.\ A. Olinto, J. Frieman, and D.~N. Schramm, (World
Scientific, 1997).

\bibitem{bps_moriond} J. R. Bond, D. Pogosyan and T. Souradeep, in:
{\it
Proc. of XXXIIIrd Recontre de Moriond}, ``Fundamental Parameters in
Cosmology'', Jan. 17-24, 1998, Les Arc, France.

\bibitem{bps_cwru} J. R. Bond, D. Pogosyan and T. Souradeep,
Class. Quant. Grav. {\bf 15}, 2671, (1998).


\bibitem{chavel} I. Chavel, {\it Eigenvalues in Riemannian geometry},
(Academic Press, 1984).


\bibitem{berard} P. H. Berard, {\it Spectral Geometry: Direct and
Inverse Problems}, Lec. Notes in Mathematics, {\bf 1207},
(Springer-Verlag, Berlin, 1980).

\bibitem{morehf} H.V. Fagundes, Ap. J. {\bf 470} 43 (1996).

\bibitem{morehf2} H.V. Fagundes, astro-ph/0007443.

\bibitem{cas}  L. Cay\'on and G. Smoot, Ap. J. {\bf 452} 494 (1995).

\bibitem{hw}  C.D. Hodgson and J.R.Weeks, Experim. Math. 3, 261 (1994).

\bibitem{cornweeks} N.J. Cornish and J. Weeks, Notices of the American
Mathematical Society, astro-ph/9807311.

\bibitem{roukemaedge}
B.F. Roukema and A.C. Edge, MNRAS {\bf 292}, 105 (1997).

\bibitem{roukemablanloeil}
B.F. Roukema and V. Blan loeil, Class. Quant. Grav., {\bf 15}, 2645 (1998).

\bibitem{best} L.A.Best, Can. J. Math. {\bf 23} No. 3, 451 (1971).

\bibitem{jdbi} J.D. Barrow and J. Levin, Phys. Rev. A (2001).

\bibitem{jefftwin} J. Weeks, unpublished.

\bibitem{uzan}
J.P. Uzan, J.P. Luminet, R. Lehoucq, and P. Peter, physics/0006039.

\bibitem{peters} P.C. Peters, Am. J. Phys. {\bf 51} (1983) 791;
P.C. Peters, Am. J. Phys. {\bf 54} (1986) 334.

\bibitem{lh} J.R. Lucas and P.E. Hodgson, {\it Spacetime and Electromagnetism}
(Oxford University Press: 1990) pp 76-83.

\bibitem{carlip}  S. Carlip, Class. Quant. Grav. {\bf 15}
2629 (1998) and references therein.

\bibitem{gibbons} G. W. Gibbons, Nucl. Phys. {bf B 472}, 683, (1996);
G. W. Gibbons, Class. Quantum Grav. {\bf 15}, 2605, (1998).

\bibitem{hfq} H.V. Fagundes and S.S. e Costa, General Relativity and
Gravitation {\bf 31} 863 (1999).

\bibitem{hfq2}  
S.S. e Costa and H.V. Fagundes, General Relativity and
Gravitation {\bf 33} in press (2001).

\bibitem{randallco}  L. Randall and R. Sundrum,
Phys. Rev. Lett. {\bf 83} 3370 (1999).

\bibitem{kk}  Th. Kaluza, Situngsber. Preuss. Akad. Wiss. Phys. Math. Kl.
966 (1921); O. Klein, Z. Phys. {\bf 37} 895 (1929).


\bibitem{glennco}  N. Kaloper, J. March-Russell, G.D. Starkman,
and M. Trodden,
Phys. Rev. Lett. {\bf 85} 928 (2001).


\bibitem{hormar} G. T. Horowitz and D. Marolf, J.High Energy
Phys. {\bf 9807}, 014, (1998).

\bibitem{wigner} E.P. Wigner, Phys. Rev. {\bf 40} (1932) 749.

\bibitem{heller}  E.J. Heller, {\it Phys. Rev. Lett.} {\bf 53}
(1984) 1515.


\bibitem{lockhart} C.M. Lockhart, B. Misra, I. Prigogine,
Phys. Rev. D. {\bf 25} 921 (1982).

\bibitem{gott} J.Richard Gott, III Mon. Not. R. astr. Soc. {\bf 192}
153 (1980).

\bibitem{detchaos}{\it Deterministic Chaos in General Relativity}
eds. D. Hobill et al., Plenum Press, Ne w York, 1994.

\bibitem{ellistavakol} G.F.R. Ellis and R. Tavakol, Class. Quantum Grav.  {\bf 11}
675 (1994).

\bibitem{ellisschreiber}  G.F.R. Ellis and G. Schreiber, Phys. Lett. A {\bf
115} 97 (1986).



\end{thebibliography}
\end{document}